\documentclass[a4paper,fleqn]{cas-sc}

\definecolor{cite_color}{rgb}{0.0, 0.58, 0.71}
\usepackage{natbib}
\setcitestyle{authoryear,open={},close={}} %Citation-related commands
\usepackage{hyperref}
\hypersetup{colorlinks=true,citecolor=cite_color,linkcolor=cite_color}
\usepackage{amsthm}
\usepackage{subcaption}
\usepackage{lineno}
\usepackage{xcolor}
\definecolor{db}{rgb}{0.0, 0.2, 0.7}

\renewcommand{\figurename}{Fig.}

\makeatletter
\renewcommand*{\fnum@figure}[1]{\figurename~\thefigure.}
\makeatother
\def\tsc#1{\csdef{#1}{\textsc{\lowercase{#1}}\xspace}}
\tsc{WGM}
\tsc{QE}
\tsc{EP}
\tsc{PMS}
\tsc{BEC}
\tsc{DE}
\begin{document}
\let\WriteBookmarks\relax
\def\floatpagepagefraction{1}
\let\printorcid\relax % Remove ORCID footnote

\def\textpagefraction{.001}
\shorttitle{}
\shortauthors{Sixu Li et~al.}
%\begin{frontmatter}

\title [mode = title]{Beyond 1D and oversimplified kinematics: A generic analytical framework for surrogate safety measures}

\author[1]{\textcolor{black}{Sixu Li}}[]

\credit{Conceptualization, Methodology, Writing – original draft,  Software, Writing – review \& editing}

\address[1]{Zachry Department of Civil $\&$ Environmental Engineering, Texas A$\&$M University, College Station, TX 77843, USA}

\author[1]{\textcolor{black}{Mohammad Anis}}[]
\credit{Writing – original draft,  Writing – review \& editing}

\author[1]{\textcolor{black}{Dominique Lord}}[]
\credit{Conceptualization, Writing – review \& editing, Supervision}

\author[1]{\textcolor{black}{Hao Zhang}}[]
\credit{Writing – original draft}

\address[2]{Department of Landscape Architecture $\&$ Urban Planning, Texas A$\&$M University, College Station, TX 77843, USA}

\author%
[1]
{\textcolor{black}{Yang Zhou}}
\cormark[1]
\ead{yangzhou295@tamu.edu}
\credit{Conceptualization, Methodology, Writing – review \& editing, Supervision}

\author%
[2]
{\textcolor{black}{Xinyue Ye}}
\credit{Conceptualization, Writing – review \& editing, Supervision}

\cortext[cor1]{Corresponding author}

\begin{abstract}
This paper presents a generic analytical framework tailored for surrogate safety measures (SSMs) that is versatile across various highway geometries, capable of encompassing vehicle dynamics of differing dimensionality and fidelity, and suitable for dynamic, real-world environments. The framework incorporates a generic vehicle movement model, accommodating a spectrum of scenarios with varying degrees of complexity and dimensionality, facilitating the estimation of future vehicle trajectory evolution. It establishes a generic mathematical criterion to denote potential collisions, characterized by the spatial overlap between a vehicle and any other entity. A collision risk is present if the collision criterion is met at any non-negative estimated future time point, with the minimum threshold representing the remaining time to collision. The framework's proficiency spans from conventional one-dimensional (1D) SSMs to extended multi-dimensional, high-fidelity SSMs. Its validity is corroborated through simulation experiments that assess the precision of the framework when linearization is performed on the vehicle movement model. The outcomes showcase remarkable accuracy in estimating vehicle trajectory evolution and the time remaining before potential collisions occur, comparing to high-accuracy numerical integration solutions. The necessity of higher-dimensional and higher-fidelity SSMs is highlighted through a comparison of conventional 1D SSMs and extended three-dimensional (3D) SSMs. The results showed that using 1D SSMs over 3D SSMs could be off by 300\% for Time-to-Collision (TTC) values larger than 1.5 seconds and about 20\% for TTC values below 1.5 seconds. In other words,
conventional 1D SSMs can yield highly inaccurate and unreliable results when assessing collision
proximity and substantially misjudge the count of conflicts with predefined threshold (e.g., below 1.5s). Furthermore, the framework's practical application is demonstrated through a case study that actively evaluates all potential conflicts, underscoring its effectiveness in dynamic, real-world traffic situations.
\end{abstract}

\begin{keywords}
surrogate safety measures \sep 
multi-dimension \sep
high-fidelity \sep
vehicle dynamics \sep linear/linearized control systems \sep 
active safety
\end{keywords}

\maketitle

\section{Introduction}

Highways, which are crucial components of the transportation network for road users, require robust safety measures to ensure collision-free travel (in theory or ideally). These measures are important not only for individual well-being but also for broader societal harmony. The substantial social and economic costs of road crashes underscore the urgency of addressing this issue. In 2021, the US witnessed 39,508 fatal crashes, translating to 1.37 fatalities per 100 million vehicle miles traveled (\citep{NHTSA}). This alarming statistic underscores the magnitude of the problem, further emphasized by the \$498.3 billion annual economic burden of crashes (\citep{nsc}). As transportation networks become increasingly intricate, upholding highway safety becomes even more crucial.

Highway safety serves as a critical gauge of our transportation system's efficiency and reliability. Multiple facets influence this vital metric, encompassing infrastructure, vehicles, and, most importantly, driver behavior. Human error is recognized as the primary catalyst for road crashes, accounting for a staggering 94\% of road crashes (\citep{zhu2023investigation,yue2018assessment,singh2018critical}). Addressing escalating traffic density necessitates innovative solutions, including the integration of connected and autonomous vehicles (CAVs). These vehicles exhibit significant potential for enhancing highway safety through improved perception, prompt response to surroundings, and immunity to fatigue or distraction (\citep{dai2023explicitly}). However, despite such advancements, driver vigilance and responsible behavior remain crucial, particularly as human-driven and autonomous vehicles coexist on highways.

Traditional highway safety models, reliant on reactive approaches (\citep{arun2021systematic}), leverage historical crash data to predict future occurrences within specific road infrastructures (\citep{lee2017intersection,pei2011joint}). However, these models face inherent limitations (\citep{lord2021highway}) due to the rarity of crashes, time-intensive data collection (\citep{mannering2020big}), and inherent limitations in historical data reliability (e.g., underreporting, flawed collection, subjectivity) (\citep{tarko2018estimating,arun2021systematic}). These limitations hinder accurate facility-specific prediction and safety assessment (\citep{mannering2016unobserved,tarko2018surrogate}). Furthermore, traditional methods lack the ability to capture potential crashes, impeding progress in highway safety research (\citep{wang2021review}). A promising alternative to address these limitations lies in utilizing near-crash or traffic conflict data  (\citep{wang2021review,tarko2018estimating}). Traffic conflicts, occurring far more frequently than crashes, can provide a valuable means to estimate safety performance of entities on the road network (\citep{tarko2018estimating,davis2011outline}). Additionally, numerous studies have established a statistically significant association between traffic conflicts and observed crashes  (\citep{stipancic2018surrogate,li2021using,arun2021systematic}). This shift towards near-crash data offers a powerful tool for advancing highway safety research and interventions.

Surrogate safety measures (SSMs) have significantly advanced transportation safety across diverse domains, including infrastructure assessment, user behavior analysis, and emerging technology evaluation (\citep{oikonomou2023conflicts}), ultimately informing policy and strategy development (\citep{arun2021systematic}). Recently, SSMs have gained particular prominence in conflict-based analyses, leveraging their ability to identify the temporal or spatial proximity of road users. This facilitates the detection, evaluation, and severity assessment of conflicts or near-misses (\citep{abdel2023advances}). In modern research, SSM is crucial for evaluating CAV safety during their developmental phases. Since the scarcity of historical and generalizable safety data on CAVs, microsimulation techniques extract vehicle trajectories and identify traffic conflicts effectively by SSM, addressing data scarcity while ensuring robust safety assessment methodologies for emerging autonomous technologies (\citep{oikonomou2023conflicts}). The SSM-based conflict studies, dating back to the early 1970s (\citep{hayward1971near}), have continuously yielded valuable insights for highway safety research. Within the SSM framework, three primary sub-categories exist: time-based, deceleration-based, and energy-based.

Time-based SSMs are widely used in highway safety research, with Time-to-Collision (TTC) being one of the most prominent examples. Introduced by Hayward (\citeyear{hayward1972near}), TTC predicts the remaining time until a collision by analyzing future vehicle trajectories. This metric assumes constant collision pair direction and speed and relies on probability estimates of a driver's inability to evade a collision  (\citep{wang2014evaluation}). In response to TTC's limitations in real-world scenarios where evasive maneuvers are common, Time to Accident (TA) was introduced by Perkins and Harris (\citeyear{perkins1967traffic}). Building upon TTC, Minderhoud and Bovy (\citeyear{minderhoud2001extended}) proposed advanced SSMs: Time-Exposed (TET) and Time-Integrated Time to Collision (TIT). TET measures the time spent below a threshold TTC, while TIT calculates the area between the threshold and the TTC curve during hazardous driving conditions. Both require continuous TTC calculations. Other notable time-based SSMs include: time headway (\citep{vogel2003comparison}), time to zebra (TTZ) (\citep{varhelyi1998drivers}), time-to-lane crossing (TTL), modified time to collision (MTTC) (\citep{ozbay2008derivation}), and T2. MTTC considers variable accelerations in car-following conflicts, while post-encroachment time (PET) (\citep{allen1978analysis}) measures the time from one vehicle's departure from a conflict point to the other's approach without relying on assumptions about speed, direction, or collision. Despite their widespread use, time-based SSMs have limitations. These include challenges in handling crossing/angle interactions, limited adaptability to dynamic changes, and an inability to account for lateral movement in overtaking and lane-changing scenarios.

Deceleration-based SSMs aim to prevent conflicts by analyzing vehicle deceleration rates. The most widely used measure is the Deceleration Rate to Avoid a Crash (DRAC), which determines interaction severity based on the minimum braking rate required to prevent a collision (\citep{mahmud2017application}). Later expansions of DRAC include the Crash Potential Index (CPI) (\citep{cunto2008assessing}), which considers braking capacity or maximum deceleration rate (MADR) to evaluate the probability of DRAC exceeding MADR. Based on the MADR assumption, the Rear-end Collision Risk Index (RCRI) is another deceleration-based SSM that compares emergency maneuver stopping distances to the lead vehicle's maximum deceleration rate to identify hazardous scenarios (\citep{oh2006method}). However, DRAC, RCRI, and CPI are limited to longitudinal conflict scenarios and have specific boundaries for determining interaction severity.

Energy-based SSMs prioritize the severity of potential conflicts over their proximity. The most prominent measure is DeltaV, which assesses the change in impact velocity during collisions using speed, mass, and angle (\citep{shelby2011delta}). Subsequent models like Conflict Severity (CS) (\citep{bagdadi2013estimation}) and extended DeltaV (\citep{laureshyn2017search}) further leverage DeltaV for comprehensive evaluations. However, DeltaV does not consider evasive maneuvers or pre-to-post-crash changes in speed and direction.  To address this limitation, alternative measures like the Crash Index (CAI) (\citep{ozbay2008derivation}) and conflict index (CFI) (\citep{alhajyaseen2015integration}) utilize kinetic energy terms instead of DeltaV. CFI integrates PET with vehicle speeds, masses, and angles to predict collision kinetic energy, while CAI utilizes acceleration, speed, and MTTC.

While SSM are widely used to study traffic conflicts, their reliance on one-dimensional (1-D) or longitudinal conflict scenarios (\citep{dai2023explicitly,das2022longitudinal,rahman2018longitudinal}) limits their capacity in real-world scenarios. The complex interactions between a vehicle and its surroundings (vehicles, obstacles, and road boundaries) particularly during maneuvers such as merging, lateral movement, and overtaking, demand consideration of both longitudinal and lateral dimensions. This complexity is amplified in diverse situations such as diversions, work zones, intersections, and roundabouts, increasing the likelihood of conflicts. Though some studies (\citep{lu2022learning,zhang2020safety,st2013automated}) consider longitudinal and lateral vehicle movements, raised concerns using SSM about accurately and objectively characterizing different facilities in recent reviews (\citep{wang2021review}), motivate the need for a generic and comprehensive analytical framework. This framework should be generic, applicable to diverse situations, and high-fidelity, ensuring an accurate representation of complex traffic dynamics.

To address this gap, our study introduces a novel analytical framework based on dynamics and control principles to provide a standard process in deriving SSMs. This framework incorporates a generic vehicle movement model, enabling the integration of complex vehicular dynamics within higher-dimensional motion scenarios. Transcending the limitations of 1D SSMs, our framework accommodates a broader spectrum of scenarios with varying degrees of complexity and dimensionality. This allows for accurate estimations of future vehicle trajectories, significantly enhancing safety analyses. Furthermore, our framework is capable of deriving SSMs that consider grade changes and complex horizontal alignments, commonly encountered in mountainous terrain, freeways, highways, and ramps. As many of these curves are non-planar  (\citep{bharat2023vehicle}) and may overlap horizontally or vertically, traditional 1D SSMs fall short of accurately capturing these complexities. Our framework overcomes this limitation by deriving not only conventional 1D SSMs but also more sophisticated multi-dimensional, high-fidelity SSMs. These SSMs incorporate both cartesian and path coordinate systems, providing a more versatile and adaptable approach. These systems, coupled with high-dimensional vehicle movement models, rigorously validated through simulation experiments that assess their precision under linearization, provide the foundation for our framework. The framework's real-world applicability is exemplified through a case study demonstrating its potential to revolutionize safety analyses in the evolving landscape of transportation safety assessment. Subsequent sections will delve deeper into the framework's specifics, highlighting its capacity to enhance our understanding of safety dynamics within transportation systems.

The structure of this study is as follows: In Section \ref{sec2}, we present the proposed generic analytical framework. Subsequently, in Section \ref{sec3}, we demonstrate instances of the proposed method which are equivalent to conventional SSMs. The extension of the proposed framework from one-dimensional to higher-dimensional and higher-fidelity SSMs are discussed in detail in Section \ref{sec4}. Case studies illustrating the application of the method are provided in Section \ref{sec5}. Lastly, we draw conclusions from our study in Section \ref{sec6}.

\section{The generic analytical framework \label{sec2}}
In this section, we propose a generic analytical framework capable of deriving both conventional one-dimensional SSMs and extended higher-dimensional, higher-fidelity  SSMs. Here, 'higher-dimensional' signifies an extension in the dimensions of movement and vehicle dynamics, while 'higher-fidelity' denotes a more sophisticated modeling approach that more accurately reflects the characteristics of vehicles. Our framework is founded on principles of dynamics and control, focusing on vehicular motion. We employ a generic vehicle movement model that encompasses various instances, each differing in complexity and dimensions, to capture vehicle dynamics and estimate future vehicle trajectory evolution. The framework introduces a generic mathematical condition to represent collisions, defined by the overlap of a vehicle's spatial occupation with another object. A collision is predicted when there exists a non-negative time at which this collision condition is fulfilled. The minimum of these times is identified as the remaining time before the collision occurs. Specifically, Subsection \ref{sec2.1} gives the formulation of the framework, Subsection \ref{sec2.2} provides analytical vehicle trajectory solutions of cases when the model is linear time-invariant (LTI) while diagonalizable, and LTI while non-diagonalizable, which cover the majority of cases. 

\subsection{Formulation \label{sec2.1}}
Consider a generic vehicle movement model for the vehicle with index $i$, described by:
\begin{equation}
\dot{\chi_i}(t) = f(\chi_i(t), u_i(t)) \label{general vehicle model}
\end{equation}
where $i\in V$ and \(V\) is the set of all vehicle indices. \(\chi_i(t) \in \mathbb{R}^n\) represents the state of the vehicle with index $i$ at time \(t\), encompassing characteristics such as position and direction of movement that describe the vehicle's spatial status. $\dot{\chi}(t)=\frac{d\chi(t)}{dt}$ represents the derivative of $\chi(t)$, for brevity, we will use this representation in most places within this paper. The control input of vehicle $i$ at time \(t\) is represented by \(u_i(t) \in \mathbb{R}^m\), which includes factors like acceleration and steering angle that can be controlled to impact the vehicle states. $n$ and $m$ denote the number of states and control inputs in the model, respectively. Eq. (\ref{general vehicle model}) is a generic representation for describing how vehicle states change in time based on current states and control and is a common representation for dynamical systems (\cite{brogan1991modern}). As the dimension and fidelity of the model increase, both $n$ and $m$ typically becom larger.  Given that SSM are typically computed on a rolling horizon basis, for the sake of simplicity and without loss of generality, we designate the time instance at which the SSM is evaluated as $t=0$ throughout this paper. Consequently, all $t>0$ refer to future or forecasted time. Given an initial state \(\chi_i(0)\) and a specified or assumed control input \(u_i(t)\) for all \(t > 0\), the future trajectory \(\chi_i(t)\) of the vehicle can be determined for all \(t > 0\). It should be noted that understandings of specific vehicle/driver behaviors can be incorporated into our proposed framework by substituting the control variable $u_i(t)$ in Eq. (\ref{general vehicle model}) with the evolving equations of specific behaviors (e.g., adaptive cruise controllers (\cite{zhou2020stabilizing}), lane-keeping controllers (\cite{li2023sequencing2}), car-following laws (\cite{brackstone1999car})). However, specific vehicle/driver behaviors extend beyond the scope of this paper, therefore, here we only consider simple vehicle/driver behaviors that are unchanging or predefined (e.g., maximum deceleration, invariant wheel torque, invariant steering wheel angle, etc.).

Collisions are essentially scenarios where a vehicle's shape overlaps with another object (e.g., another vehicle, an obstacle, or a road boundary) at a specific time \(t_c\). Different methods can be used to define the boundary of a vehicle's or obstacle's shape, such as its length or bounding shapes like circles or rectangles. Let a generic function \(g_v(\cdot)\) characterize the minimum distance between the vehicle and another vehicle, with examples including the longitudinal gap and minimal distance between two bounding shapes. Similarly, we define a generic function \(g_r(\cdot)\) to characterize the minimal distance between the vehicle and an obstacle or a road boundary, with examples including the minimal distance between two bounding shapes and minimal distance between the vehicle's bounding shape and the road boundary.

For vehicle-to-vehicle collisions, the SSM can be derived by solving for \(t_c\) in the equation:
\begin{equation}
 g_v(\chi_i(t_c), \chi_j(t_c)) = 0 \label{general vehicle collide}
 \end{equation}
where \(i \neq j\) and \(i, j \in V\). Here, \(\chi_i\) and \(\chi_j\) are the trajectories of vehicles indexed by \(i\) and \(j\).

For vehicle-to-road or vehicle-to-obstacle collisions, the SSM is obtained by solving for \(t_c\) in:
\begin{equation}
g_r(\chi_i(t_c), R_k) = 0 \label{general road collide}
\end{equation}
where \(i \in V\) and \(k \in O\). In this case, $O$ denotes the set of boundary and obstacle indices, and $R_k$ represents the spatial occupation of the object with index $k$.

It should be noted that $g_v(\cdot)$ and $g_r(\cdot)$ are generic representations here and can be freely designed. When modeling the vehicle movement in Eq. (\ref{general vehicle model}), the authors suggest the use of a path (or Frenet) coordinate system for evaluating vehicle-to-road collisions and a Cartesian coordinate system for evaluating vehicle-to-vehicle and vehicle-to-obstacle collisions, this will be further discussed and showcased in Section \ref{sec4}.

\begin{figure}[h]
    \centering
    \setlength{\abovecaptionskip}{0pt}
    \subcaptionbox{}
    {\includegraphics[width=0.45\textwidth]{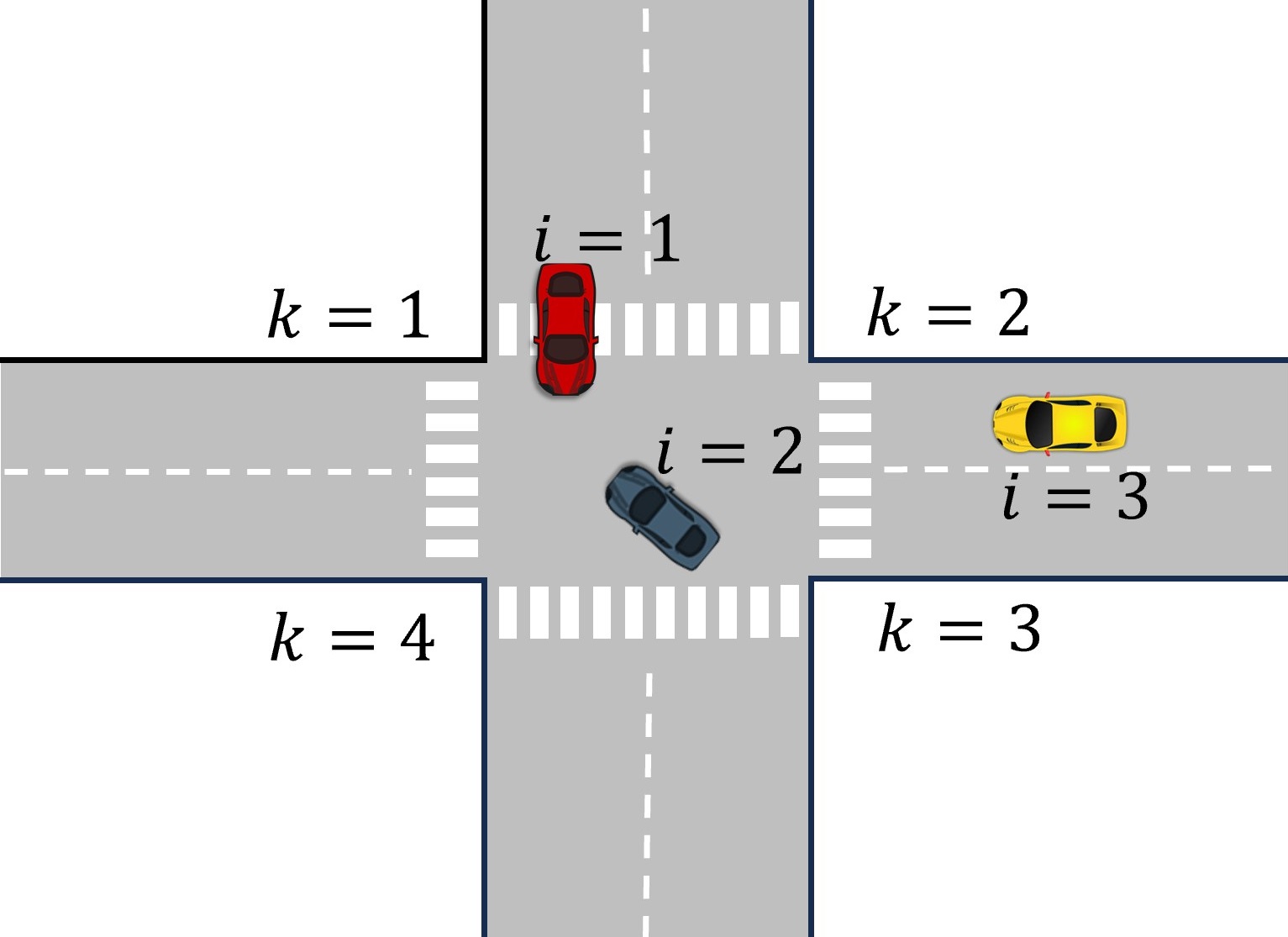}}
    \hspace{15mm}
    \subcaptionbox{}{\includegraphics[width=0.38\textwidth]{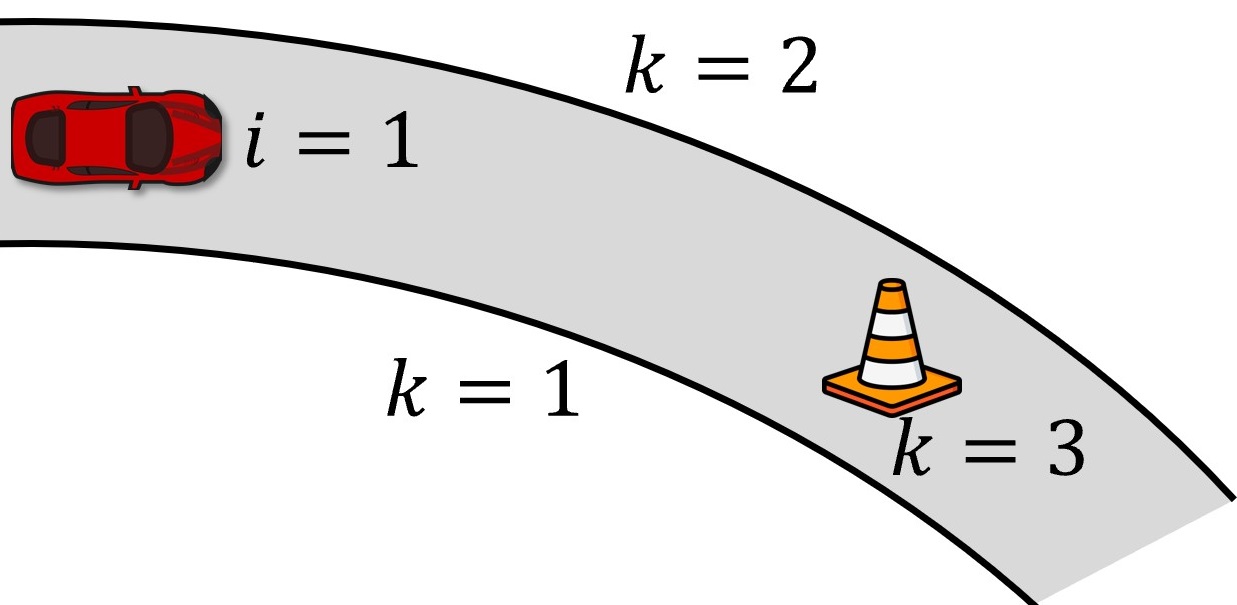}}
    \caption{Examples of index $i$ and $k$: (a) intersection scenario; (b) curved road scenario with obstacle}
    \label{index example}
\end{figure}

Examples of how $i\in V$ and $k\in O$ are indexed are depicted in Fig. \ref{index example}, with Fig. \ref{index example}(a) illustrating an intersection scenario while Fig. \ref{index example}(b) demonstrating a curved road scenario with an obstacle positioned on the road.

If non-negative solutions for \( t_c \) exist for Eq. (\ref{general vehicle collide}) and/or Eq. (\ref{general road collide}), the earliest one, denoted as $t_c^*$, indicates when the vehicle will collide. Otherwise, the vehicle is not at risk of collision.

When $t_c^*$ exists, with future trajectories $\chi_i(t)$ obtained based on Eq. (\ref{general vehicle model}), pre-collision values for the the calculation of energy-based SSM (e.g., velocity, angle) can be extracted from $\chi_i(t_c^*)$. Considering that this study mainly focuses on collision proximity, the extraction of energy-based SSM is only briefly illustrated in Section \ref{sec3}.

\subsection{Time-Invariant Linearization Approximation\label{sec2.2}}
Given the vehicle movement model Eq. (\ref{general vehicle model}), if it remains time-invariant, and furthermore, if it's inherently linear or has been linearized by taking its Taylor series at an operating point ($\chi_{op},u_{op})$ and neglecting the higher-order terms as shown in Eq. (\ref{eq:general linearization}), then it can be described in the form of Eq. (\ref{LTI vehicle model}).

\begin{equation}\label{eq:general linearization}
\begin{split}
\dot{{\chi}}(t) \approx 
{f}_l({\chi}(t),{u(t)},{\chi_{op}},{u_{op}})=
{f}({\chi_{op}},{u_{op}})+\left.\frac{\partial f}{\partial \chi}\right|_{\substack{\chi=\chi_{op} \\ u=u_{op}}}({\chi(t)}-{\chi_{op}})+\left.\frac{\partial f}{\partial u}\right|_{\substack{\chi=\chi_{op} \\ u=u_{op}}}({u(t)}-{u_{op}})
\end{split}
\end{equation}

\begin{equation}
 \dot{\chi}(t) = A\chi(t) + Bu(t)+C \label{LTI vehicle model}
\end{equation}

In Eq. (\ref{LTI vehicle model}), $A$ is an $n\times n$ matrix, $B$ is an $n\times m$ matrix, and $C$ is an $n\times 1$ column vector. Given \( \chi(0) \) and with a known or assumed \( u(t) \) for all \( t > 0 \), the trajectory \( \chi(t) \) can be determined analytically as follows (\citep{brogan1991modern}):
\begin{equation}
\chi(t) = e^{At} \chi(0) + \int_0^t e^{A(t-\tau)}Bu(\tau)+e^{A(t-\tau)}Cd\tau \label{LTI solution}
\end{equation}
this represents the solution for a system of order \( n \) that's linear and time-invariant. The term \( e^{At} \chi(0) \) is known as the free response, while \( \int_0^t e^{A(t-\tau)}Bu(\tau)d\tau \) is the forced response, and \( \int_0^t e^{A(t-\tau)}Cd\tau \) is a constant term. As shown by Eq. (\ref{LTI vehicle model}), the calculation of the exponential matrix, $e^{At}$, is nontrivial, it has dimensions \( n \times n \) and can be represented by its Taylor expansion:
\begin{equation}
e^{At} = I_n + At + \frac{1}{2!}A^2 t^2 + \frac{1}{3!}A^3 t^3 + \dots + \frac{1}{k!}A^k t^k + \dots \label{exponential taylor series}
\end{equation}

As shown in Eq. (\ref{exponential taylor series}), the series is infinite. If matrix $A$ is nilpotent (i.e., $A^k=0$ for some finite k), Eq. (\ref{exponential taylor series}) becomes finite and $e^{At}$ can be directly calculated from Eq. (\ref{exponential taylor series}). Otherwise, a more complex process based on matrix theory is needed for the calculation of $e^{At}$, with the following general steps (\cite{strang2012linear}): 

(a) find a linear transformation $\chi^*(t)=T^{-1}\chi(t)$, such that the transformed version of Eq. (\ref{LTI vehicle model}), $\dot{\chi^*}(t) = \bar{A} \chi^*(t) + \bar{B} u(t) +\bar{C}$ contains a $\bar{A}$ matrix whose exponential, $e^{\bar{A}t}$, can be easily calculated ($\bar{A}$ in diagonal form if $A$ is diagonalizable and in Jordan canonical form if $A$ is not diagonalizable);

(b) obtain $e^{\bar{A}t}$ following standard processes for matrices in diagonal form or Jordan canonical form (\citep{strang2012linear});

(c) obtain the desired $e^{At}$ with inverse transformation, $e^{At} = T e^{\bar{A} t} T^{-1}$.

More details for cases where $A$ is diagonalizable and non-diagonalizable are given below.

If matrix \( A \) is diagonalizable (i.e., if all eigenvalues of $A$ are distinct, or in cases where repeated eigenvalues each have a corresponding set of independent eigenvectors, equal in number to the multiplicity of the eigenvalue), \( e^{At} \) can be obtained through the process of diagonalization. To determine the vehicle's movement, we first seek a nonsingular matrix \( T \). Define \( \chi^*(t) \) as \( T^{-1} \chi(t) \), on substituting this into our primary equation Eq. (\ref{LTI vehicle model}), we obtain:

\begin{equation}
\dot{\chi^*}(t) = T^{-1} A T \chi^*(t) + T^{-1} B u(t) +T^{-1}C
\end{equation}

Here, \( T^{-1} A T \) simplifies to a diagonal matrix, \( \Lambda \), represented as 

\begin{equation}
\Lambda = \text{diag}{\{\lambda_1, \lambda_2, \ldots, \lambda_n\}}
\end{equation}

This makes \( e^{\Lambda t} \) a diagonal matrix given by:

\begin{equation}
e^{\Lambda t} = \text{diag}{\{e^{\lambda_1 t}, e^{\lambda_2 t}, \ldots, e^{\lambda_n t}\}}
\end{equation}

By performing an inverse transformation, we obtain:

\begin{equation}
e^{At} = T e^{\Lambda t} T^{-1}
\end{equation}

It is well known that in the above process, \( \lambda_1, \lambda_2, \ldots, \lambda_n \) are the eigenvalues of matrix \( A \), while $T = [t_1, t_2, \ldots, t_n]$ is a collection of the corresponding eigenvectors, \( t_1, t_2, \ldots, t_n \), as columns (\citep{strang2012linear}). With $e^{At}$ obtained, the future vehicle trajectory can be calculated by Eq. (\ref{LTI solution}).

In some scenarios, matrix \(A\) is not diagonalizable. This happens when there are repeated eigenvalues without sufficient linearly independent eigenvectors for matrix \(T\). In such cases, \(A\) can be transformed into its Jordan canonical form using the equation:

\begin{equation}
T^{-1} A T = J = \text{blockdiag}\{J_1, J_2, \dots, J_q\}
\end{equation}
where \(q\) represents the number of Jordan blocks in $J$. Each \(J_i\) is a Jordan block of size \(n_i\) corresponding to an eigenvalue \(\lambda_j\) and is defined as:

\begin{equation}
J_i = \begin{bmatrix}
\lambda_j & 1 & & \\
 & \lambda_j & \ddots & \\
 & & \ddots & 1 \\
 & & & \lambda_j 
\end{bmatrix}
\end{equation}

It's important to note that an eigenvalue can be linked to multiple Jordan blocks. This association is based on the number of times that the eigenvalue is repeated and the count of its linearly independent eigenvectors. Here, matrix \(T\) is constructed as:

\begin{equation}
T = [T_1, T_2, \dots, T_q]
\end{equation}
where each \(T_i \in \mathbb{C}^{n \times n_i}\) are columns of \(T\) linked to the i-th Jordan block, \(J_i\). Specifically, for each \(T_i\), we have:

\begin{equation}
T_i = [t_{i,1}, t_{i,2}, \dots, t_{i,n_i}]
\end{equation}

Here, \(t_{i,1}\) is an eigenvector associated with the eigenvalue linked to $J_i$, and for \(j = 2, 3, \dots, n_i\), the relationship is defined as:

\begin{equation}
A t_{i,j} = t_{i,j-1} + \lambda_i t_{i,j}
\end{equation}

The vectors \(t_{i,1}, t_{i,2}, \dots, t_{i,n_i}\) are termed generalized eigenvectors. For a deeper understanding of the Jordan canonical form and generalized eigenvectors, readers are directed to \citep{brogan1991modern} and \citep{strang2012linear}. With $T$ calculated, closed-form solutions can be given for $e^{Jt}$, and similarly, by the inverse transformation $e^{At}=Te^{Jt}T^{-1}$, $e^{At}$ can be obtained. Consequently, the future vehicle trajectory can be calculated by Eq. (\ref{LTI solution}).

Upon computation of the vehicle trajectory, an SSM is constructed through the determination of $t_c$ from Eqs. (\ref{general vehicle collide}) and/or (\ref{general road collide}). The existence of non-negative real values for $t_c$ indicates a probable collision event within the forecasted trajectory. In such cases, the minimal non-negative solution of $t_c$ represents the remaining time before the collision. Conversely, the absence of non-negative solutions for $t_c$ suggests that a collision is not anticipated based on the current estimations of trajectory evolution. Analytically, 1-D SSM often permit the derivation of closed-form expressions for $t_c$, facilitating straightforward analytical estimations of future vehicle state evolution. However, for higher-dimensional SSM, the complexity typically necessitates the application of numerical techniques to resolve the values of $t_c$.

\section{Instances equivalent to conventional SSMs \label{sec3}}

In this section, to demonstrate the generality of the proposed framework, we derive instances that acquire conventional SSMs, with appropriate selections of the vehicle movement model Eq. (\ref{general vehicle model}) and the collision condition Eq. (\ref{general vehicle collide}). Specifically, Subsections \ref{sec3.1} and \ref{sec3.2} showcase the derivation of a conventional time-based SSM and a deceleration-based SSM, respectively. These are the two main types of conventional SSM that measure the collision proximity (\citep{wang2021review}). Furthermore, for both subsections, we show that energy-based SSM, which measures the severity of a collision, can be extracted from the solution of this framework.

\subsection{Conventional time-based SSM\label{sec3.1}}
\begin{figure}[h]
    \centering
    \setlength{\abovecaptionskip}{0pt}
    \includegraphics[width=0.6\textwidth]{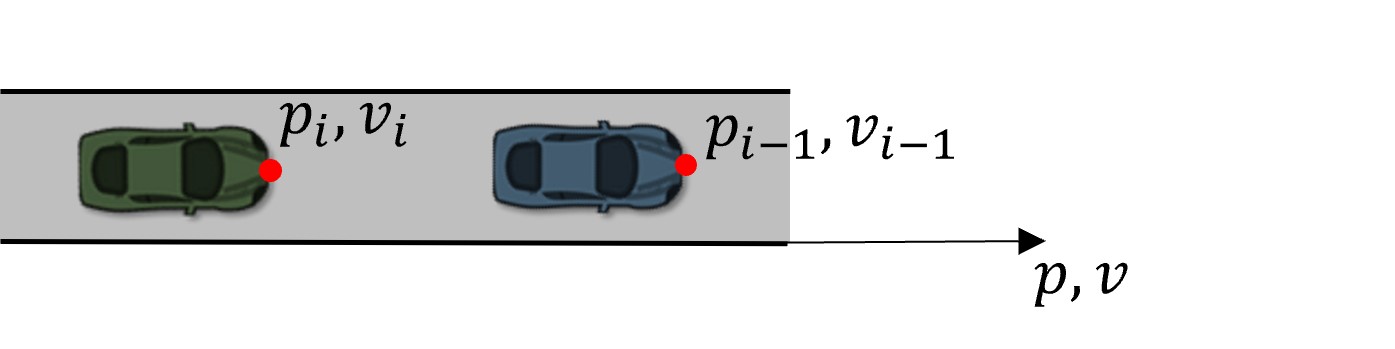}
    \caption{Direction and measure points of position and velocity}
    \label{fig:TTC measure}
\end{figure}

The time-based SSM quantifies the risk associated with an interaction by evaluating the temporal proximity of the interaction to a potential collision. For illustration purpose, we specifically consider TTC (\citep{hayward1971near}), which stands as a predominant time-based SSM and serves as a foundational element in numerous subsequent research (\citep{minderhoud2001extended}), and demonstrate the proposed framework obtains a solution identical to TTC with appropriate model and distance representation selections. Mathematically, TTC is defined as:
\begin{equation}
    TTC_i=\frac{p_{i-1}(0)-p_i(0)-l_i}{v_i(0)-v_{i-1}(0)} \label{TTC}
\end{equation}
where $TTC_i$ denotes the TTC of the i-th vehicle (the following vehicle), $p_{i-1}(0)$ and $p_i(0)$ represent the longitudinal position of the (i-1)-th vehicle (the leading vehicle) and the i-th vehicle at $t=0$, which is the time instance at which the SSM is evaluated, respectively. Similarly, $v_{i-1}(0)$ and $v_i(0)$ represent the longitudinal velocity of the (i-1)-th vehicle and the i-th vehicle at $t=0$, respectively. $l_i$ denotes the length of the i-th vehicle. The direction of position and velocity are shown by the arrow and the measure points are indicated by the red points in Fig. \ref{fig:TTC measure}. 

In the context of the framework proposed in Section \ref{sec2}, by assuming that the involved vehicles maintain an invariant moving direction on a straight lane, which is consistent with one of the primary assumptions of TTC, the vehicular motion is reduced to a 1-D representation, specifically in the longitudinal direction. Consequently, for equation Eq. (\ref{general vehicle model}), we employ a rudimentary model describing the position-velocity relationship:
\begin{equation}
\dot{p}(t) = v(t)
\label{TTC model}
\end{equation}
where $p(t)$ represents the vehicle longitudinal position at time $t$ and $v(t)$ represents the vehicle longitudinal velocity at time $t$. It is evident that this model is naturally LTI in the form of Eq. (\ref{LTI vehicle model}), with $\chi(t)=p(t)$, $u(t)=v(t)$, $A=[0]$, $B=[1]$, and $C=[0]$. Note that with the specific model Eq. (\ref{TTC model}), matrices A and B, and vector C degenerate to scalars as the model only contains one state and one control variable (i.e., $m=n=1$ in Eq. (\ref{LTI vehicle model})). However, to keep the consistency of their definition, we represent them as $1\times1$ matrices/vectors here.

By further assuming that the involved vehicles maintain constant speeds (i.e., $u(t)=u(0)$ for all $t>0$), which is another major assumption of TTC, and based on Eq. (\ref{LTI solution}), the future trajectory of a vehicle can be obtained as:
\begin{equation}
\chi(t) = e^0 \chi(0) + \int_0^t e^0 \times 1 \times u(0) d\tau = p(0) + v(0)t
\end{equation}

With vehicle indices that are consistent with Eq. (\ref{TTC}), the left-hand-side of Eq. (\ref{general vehicle collide}) can be expressed as a function describing the gap between the i-th and (i-1)-th vehicle:
\begin{equation}
g_v(\chi_i(t_c),\chi_{i-1}(t_c)) = \chi_{i-1}(t_c) - \chi_i(t_c) - l_i = p_{i-1}(0) + v_{i-1}(0)t_c - (p_i(0) + v_i(0)t_c) - l_i \label{TTC_distance eq}
\end{equation}

By setting Eq. (\ref{TTC_distance eq}) to 0 and solving for $t_c$, a closed-form solution is obtained as:
\begin{equation}
t_c = \frac{p_{i-1}(0) - p_i(0) - l_i}{v_i(0) - v_{i-1}(0)} \label{TTC t_c}
\end{equation}
which is identical to Eq. (\ref{TTC}). When $t_c\geq0$, $t_c$ equals $t_c^*$ and represents the remaining time before the two vehicles collide, otherwise, there is no risk of collision.

To demonstrate how energy-based SSM can be extracted from the solution, we reference DeltaV (\citep{shelby2011delta}). Recognized as a significant energy-based SSM (\citep{wang2021review}), DeltaV quantifies the velocity alteration experienced by road users due to a collision. Building upon the assumptions inherent to TTC, when Eq. (\ref{TTC t_c}) indicates a risk of collision, DeltaV is computed as follows:
\begin{equation}
\Delta v_i = \frac{m_{i-1}}{m_i + m_{i-1}} (v_{i-1}(t_c^*) - v_i(t_c^*)) = \frac{m_{i-1}}{m_i + m_{i-1}} (v_{i-1}(0) - v_i(0))
\end{equation}

\begin{equation}
\Delta v_{i-1} = \frac{m_i}{m_{i-1} + m_i} (v_i(t_c^*) - v_{i-1}(t_c^*)) = \frac{m_i}{m_{i-1} + m_i} (v_i(0) - v_{i-1}(0))
\end{equation}
where $m_{i-1}$ and $m_i$ are the mass of the (i-1)-th and i-th vehicle, respectively. $v_{i-1}(t_c^*)=v_{i-1}(0)$ and $v_i(t_c^*)=v_i(0)$ represent the pre-collision velocities.

\subsection{Conventional deceleration-based SSM\label{sec3.2}}
Being slightly different from time-based SSMs, deceleration-based SSMs emphasize the role of vehicular deceleration in averting potential collisions. For illustration purpose, this study delves into the RCRI (\citep{oh2006method}). We elucidate how an identical solution can be derived within the context of our proposed framework, and subsequently present an equivalent time-based indicator of safe/unsafe as an inherent outcome of our analytical approach. Additionally, our framework provides the remaining time before collision (if any).  Mathematically, the RCRI is defined as (\citep{oh2006method,wang2021review}):
\begin{equation}
    SSD_{i-1}=S+\frac{v_{i-1}^2(0)}{2d_m} \label{RCRI1}
\end{equation}
\begin{equation}
    SSD_i=v_i^2(0)t_d+\frac{v_i^2(0)}{2d_m} \label{RCRI2}
\end{equation}
\begin{equation}
RCRI=
\begin{cases} 
0\text{(safe)} & \text{if $SSD_{i-1}-SSD_i>0$} \\
1\text{(dangerous)} & \text{otherwise}
\end{cases}\label{RCRI3}
\end{equation}
where $SSD_{i-1}$ and $SSD_i$ represent the stopping distances for the leading and following vehicles, respectively. $v_i(0)$ denotes the current speed of the following vehicle, while $v_{i-1}(0)$ corresponds to the current speed of the leading vehicle. The term $t_d$ indicates the time delay, $S$ is defined as the clearance distance between vehicles, and $d_m$ is the maximum deceleration rate that the vehicles can achieve.

In the context of the framework proposed in Section \ref{sec2}, the vehicular motion is also a simplified 1-D representation in the longitudinal direction. For equation Eq. (\ref{general vehicle model}), we select a second-order longitudinal model:
\begin{equation}
\begin{bmatrix}
\dot{p}(t) \\
\dot{v}(t)
\end{bmatrix}
=
\begin{bmatrix}
0 & 1 \\
0 & 0 
\end{bmatrix}
\begin{bmatrix}
p(t) \\
v(t)
\end{bmatrix}
+
\begin{bmatrix}
0 \\
1
\end{bmatrix}
a(t)
\end{equation}
where $a(t)$ represents the vehicle acceleration at time $t$. This is also a natural LTI model in the form of Eq. (\ref{LTI vehicle model}), with $\chi(t)=\begin{bmatrix}
p(t) \\
v(t)
\end{bmatrix}$, $u(t)=a(t)$, $A=\begin{bmatrix}
0 & 1 \\
0 & 0 
\end{bmatrix}$, $B=\begin{bmatrix}
0 \\
1
\end{bmatrix}$, and $C=\mathbf{0}_{2 \times 1}$.

Here, $A$ is a Nilpotent matrix such that $A^2=\mathbf{0}_{2 \times 2}$, therefore, instead of going through the process shown in Subsections \ref{sec2.2}, its exponential can be easily obtained by Eq. (\ref{exponential taylor series}):
\begin{equation}
e^{At} = I_2 + At + 0 + 0 + \dots = 
\begin{bmatrix}
1 & t \\
0 & 1
\end{bmatrix}
\end{equation}

With vehicle indices consistent with Eqs. (\ref{RCRI1})-(\ref{RCRI3}), for the scenario where the leading vehicle decelerates at the maximum rate (i.e., $u_{i-1}(t)=-d_m$ for all $t>0$) and the following vehicle decelerates at the maximum rate after a delay $t_d$ (i.e., $u_i(t)=0$ for $0\leq t<t_d$ and $u_i(t)=-d_m$ for $t\geq t_d$), the trajectories of the involved vehicles are obtained as:
\begin{equation}
\chi_{i-1}(t) = 
\begin{bmatrix}
{p}_{i-1}(t) \\
{v}_{i-1}(t)
\end{bmatrix}
=
\begin{bmatrix}
1 & t \\
0 & 1
\end{bmatrix}
\begin{bmatrix}
p_{i-1}(0) \\
v_{i-1}(0)
\end{bmatrix}
+
\begin{bmatrix}
1 & t \\
0 & 1
\end{bmatrix}
\int_0^t -d_m \cdot 
\begin{bmatrix}
1 & -\tau \\
0 & 1
\end{bmatrix}
\begin{bmatrix}
0 \\
1
\end{bmatrix}
d\tau
=
\begin{bmatrix}
p_{i-1}(0) + v_{i-1}(0)t - \frac{1}{2} d_m t^2 \\
v_{i-1}(0) - d_m t
\end{bmatrix} \label{RCRI veh traj1}
\end{equation}

\begin{align}
\chi_i(t) = 
\begin{bmatrix}
{p}_i(t) \\
{v}_i(t)
\end{bmatrix}
&=
\begin{bmatrix}
1 & t \\
0 & 1
\end{bmatrix}
\begin{bmatrix}
p_i(0) \\
v_i(0)
\end{bmatrix}
+ 
\begin{bmatrix}
1 & t \\
0 & 1
\end{bmatrix}
\left( \int_0^{t_d} 0 \cdot 
\begin{bmatrix}
1 & -\tau \\
0 & 1
\end{bmatrix}
\begin{bmatrix}
0 \\
1
\end{bmatrix}
d\tau + \int_{t_d}^t -d_m \cdot 
\begin{bmatrix}
1 & -\tau \\
0 & 1
\end{bmatrix}
\begin{bmatrix}
0 \\
1
\end{bmatrix}
d\tau \right) \nonumber \\
&=
\begin{bmatrix}
p_i(0) + v_i(0)t - \frac{1}{2} d_m (t^2 + t_d^2 - 2t_d t) \\
v_i(0) - d_m (t - t_d)
\end{bmatrix}\label{RCRI veh traj2}
\end{align}

We first demonstrate how the exact RCRI can be derived from Eqs. (\ref{RCRI veh traj1}) and (\ref{RCRI veh traj2}). By setting $v_{i}(t)=0$ in Eq. (\ref{RCRI veh traj2}), the time needed for the following vehicle to fully stop, $t_{i,stop}$, can be solved as:
\begin{equation}
t_{i,stop} = \frac{v_i(0)}{d_m} + t_d
\end{equation}

By substituting $t_{i,stop}$ into Eq. (\ref{RCRI veh traj2}), the position of the following vehicle when fully stopped is obtained:
\begin{equation}
p_i(t_{i,stop}) = p_i(0) + \frac{v_i^2(0)}{2d_m} + v_i(0) t_d
\end{equation}

Similarly, for the leading vehicle, we obtain:
\begin{equation}
t_{i-1,stop} = \frac{v_{i-1}(0)}{d_m}
\end{equation}

\begin{equation}
p_{i-1}(t_{i-1,stop}) = p_{i-1}(0) + \frac{v_{i-1}^2(0)}{2d_m}
\end{equation}

Denoting the gap between the two vehicles when both are fully stopped as $\Delta p$, it is calculated as:
\begin{align}
\Delta p = p_{i-1}(t_{i-1,stop}) - p_i(t_{i,stop}) - l_i &= (p_{i-1}(0) - p_i(0) - l_i) - v_i(0) t_d + \frac{v_i^2(0)}{-2d_m} - \frac{v_{i-1}^2(0)}{-2d_m} \nonumber \\
&= S - v_i(0) t_d - \frac{v_i^2(0)}{2d_m} + \frac{v_{i-1}^2(0)}{2d_m}
\end{align}
which is identical to $(SSD_{i-1}-SSD_i)$ in Eq. (\ref{RCRI3}), and $\Delta p>0$ means there is no risk of collision, and otherwise, there exists a potential collision.

We now demonstrate that the solution derived from our proposed framework, despite its distinct formulation, yields an equivalent indicator for safe/dangerous. In addition, this solution offers a quantification of the remaining duration before an impending collision. Utilizing the closed-form solutions given by Eqs. (\ref{RCRI veh traj1})-(\ref{RCRI veh traj2}) for the vehicle trajectories, Eq. (\ref{general vehicle collide}) is split into two distinct equations. These correspond to cases where: (i) both vehicles are in motion, and (ii) one of the vehicles has come to a complete halt. It should be noted that this split arises from the exclusion of scenarios where vehicles engage in reverse motion (i.e., the scenario where vehicles start to move backwards after coming to a complete stop is not considered). The resultant functions are presented below:

When both vehicles are still in motion, the longitudinal trajectories of the vehicles can be represented by $p_{i-1}(t)$ in Eq. (\ref{RCRI veh traj1}) and $p_i(t)$ in Eq. (\ref{RCRI veh traj2}), respectively. In this scenario, the function $g_v(\cdot)$ in Eq. (\ref{general vehicle collide}) is specified as:
\begin{align}
g_{v1} \left( \chi_i(t_c), \chi_{i-1}(t_c) \right) &= p_{i-1}(t_c) - p_i(t_c) - l_i = p_{i-1}(0) - p_i(0) + \left( v_{i-1}(0) - v_i(0) \right) t_c - \frac{1}{2} d_m \left( t_R^2 - 2t_R t_c \right) - l_i, \nonumber \\&\quad t_c \in \left[ 0, \min(t_{i,\text{stop}}, t_{i-1,\text{stop}}) \right] \label{SSD g1}
\end{align}

After a vehicle has come to a complete halt, its position remains unchanged (i.e., $p_{i-1}(t)\equiv p_{i-1}(t_{i-1,stop})$ if $t\geq t_{i-1,stop}$ and $p_{i}(t)\equiv p_{i}(t_{i,stop})$ if $t\geq t_{i,stop}$). We can therefore divide this scenario into two: (1) the leading vehicle has stopped and the following vehicle is still in motion; (2) the following vehicle has stopped and the leading vehicle is still in motion. It should be noted that in practice, a collision cannot happen in case (2), nevertheless, we keep the case in the following discussions for the completeness of deduction.

\begin{equation}
g_{v2}(\chi_i(t_c), \chi_{i-1}(t_c)) = 
\begin{cases}
\begin{aligned}
 p_{i-1}(t_{i-1,\text{stop}}) - p_i(t_c) - l_i =& p_{i-1}(0) + \frac{v_{i-1}^2(0)}{-2d_m} - \left( p_i(0) + v_i(0)t_c + \frac{1}{2}d_m(t_c^2 + t_R^2 - 2t_R t_c) \right) - l_i, \\
& t_c \in [t_{i-1,\text{stop}}, t_{i,\text{stop}}] \quad\quad\quad\quad\text{ if } t_{i-1,\text{stop}} < t_{i,\text{stop}} 
\end{aligned}
\\
\begin{aligned}
p_{i-1}(t_c) - p_i(t_{i,\text{stop}}) - l_i =& p_{i-1}(0) + v_{i-1}(0)t_c + \frac{1}{2}d_m t_c^2 - \left( p_i(0) + \frac{v_i^2(0)}{-2d_m} + v_i(0)t_R \right) - l_i, \\
& t_c \in [t_{i,\text{stop}}, t_{i-1,\text{stop}}] \quad\quad\quad\quad\text{otherwise} \label{SSD g2}
\end{aligned}
\end{cases}
\end{equation}

By setting $g_{v1}(\cdot)$ to 0 and solve for $t_c$, we check for potential collisions when both vehicle are in the motion of braking. A closed-form solution can be obtained as:
\begin{equation}
    t_c=\frac{p_{i-1}(0)+p_{i}(0)-0.5d_mt_R^2-l_i}{v_{i}-v_{i-1}+d_mt_R} \label{SSD tc sol1}
\end{equation}

Setting $g_{v2}(\cdot)$ to 0 and solving for $t_c$ as the roots of quadratic equations checks for potential collisions when one of the vehicles has stopped while the other is still decelerating. We get the closed-form solutions:
\begin{equation}
t_c = 
\begin{cases}
\frac{v_i(0) - d_m t_R \pm \sqrt{(d_m t_R - v_i(0))^2 + 2d_m (p_{i-1}(0) - \frac{v_{i-1}^2(0)}{2d_m} - p_i(0) - 0.5d_m t_R^2 - l_i)}}{-d_m} & \text{if } t_{i-1,\text{stop}} < t_{i,\text{stop}} \\
\frac{-v_{i-1}(0) \pm \sqrt{v_{i-1}^2 - 2d_m (p_{i-1}(0) + \frac{v_i^2(0)}{2d_m} - p_i(0) - v_i(0) t_R - l_i)}}{d_m} & \text{otherwise} \label{SSD tc sol2}
\end{cases}
\end{equation}

If either Eq. (\ref{SSD tc sol1}) or Eq. (\ref{SSD tc sol2}) is feasible (i.e., satisfies the corresponding restrictive range of $t_c$), it is indicated that there is a risk of collision, and the smallest feasible solution, $t_c^*$, represents the remaining time before collision. Otherwise, the vehicles do not hit each other.

Similarly, the energy-based SSM, DeltaV, can be extracted when there's a potential collision as follows:
\begin{align}
\Delta v_i &= \frac{m_{i-1}}{m_i + m_{i-1}} \left( v_{i-1}(t_c^*) - v_i(t_c^*) \right) \nonumber\\&= \begin{cases} 
\frac{m_{i-1}}{m_i + m_{i-1}} \left( v_{i-1}(0) - v_i(0) + d_m t_R \right) & \text{if } t_c^* \in [0, \min(t_{i,stop}, t_{i-1,stop})] \\
\frac{m_{i-1}}{m_i + m_{i-1}} \left( -v_i(0) - d_m (t_c^* - t_R) \right) & \text{if } t_c^* \in [t_{i-1,stop}, t_{i,stop}] \\
\frac{m_{i-1}}{m_i + m_{i-1}} \left( v_{i-1}(0) + d_m t_c^* \right) & \text{if } t_c^* \in [t_{i,stop}, t_{i-1,stop}]
\end{cases}
\end{align}

\begin{align}
\Delta v_{i-1} &= \frac{m_i}{m_{i-1} + m_i} \left( v_i(t_c^*) - v_{i-1}(t_c^*) \right) \nonumber\\&= \begin{cases} 
\frac{m_i}{m_{i-1} + m_i} \left( -v_{i-1}(0) + v_i(0) - d_m t_R \right) & \text{if } t_c^* \in [0, \min(t_{i,stop}, t_{i-1,stop})] \\
\frac{m_i}{m_{i-1} + m_i} \left( v_i(0) + d_m (t_c^* - t_R) \right) & \text{if } t_c^* \in [t_{i-1,stop}, t_{i,stop}] \\
\frac{m_i}{m_{i-1} + m_i} \left( -v_{i-1}(0) - d_m t_c^* \right) & \text{if } t_c^* \in [t_{i,stop}, t_{i-1,stop}]
\end{cases}
\end{align}

\section{Extended higher-dimensional and higher-fidelity SSMs \label{sec4}}

In this section, we extend the proposed framework to SSMs that integrate more complex (higher-fidelity) vehicular dynamics and accommodate higher-dimensional motion. The augmentation to higher dimensions necessitates a reconsideration of the coordinate systems employed to delineate vehicular motion. Primarily, the Cartesian and path (or Frenet) coordinate systems emerge as critical frameworks for this analysis. The Cartesian system is advantageous for describing convex bounding shapes like circles and rectangles, whereas the path coordinate system particularly excels in simplifying reference paths and road boundaries (\citep{reiter2023frenet}). According to this, we propose the use of the Cartesian coordinate system for examining vehicle-to-vehicle or vehicle-to-obstacle (or vehicle-to-other users) collisions, whereas the path coordinate system should be employed when analyzing vehicle-to-road collisions.

The objective of this section is to demonstrate the application of vehicle movement models that are adapted to high-dimensional dynamics and motion, as well as the implementation of disparate coordinate systems to facilitate this analysis. In particular, Subsection \ref{sec4.1} delineates the application of a 2-D kinematic model, encompassing both x and y coordinates. This model is applied within the framework of the Cartesian coordinate system to investigate vehicle-to-vehicle collisions. Subsequently, Subsection \ref{sec4.2} details the utilization of a model that combines 2-D motion principles, emphasizing alignment and orthogonality with respect to the path centerline, and incorporates 3-D dynamics, encompassing motion along the path centerline, orthogonal to the centerline, and in the vertical dimension. This model is applied within the path coordinate system to investigate vehicle-to-road collisions. Furthermore, Subsection \ref{sec4.3} discusses the cases where instead of deriving a closed-form expression of the future vehicle trajectory, numerical integration methods are employed to generate future trajectory points.

\subsection{2-D kinematic model in the Cartesian coordinate system\label{sec4.1}}
\begin{figure}[h]
    \centering
    \setlength{\abovecaptionskip}{0pt}
    \includegraphics[width=0.6\textwidth]{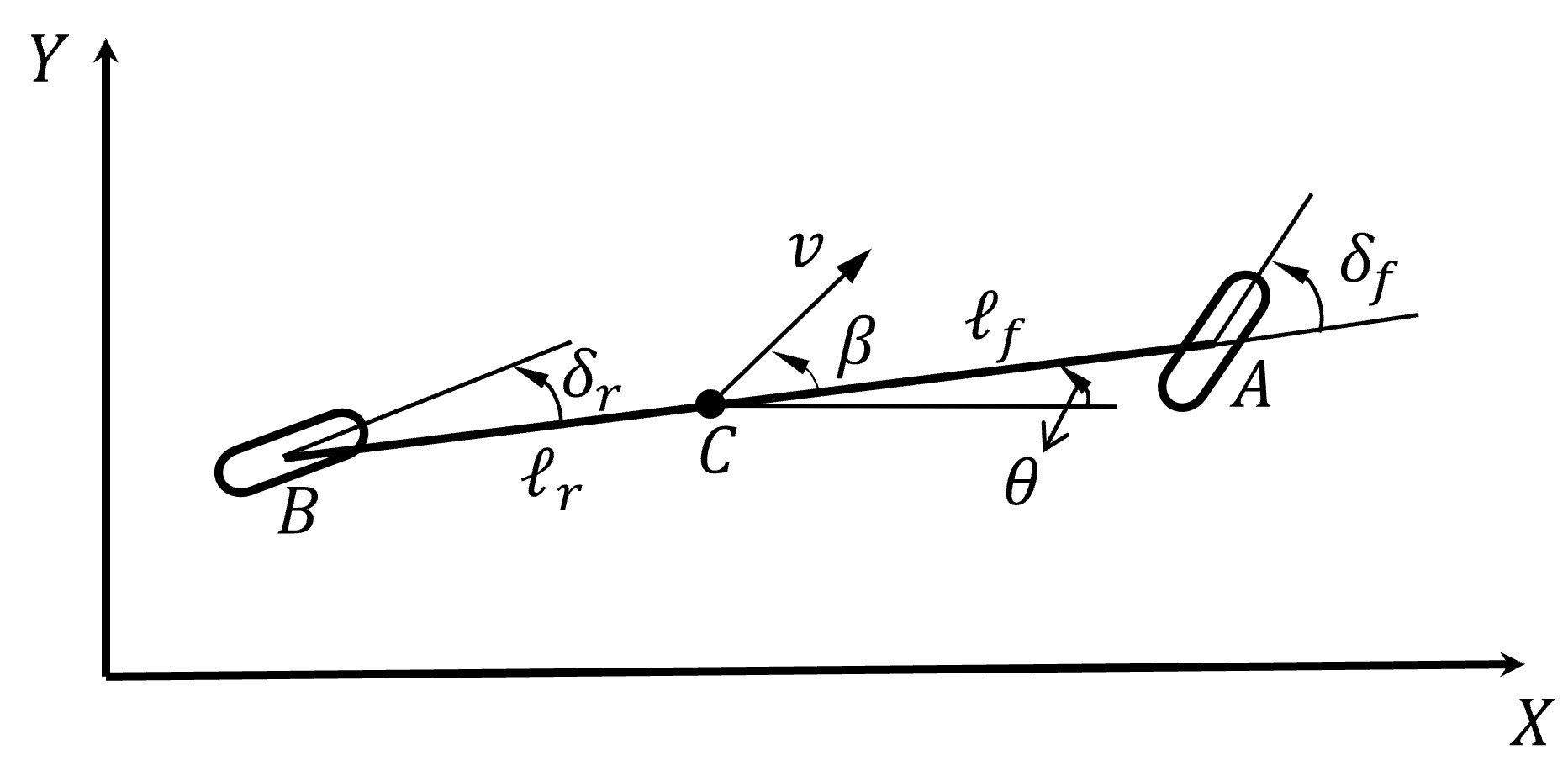}
    \caption{2-D kinematic model}
    \label{fig:kinematic}
\end{figure}

To illustrate how the proposed framework can be extended beyond 1-D, we considering a kinematic bicycle model (\citep{rajamani2011vehicle}) that describes 2-D movements (i.e., Cartesian x and y coordinates) of a vehicle, as shown in Fig. \ref{fig:kinematic}, the overall equations of motion are given as follows: 
\begin{equation}
\dot{x}=v\text{~cos}(\theta+\beta)
\end{equation}
\begin{equation}
\dot{y}=v\text{~sin}(\theta+\beta)
\end{equation}
\begin{equation}
\dot{\theta}=\frac{v\text{~cos}(\beta)}{l_f+l_r}(\text{tan}(\delta_f)-\text{tan}(\delta_r))
\end{equation}
\begin{equation}
\dot{v}=a
\end{equation}
where  the Cartesian position of the vehicle's center of gravity (C.G.) is denoted by the coordinates ($x,y$), corresponding to point C as illustrated in Fig. \ref{fig:kinematic}. The velocity vector at the C.G. is represented by $v$, forming an angle $\beta$ with the vehicle's longitudinal axis. The heading angle, $\theta$, defines the orientation of the vehicle relative to the Cartesian reference frame. The distances from the C.G. to the front and rear axles are given by $l_f$ and $l_r$, respectively. The steering angles at the front and rear wheels are represented by $\delta_f$ and $\delta_r$, respectively. Furthermore, the acceleration of the C.G. is signified by $a$.

When the vehicle model is restricted to front-steer-only configurations and the slip angle $\beta$ is assumed to be sufficiently small to warrant approximation as negligible, the system can be described by a simplified set of equations:
\begin{equation}
\dot{x}=v\text{~cos}(\theta) \label{kinematic1}
\end{equation}
\begin{equation}
\dot{y}=v\text{~sin}(\theta) \label{kinematic2}
\end{equation}
\begin{equation}
\dot{\theta}=\frac{v\text{~tan}(\delta)}{L} \label{kinematic3}
\end{equation}
\begin{equation}
\dot{v}=a \label{kinematic4}
\end{equation}
where $\delta$ is used as a simplified representation for $\delta_f$ under the assumption that $\delta_r=0$. The variable $L$, defined as $l_f+l_r$, represents the wheelbase of the vehicle.

we designate the state vector $\chi(t)=[x(t), y(t), \theta(t), v(t)]^T$ and the control input vector $u(t)=[\delta(t), a(t)]^T$. The governing equations, Eqs. (\ref{kinematic1})-(\ref{kinematic4}), describe a nonlinear model in the form of Eq. (\ref{general vehicle model}). For the purpose of deriving analytical solutions for vehicle trajectories, we perform a linearization of the nonlinear bicycle model based on Eq. (\ref{eq:general linearization}). This is executed by applying a Taylor series expansion at the initial state $\chi(0)$ and initial control input $u(0)$, and neglecting the higher-order terms:
\begin{equation}\label{eq:linearization}
\begin{split}
\dot{{\chi}}(t) \approx 
{f}_l({\chi}(t),{u(t)},{\chi(0)},{u(0)})=
{f}({\chi(0)},{u(0)})+\left.\frac{\partial f}{\partial \chi}\right|_{\substack{\chi=\chi(0) \\ u=u(0)}}({\chi(t)}-{\chi(0)})+\left.\frac{\partial f}{\partial u}\right|_{\substack{\chi=\chi(0) \\ u=u(0)}}({u(t)}-{u(0)})
\end{split}
\end{equation}

As stated in Subsection \ref{sec2.1}, specific vehicle/driver behaviors extend beyond the scope of this paper, therefore, inspired by the conventional TTC, we assume that $u(t)=u(0)$, meaning that the acceleration and steering angle are assumed to be invariant. Under this assumption, the calculation of the SSM only requires information of the current states and current control variables, as most conventional SSMs. We derive the following expression:

\begin{equation}\label{eq:kinematic linearized}
\begin{split}
\dot{{\chi}}(t) &\approx 
{f}({\chi(0)},{u(0)})+\left.\frac{\partial f}{\partial \chi}\right|_{\substack{\chi=\chi(0) \\ u=u(0)}}({\chi(t)}-{\chi(0)})\\
&=\left[
\begin{array}{c}
v_0 \cos(\theta_0) \\
v_0 \sin(\theta_0) \\
\frac{v_0 \tan(\delta_0)}{L} \\
a_0
\end{array}
\right]+\begin{bmatrix}
0 & 0 & -v_0 \sin(\theta_0) & \cos(\theta_0) \\
0 & 0 & v_0 \cos(\theta_0) & \sin(\theta_0) \\
0 & 0 & 0 & \frac{\tan(\delta_0)}{L} \\
0 & 0 & 0 & 0
\end{bmatrix}
(\chi(t) - \chi(0))\\
&=\begin{bmatrix}
0 & 0 & -v_0 \sin(\theta_0) & \cos(\theta_0) \\
0 & 0 & v_0 \cos(\theta_0) & \sin(\theta_0) \\
0 & 0 & 0 & \frac{\tan(\delta_0)}{L} \\
0 & 0 & 0 & 0
\end{bmatrix}
\chi(t) +
\left[
\begin{array}{c}
v_0 \sin(\theta_0)\theta_0 \\
-v_0 \cos(\theta_0)\theta_0 \\
0
\\
a_0
\end{array}
\right] 
\end{split}
\end{equation}
where subscript zero appended to each variable indicates evaluation at the initial time $t=0$.

The linearized model Eq. (\ref{eq:kinematic linearized}) conforms to an LTI model as depicted in Eq. (\ref{LTI vehicle model}). The first term in Eq. (\ref{eq:kinematic linearized}) corresponds to $A\chi(t)$ in Eq. (\ref{LTI vehicle model}), and the second term corresponds to $C$ in Eq. (\ref{LTI vehicle model}). Here, the matrix $B$ in Eq. (\ref{LTI vehicle model}) equals $\mathbf{0}_{4 \times 2}$. 

The matrix $A$ here also greatly benefits from the Nilpotent property that $A^3=\mathbf{0}_{4 \times 4}$, therefore, $e^{At}$ can be easily obtained by Eq. (\ref{exponential taylor series}) as follows:

\begin{equation}
e^{At} = I_4 + At + \frac{1}{2!}A^2t^2 + 0 + 0 + \dots = 
\left[
\begin{array}{cccc}
1 & 0 & -v_0 \sin(\theta_0) t & \cos(\theta_0) t - \frac{1}{2} v_0 \sin(\theta_0) \frac{\tan(\delta_0)}{L}t^2 \\
0 & 1 & v_0 \cos(\theta_0) t & \sin(\theta_0) t + \frac{1}{2} v_0 \cos(\theta_0) \frac{\tan(\delta_0)}{L}t^2 \\
0 & 0 & 1 & \frac{\tan(\delta_0)}{L}t \\
0 & 0 & 0 & 1
\end{array}
\right]
\end{equation}

Based on Eq. (\ref{LTI solution}), the future trajectory of a vehicle is obtained:
\begin{equation}
\begin{split}
\chi(t)=\begin{bmatrix}
x(t) \\
y(t) \\
\theta(t) \\
v(t)
\end{bmatrix}=&\left[
\begin{array}{cccc}
1 & 0 & -v_0 \sin(\theta_0) t & \cos(\theta_0) t - \frac{1}{2} v_0 \sin(\theta_0) \frac{\tan(\delta_0)}{L}t^2 \\
0 & 1 & v_0 \cos(\theta_0) t & \sin(\theta_0) t + \frac{1}{2} v_0 \cos(\theta_0) \frac{\tan(\delta_0)}{L}t^2 \\
0 & 0 & 1 & \frac{\tan(\delta_0)}{L}t \\
0 & 0 & 0 & 1
\end{array}
\right]
\begin{bmatrix}
x_0 \\
y_0 \\
\theta_0 \\
v_0
\end{bmatrix}\\
&+\int_0^t\left[
\begin{array}{cccc}
1 & 0 & -v_0 \sin(\theta_0) (t-\tau) & \cos(\theta_0) (t-\tau) - \frac{1}{2} v_0 \sin(\theta_0) \frac{\tan(\delta_0)}{L}(t-\tau)^2 \\
0 & 1 & v_0 \cos(\theta_0) (t-\tau) & \sin(\theta_0) (t-\tau) + \frac{1}{2} v_0 \cos(\theta_0) \frac{\tan(\delta_0)}{L}(t-\tau)^2 \\
0 & 0 & 1 & \frac{\tan(\delta_0)}{L}(t-\tau) \\
0 & 0 & 0 & 1
\end{array}
\right]
\left[
\begin{array}{c}
v_0 \sin(\theta_0)\theta_0 \\
-v_0 \cos(\theta_0)\theta_0 \\
0
\\
a_0
\end{array}
\right] d\tau
\\=&
\begin{bmatrix}
-\frac{1}{12} a_0 v_0 \sin(\theta_0) \frac{\tan(\delta_0)}{L} t^3 + \frac{1}{2} \left(a_0 \cos(\theta_0)-v_0^2\sin(\theta_0)\frac{\tan(\delta_0)}{L}\right) t^2+ v_0\cos(\theta_0)t + x_0 \\
\frac{1}{12} a_0 v_0 \cos(\theta_0) \frac{\tan(\delta_0)}{L} t^3 + \frac{1}{2} \left(a_0 \sin(\theta_0)+v_0^2\cos(\theta_0)\frac{\tan(\delta_0)}{L}\right) t^2 + v_0\sin(\theta_0)t + y_0 \\
a_0 \frac{\tan(\delta_0)}{2L} t^2 +\frac{v_0\tan(\delta_0)}{L}t+ \theta_0 \\
a_0 t + v_0
\end{bmatrix}
\end{split}
\end{equation}

In the context of analyzing vehicle-to-vehicle collisions, we consider two vehicles indexed by $i$ and $j$. For the brevity of the worked-out examples in this paper, we only consider a simplified geometric representation of these vehicles and approximate them by circumscribed circles. This approximation involves determining the minimum radius for each vehicle, such that the entire area of the vehicle is enclosed within the circle. These radii are denoted as $r_i$ and $r_j$ for vehicles $i$ and $j$, respectively. This method provides a simplified yet effective geometric representation for assessing potential collision scenarios between the two vehicles. 

\begin{figure}[h]
    \centering    \setlength{\abovecaptionskip}{0pt}    \includegraphics[width=0.4\textwidth]{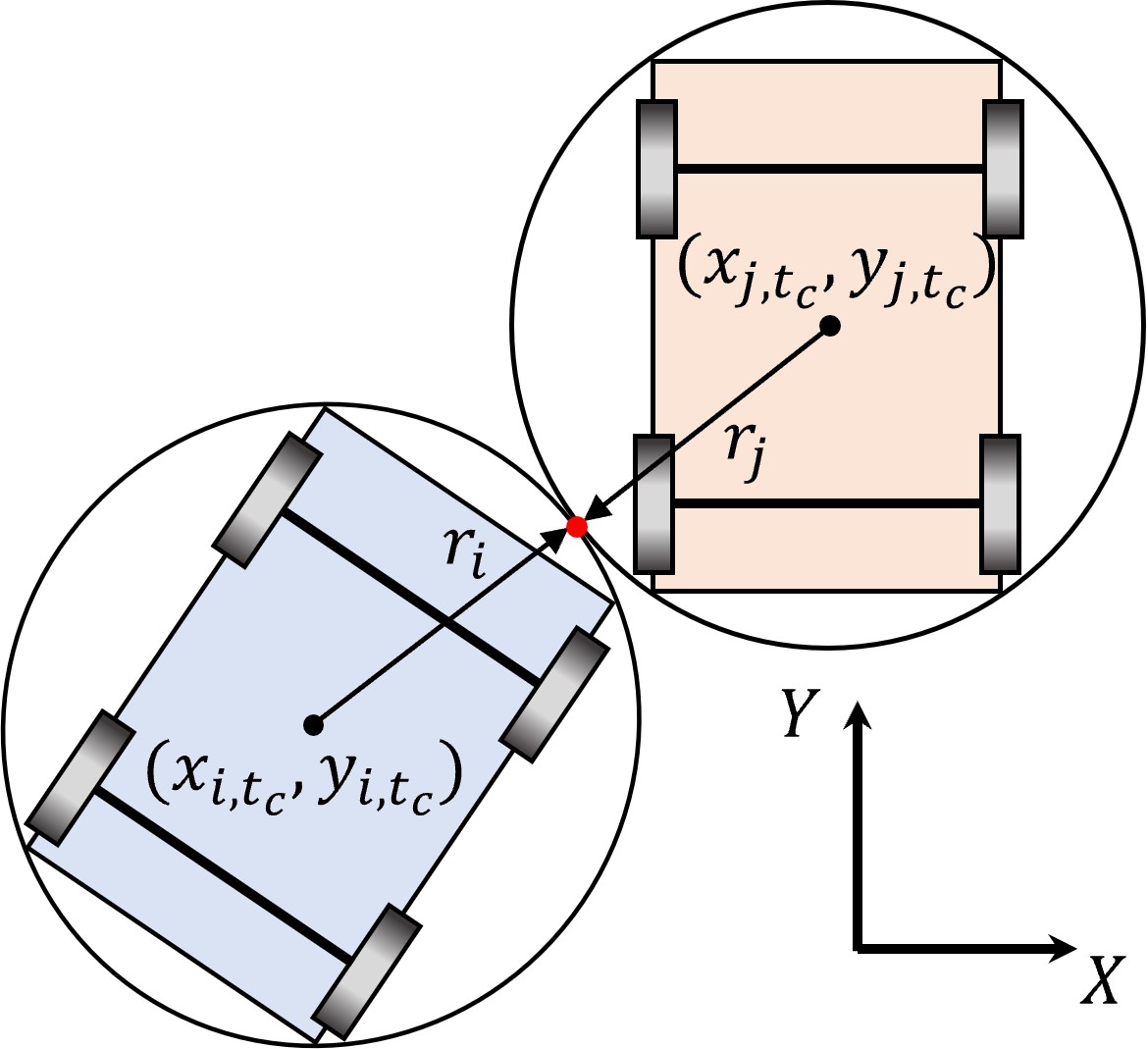}
    \caption{Visualization of bounding circles being tangential in 2-D space}
    \label{fig:kinematic collide}
\end{figure}

The instance of Eq. (\ref{general vehicle collide}) here quantitatively formulates the Euclidean distance between the C.G. of the two vehicles located at coordinates ($x_i,y_i$) and ($x_j,y_j$). This equation sets the distance equal to the sum of the radii $r_i+r_j$, which geometrically implies that the perimeters of the two circles are tangential to one another, intersecting at precisely one point. This condition is indicative of the critical threshold for an imminent collision between the two vehicles, assuming each vehicle is encapsulated within its respective bounding circle. The function $g_v(\cdot)$ is formulated as follows:

\begin{align}
    &g_v(\chi_i(t_c),\chi_{j}(t_c)) \nonumber\\
    =&(x_{i,t_c}-x_{j,t_c})^2+(y_{i,t_c}+y_{j,t_c})^2-(r_i+r_j)^2\nonumber\\
=&\biggl(-\frac{1}{12} \left(a_{i,0} v_{i,0} \sin(\theta_{i,0}) \frac{\tan(\delta_{i,0})}{L_i} - a_{j,0} v_{j,0} \sin(\theta_{j,0}) \frac{\tan(\delta_{j,0})}{L_j} \right) t_c^3 + \frac{1}{2} \biggl(a_{i,0} \cos(\theta_{i,0})-v_{i,0}^2\sin(\theta_{i,0})\frac{\tan(\delta_{i,0})}{L_i} \nonumber\\
&- a_{j,0} \cos(\theta_{j,0})+ v_{j,0}^2\sin(\theta_{j,0})\frac{\tan(\delta_{j,0})}{L_j}\biggr) t_c^2 + \left(v_{i,0}\cos(\theta_{i,0})-v_{j,0}\cos(\theta_{j,0})\right)t_c +(x_{i,0} - x_{j,0})\biggr)^2 \nonumber \\
&+ \biggl(\frac{1}{12} \biggl(a_{i,0} v_{i,0} \cos(\theta_{i,0}) \frac{\tan(\delta_{i,0})}{L_i} - a_{j,0} v_{j,0} \cos(\theta_{j,0}) \frac{\tan(\delta_{j,0})}{L_j} \biggr) t_c^3 \nonumber+ \frac{1}{2} \biggl(a_{i,0} \sin(\theta_{i,0}) +v_{i,0}^2\cos(\theta_{i,0})\frac{\tan(\delta_{i,0})}{L_i} \nonumber \\
&- a_{j,0} \sin(\theta_{j,0})-v_{j,0}^2\cos(\theta_{j,0})\frac{\tan(\delta_{j,0})}{L_j}\biggr) t_c^2 + \left(v_{i,0}\sin(\theta_{i,0})-v_{j,0}\sin(\theta_{j,0})\right)t_c+ (y_{i,0} - y_{j,0})\biggr)^2 - (r_i + r_j)^2  \label{kinematic collide}
\end{align}
where the subscripts $i$ and $j$ are utilized to designate the indices corresponding to individual vehicles, whereas the subscript $t_c$ denotes the temporal parameter at which a collision occurs. The vehicular interaction encapsulated by Eq. (\ref{kinematic collide}) is visualized in Fig. \ref{fig:kinematic collide}, wherein the regions between the vehicle body and the bounding circle serve as safety buffers. For an enhancement in the geometric fidelity of the shape of vehicles, the literature suggests the employment of alternative bounding constructs. These include using multiple, smaller bounding circles (\citep{reiter2023frenet}), or using a bounding rectangle (\citep{li2015unified}). It should be noted that using a bounding rectangle representation can in fact introduce much more complexity for the mathematical derivation (\cite{reiter2023frenet}). Therefore, the authors would suggest the multiple, small circle representation for improving geometric fidelity while keeping the derivation relatively simple. This can be easily generalized from the one-bounding circle representation that we exemplify in this paper.

cEq. (\ref{kinematic collide}) constitutes a sextic polynomial in $t_c$. Due to the polynomial's sixth-degree nature, extracting an closed-form solution for its roots presents a significant challenge. Nonetheless, the application of numerical algorithms, such as the companion matrix method (\citep{edelman1995polynomial}) and the Brent-Dekker method  (\citep{dekker1969finding,brent2013algorithms}), facilitates the acquisition of the roots in a numerical context. 

In the event that the set of solutions to Equation (\ref{kinematic collide}), obtained by equating it to zero, includes one or more non-negative values of $t_c$, the minimum non-negative value among these solutions, $t_c^*$, is the remaining time before an anticipated collision. Otherwise, no collision is anticipated.

The analysis of vehicle-to-obstacle collisions closely parallels. When dealing with an obstacle indexed as $k$ and enclosed by a circumscribed circle with a radius of $r_k$, the primary distinction lies in the fact that, instead of deriving a trajectory as a function of time $t$, the $x$ and $y$ coordinates of the obstacle remain constant throughout time.

\subsection{2-D motion and 3-D dynamics model in the path coordinate system\label{sec4.2}}
To further demonstrate how the proposed framework's capability of integrating with more sophisticated vehicle dynamics, thereby enhancing the fidelity and applicability of the model in real-world contexts, we propose its application to a scenario that incorporates 3-D dynamic modeling of a vehicle. Furthermore, the path coordinate system can be especially useful for investigating vehicle-to-road collisions. Therefore, for illustration, we transpose the model into the path coordinate system. Considering the complexity of this case, we decouple the vehicle movement model Eq. (\ref{general vehicle model}) into longitudinal and lateral vehicle dynamics.

\begin{figure}[h]
    \centering    \setlength{\abovecaptionskip}{0pt}    \includegraphics[width=0.6\textwidth]{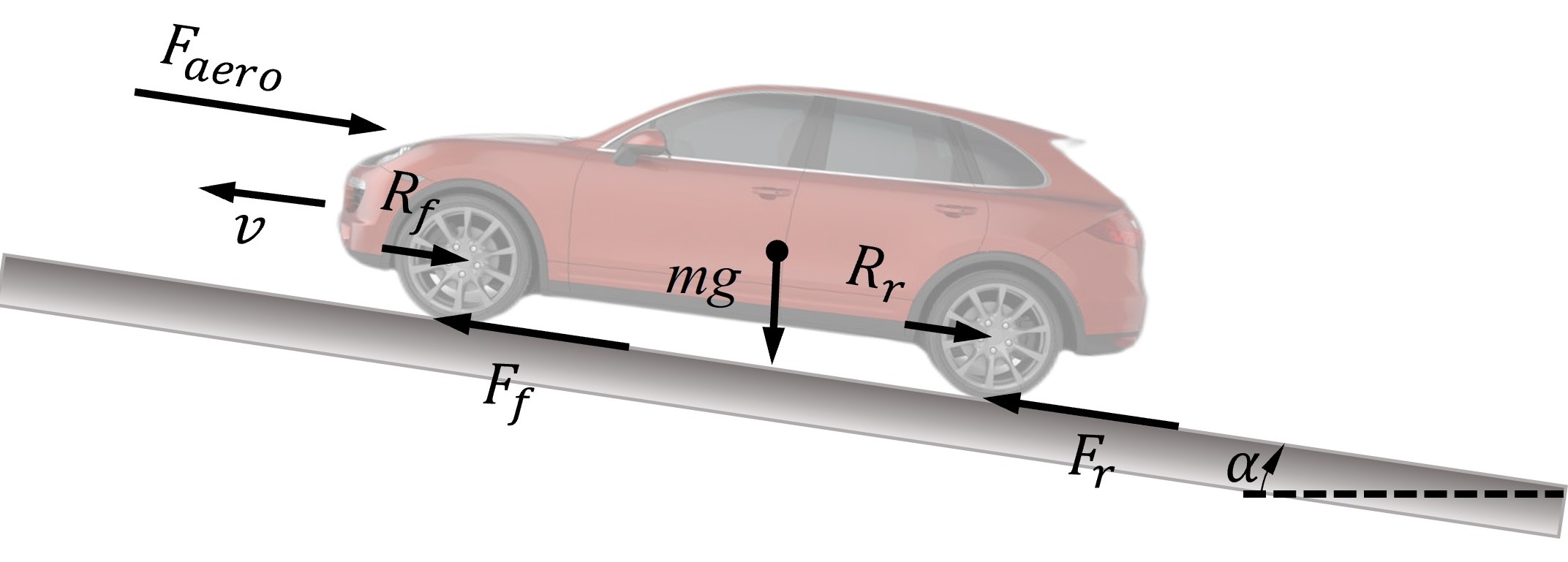}
    \caption{Longitudinal forces acting on a vehicle moving on an inclined road}
    \label{fig:longitudinal forces}
\end{figure}

We first introduce the longitudinal vehicle dynamics (\citep{rajamani2011vehicle}), which considers the balance of longitudinal forces acting on a vehicle, as depicted in Fig. \ref{fig:longitudinal forces}. The overall force balance equation is given as:
\begin{equation}
    ma = F_{f} + F_{r} - F_{aero} - R_{f} - R_{r} - mg\sin(\alpha) \label{eq:vehicle_dynamics}
\end{equation}
where \( F_{f} \) denotes the longitudinal force exerted by the front tires, which is a result of the tire-road interaction during vehicle motion. Similarly, \( F_{r} \) represents the analogous longitudinal force generated by the rear tires. \( F_{aero} \) corresponds to the longitudinal aerodynamic drag force, which is a function of the vehicle's velocity, shape, and the density of the air. \( R_{f} \) and \( R_{r} \) are forces attributable to the rolling resistance at the front and rear tires, respectively. \( m \) signifies the mass of the vehicle. \( g \) stands for the acceleration due to gravity. Lastly, \( \alpha \) is defined as the angle of the road's inclination relative to the horizontal, which influences the component of gravitational force acting along the vehicle's trajectory.

Defining the vehicle traction force, $F_{trac}$, as the sum of $F_f$ and $F_r$, it is calculated as $F_{trac}=\frac{T_{whl}}{r_{whl}}$ (\citep{xu2021energy}), where $T_{whl}$ is the wheel torque and $r_{whl}$ is the wheel radius. Similarly, defining the rolling resistance force, $F_{roll}$, as the sum of $R_f$ and $R_r$, it is calculated as $F_{roll}=f_{roll}\cos(\alpha)mg$ (\citep{xu2021energy}), where $f_{roll}$ is the rolling resistance coefficient. Furthermore, $F_{aero}$ is calculated as $F_{aero}=\frac{1}{2}\rho C_dSv^2$, where $\rho$ represents the ambient air density, $C_d$ denots the air drag coefficient, and $S$ represents the vehicle front projection area. (\citep{li2023sequencing}). Defining the longitudinal state $\chi_{lon}(t)=v(t)$ and the longitudinal control input being $T_{whl}(t)$. The governing equation is obtained as follows:
\begin{equation}
    \dot{v}(t)=a(t)=\frac{T_{whl}(t)}{mr_{whl}}-\frac{1}{2m}\rho C_dSv^2(t)-f_{roll}\cos(\alpha(t))g-g\sin(\alpha(t)) \label{longitudinal dy}
\end{equation}

Assuming the inclination $\alpha(t)$ is a known function of time, it can also be viewed as an input to the system. Therefore, the longitudinal control input vector becomes $u_{lon}(t)=[T_{whl}(t), \alpha(t)]^T$. For the purpose of deriving analytical solutions for vehicle trajectories, we perform a linearization based on Eq. (\ref{eq:general linearization}) at the initial state $\chi_{lon}(0)=v_0$ and initial control $u_{lon}(0)=[T_{whl,0}, \alpha_0]^T$:

\begin{equation}\label{eq:3-D lon linearized}
\begin{split}
\dot{\chi}_{lon}(t) \approx& 
{f}({\chi_{lon}(0)},{u_{lon}(0)})+\left.\frac{\partial f}{\partial \chi_{lon}}\right|_{\substack{\chi_{lon}=\chi_{lon}(0) \\ u_{lon}=u_{lon}(0)}}({\chi_{lon}(t)}-{\chi_{lon}(0)})+\left.\frac{\partial f}{\partial u_{lon}}\right|_{\substack{\chi_{lon}=\chi_{lon}(0) \\ u_{lon}=u_{lon}(0)}}({u_{lon}(t)}-{u_{lon}(0)})\\
=&\left(\frac{T_{whl,0}}{mr_{whl}}-\frac{1}{2m}\rho C_dSv^2_0-f_{roll}\cos(\alpha_0)g-g\sin(\alpha_0)\right)-\frac{\rho C_dSv_0}{m}({\chi_{lon}(t)}-{\chi_{lon}(0)})\\
&+\left[ \begin{array}{cc}
\frac{1}{m r_{whl}} &f_{roll} \sin(\alpha_0) g - g \cos(\alpha_0)
\end{array} \right]({u_{lon}(t)}-{u_{lon}(0)})\\
=&-\frac{\rho C_dSv_0}{m}{\chi_{lon}(t)}+\left[ \begin{array}{cc}
\frac{1}{m r_{whl}} &f_{roll} \sin(\alpha_0) g - g \cos(\alpha_0)
\end{array} \right]{u_{lon}(t)}+\biggl(\frac{1}{2m}\rho C_dSv^2_0\\
&-f_{roll}\cos(\alpha_0)g-g\sin(\alpha_0)-\alpha_0f_{roll}\sin(\alpha_0) g + \alpha_0g \cos(\alpha_0)\biggr)
\end{split}
\end{equation}
where the first term specifically represents $A\chi(t)$ in Eq. (\ref{LTI vehicle model}), the second term corresponds to $Bu(t)$ in Eq. (\ref{LTI vehicle model}), and the third term represents $C$ in Eq. (\ref{LTI vehicle model}).

Similarly, as we do not consider specific vehicle/driver behaviors in this paper, we assume that $T_{whl}$ remains invariant so that $T_{whl}(t)=T_{whl,0}$ for all $t>0$, inspired by the conventional TTC. Based on this assumption, we only need information on the current states and control variables to calculate the SSM. Based on Eq. (\ref{LTI solution}), the longitudinal velocity trajectory is obtained as follows:

\begin{equation}
\begin{split}
    \chi_{lon}(t)=v(t)=&e^{-\frac{\rho C_dSv_0}{m}t}v_0+\int_0^te^{-\frac{\rho C_dSv_0}{m}(t-\tau)}\left[\begin{array}{cc}
\frac{1}{m r_{whl}} &f_{roll} \sin(\alpha_0) g - g \cos(\alpha_0)
\end{array} \right] \left[\begin{array}{c}
T_{whl,0} \\ \alpha(\tau)
\end{array} \right]d\tau\\
&+\int_0^te^{-\frac{\rho C_dSv_0}{m}(t-\tau)}\biggl(\frac{1}{2m}\rho C_dSv^2_0-f_{roll}\cos(\alpha_0)g-g\sin(\alpha_0)-\alpha_0f_{roll}\sin(\alpha_0) g \\
&+ \alpha_0g \cos(\alpha_0)\biggr)d\tau\\
=&e^{-\frac{\rho C_dSv_0}{m}t}v_0-\frac{me^{-\frac{\rho C_dSv_0}{m}t}-m}{\rho C_dSv_0}\biggl(\frac{T_{whl,0}}{mr_{whl}}+\frac{1}{2m}\rho C_dSv^2_0-f_{roll}\cos(\alpha_0)g-g\sin(\alpha_0)\\
&-\alpha_0f_{roll}\sin(\alpha_0) g + \alpha_0g \cos(\alpha_0)\biggr)+\int_0^te^{-\frac{\rho C_dSv_0}{m}(t-\tau)}\left(f_{roll} \sin(\alpha_0) g - g \cos(\alpha_0)\right)\alpha(\tau)d\tau \label{3-D lon trajectory}
\end{split}
\end{equation}

As depicted by Eq. (\ref{3-D lon trajectory}), the longitudinal velocity trajectory depends on the specific expression of $\alpha(t)$.

\begin{figure}[h]
    \centering    \setlength{\abovecaptionskip}{0pt}    \includegraphics[width=0.5\textwidth]{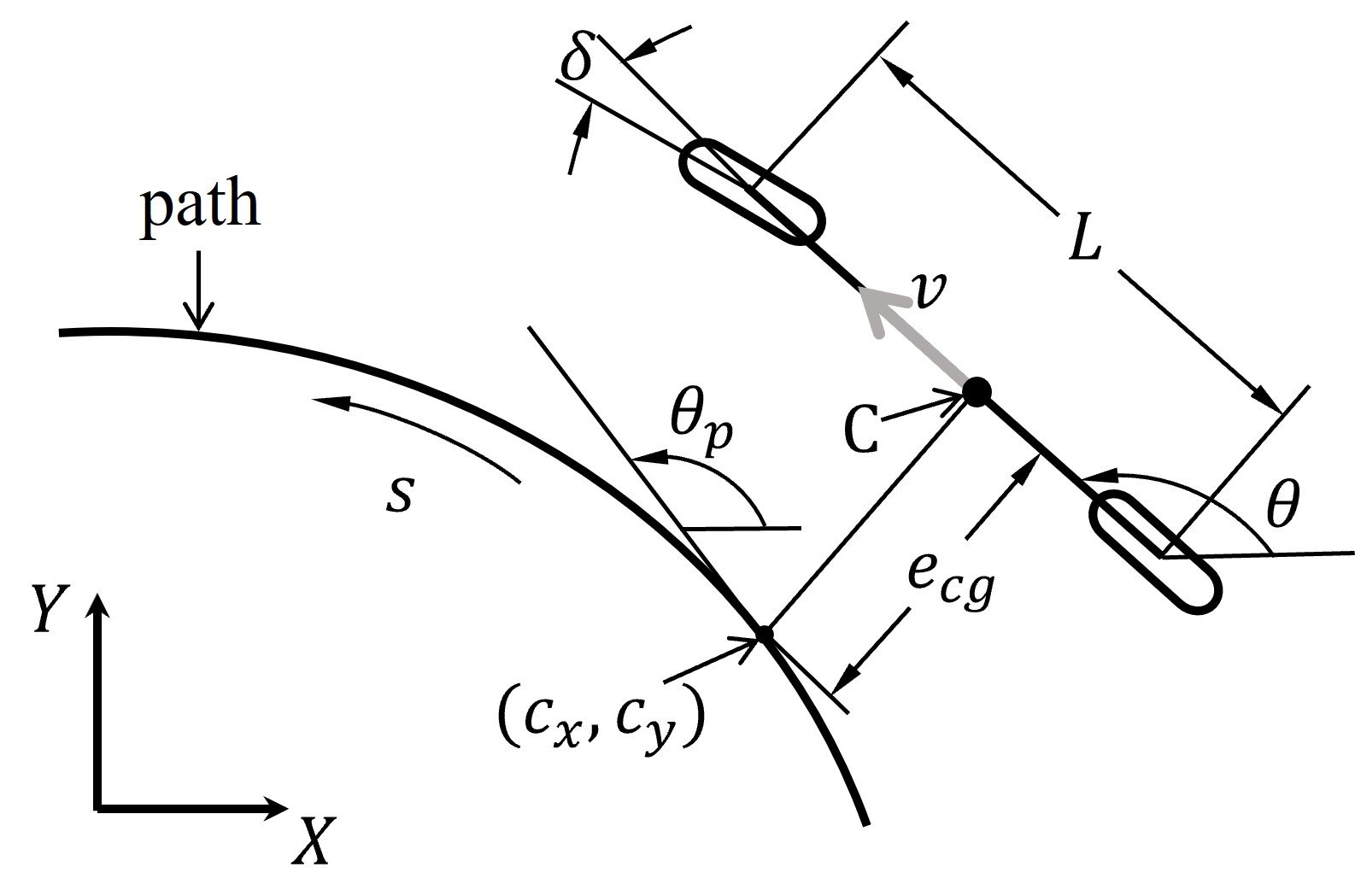}
    \caption{Lateral kinematic bicycle model in the path coordinate system}
    \label{fig:kinematic_path}
\end{figure}

Upon determination of the longitudinal velocity trajectory, it can be integrated as one of the inputs of the lateral dynamic model. Subsequently, we proceed to introduce the lateral kinematic bicycle model within the path coordinate system (\citep{snider2009automatic}), which transposes the model described by Eqs. (\ref{kinematic1})-(\ref{kinematic3}) into the framework of the path, as shown in Fig. \ref{fig:kinematic_path}. The overall equations of motion within the path coordinate system are:
\begin{equation}
\dot{s} = v \left( \frac{\cos(\theta_e)}{1 - e_{cg} \kappa(s)} \right)
\end{equation}
\begin{equation}
\dot{e}_{cg} = v \sin(\theta_e)
\end{equation}
\begin{equation}
\dot{\theta}_e = v \left( \frac{\tan(\delta)}{L} - \frac{\kappa(s) \cos(\theta_e)}{1 - e_{cg} \kappa(s)} \right)
\end{equation}
where $s$ designates the arc length parameter describing the vehicle's progression along the path. $e_{cg}$ quantifies the perpendicular deviation of the vehicle's C.G. from the path. The term $\theta_e$, known as the heading or orientation error, measures the angular discrepancy between the vehicle's current orientation and the path's orientation, and is mathematically expressed as $\theta-\theta_p(s)$, where $\theta_p(s)$ defines the orientation of the path's tangent relative to the Cartesian $X$ axis. The path curvature, which is a function of the arc length, is denoted by $\kappa(s)$ and reflects the rate of change of the path's direction with respect to the arc length. 

It's worth emphasizing that when dealing with more intricate road geometries, like super-elevations, a more intricate model that incorporates lateral tire forces becomes essential (\citep{rajamani2011vehicle}). However, for the sake of brevity and clarity, we do not delve into such scenarios within this paper. It's worth noting that the derivation of solutions for these cases follows a similar approach as the ones presented in this paper.

For the sake of simplicity, we assume that the curvature is a known function of time, therefore denoted as $\kappa(t)$. We define the lateral state vector $\chi_{lat}(t)=[s(t), e_{cg}(t),\theta_e(t)]^T$, and the control input vector $u_{lat}(t)=[\delta(t), v(t), \kappa(t)]^T$. To obtain a closed-form solution of the lateral trajectory, we perform the linearizetion described by (\ref{eq:general linearization}) at the operating point ($\chi_{lat}(0),u_{lat}(0)$), where $\chi_{lat}(0)=[s_0,e_{cg,0},\theta_{e,0}]^T$ and $u_{lat}(0)=[\delta_0, v_0, \kappa_0]^T$:
\begin{equation}
\begin{split}
\dot{\chi}_{lat}(t) \approx& 
{f}({\chi_{lat}(0)},{u_{lat}(0)})+\left.\frac{\partial f}{\partial \chi_{lat}}\right|_{\substack{\chi_{lat}=\chi_{lat}(0) \\ u_{lat}=u_{lat}(0)}}({\chi_{lat}(t)}-{\chi_{lat}(0)})+\left.\frac{\partial f}{\partial u_{lat}}\right|_{\substack{\chi_{lat}=\chi_{lat}(0) \\ u_{lat}=u_{lat}(0)}}({u_{lat}(t)}-{u_{lat}(0)}) \\
=& 
\begin{bmatrix}
\frac{v_0 \cos(\theta_{e,0})}{1 - e_{cg,0} \kappa_0} \\
v_0 \sin(\theta_{e,0}) \\
v_0 \left( \frac{\tan(\delta_0)}{L} - \frac{\kappa_0 \cos(\theta_{e,0})}{1 - e_{cg,0} \kappa_0} \right)
\end{bmatrix}
+\begin{bmatrix}
0 & \frac{v_0 \cos(\theta_{e,0}) \kappa_0}{(1 - e_{cg,0} \kappa_0)^2} & -\frac{v_0 \sin(\theta_{e,0})}{1 - e_{cg,0} \kappa_0} \\
0 & 0 & v_0 \cos(\theta_{e,0}) \\
0 & -\frac{v_0 \cos(\theta_{e,0}) \kappa_0^2}{(1 - e_{cg,0} \kappa_0)^2} & \frac{v_0 \kappa_0 \sin(\theta_{e,0})}{1 - e_{cg,0} \kappa_0}
\end{bmatrix}({\chi_{lat}(t)}-{\chi_{lat}(0)}) \\
&+\begin{bmatrix}
0 & \frac{\cos(\theta_{e,0})}{1 - e_{cg,0} \kappa_0} & \frac{v_0 \cos(\theta_{e,0}) e_{cg,0}}{(1 - e_{cg,0} \kappa_0)^2} \\
0 & \sin(\theta_{e,0}) & 0 \\
\frac{v_0}{L\cos^2(\delta_0)} & \frac{\tan(\delta_0)}{L} - \frac{\kappa_0 \cos(\theta_{e})}{1 - e_{cg,0} \kappa_0} & -\frac{v_0 \cos(\theta_{e,0})}{(1 - e_{cg,0} \kappa_0)^2}
\end{bmatrix}
\left(u_{lat}(t) - u_{lat}(0)\right) \\
=&\begin{bmatrix}
0 & \frac{v_0 \cos(\theta_{e,0}) \kappa_0}{(1 - e_{cg,0} \kappa_0)^2} & -\frac{v_0 \sin(\theta_{e,0})}{1 - e_{cg,0} \kappa_0} \\
0 & 0 & v_0 \cos(\theta_{e,0}) \\
0 & -\frac{v_0 \cos(\theta_{e,0}) \kappa_0^2}{(1 - e_{cg,0} \kappa_0)^2} & \frac{v_0 \kappa_0 \sin(\theta_{e,0})}{1 - e_{cg,0} \kappa_0}
\end{bmatrix}{\chi_{lat}(t)}+\begin{bmatrix}
0 & \frac{\cos(\theta_{e,0})}{1 - e_{cg,0} \kappa_0} & \frac{v_0 \cos(\theta_{e,0}) e_{cg,0}}{(1 - e_{cg,0} \kappa_0)^2} \\
0 & \sin(\theta_{e,0}) & 0 \\
\frac{v_0}{L\cos^2(\delta_0)} & \frac{\tan(\delta_0)}{L} - \frac{\kappa_0 \cos(\theta_{e})}{1 - e_{cg,0} \kappa_0} & -\frac{v_0 \cos(\theta_{e,0})}{(1 - e_{cg,0} \kappa_0)^2}
\end{bmatrix}
u_{lat}(t)\\
&+\begin{bmatrix}
-\frac{2v_0 \cos(\theta_{e,0}) \kappa_0 e_{cg,0}}{(1 - e_{cg,0} \kappa_0)^2} + \frac{v_0 \sin(\theta_{e,0}) \theta_{e,0}}{1 - e_{cg,0} \kappa_0} \\
v_0 \cos(\theta_{e,0}) \theta_{e,0} \\
\frac{v_0 \cos(\theta_{e,0}) \kappa_0^2 e_{cg,0}}{(1 - e_{cg,0} \kappa_0)^2} - \frac{v_0 \kappa_0 \sin(\theta_{e,0}) \theta_{e,0}}{1 - e_{cg,0} \kappa_0} - \frac{v_0 \delta_0}{L\cos^2(\delta_0)} + \frac{v_0 \cos(\theta_{e,0}) \kappa_0}{(1 - e_{cg,0} \kappa_0)^2}
\end{bmatrix}
 \label{3_D lat linearize}
\end{split}
\end{equation}

The linearized model Eq. (\ref{3_D lat linearize}) conforms to an LTI model as depicted in Eq. (\ref{LTI vehicle model}). The first, second, and third term in Eq. (\ref{3_D lat linearize}) correspond to $A\chi(t)$, $Bu(t)$, and $C$ in Eq. (\ref{LTI vehicle model}), respectively.

Inspired by the conventional TTC, we assume that the steering wheel remains invariant, therefore $\delta(t)=\delta_0$. Further based on Eq. (\ref{LTI solution}), the vehicle trajectory can be calculated. However, the specific calculation of the trajectory follows the process of the non-diagonalizable case introduced in Subsection \ref{sec2.2} and depends on the specific values of the entries of matrix $A$ and the expression of $v(t)$, which is calculated by Eq. (\ref{3-D lon trajectory}), and $\kappa(t)$, which is a known external input.

With the expression of the lateral trajectory $\chi_{lat}(t)$ calculated, the trajectory of the perpendicular deviation, $e_{cg}(t)$, can be extracted. Similar to Subsection \ref{sec4.2}, we bound the i-th vehicle's shape in the path coordinate system with its circumscribed circle, with the circle's radius denoted as $r_i$. For a road with width $w$, the two road boundaries are described by $bound_1=\frac{w}{2}$ and $bound_2=-\frac{w}{2}$ in the path coordinate system, respectively. Therefore, for vehicle-to-road collisions, $g_r(\cdot)$ in Eq. (\ref{general road collide}) becomes:
\begin{equation}
    g_r(\chi_{i,lat}(t_c),R_k)=e_{i,cg}(t_c)+\text{sign}(bound_k)\times r_i-bound_k, \text{~~} k\in\{1,2\} \label{3-D g}
\end{equation}

By setting Eq. (\ref{3-D g}) equate to zero, we attempt to solve for a time $t_c$ such that the maximum absolute perpendicular deviation on the circumscribe circle reaches the road boundary. If non-negative solutions of $t_c$ are obtained, a collision with the road boundary is expected, and the remaining time before collision is the smallest non-negative solution, $t_c^*$. Otherwise, there is no risk of collision. 

\subsection{Discussion of cases beyond analytical trajectory solutions \label{sec4.3}}
The analytical solutions derived from LTI models confer distinct advantages in terms of possessing an explicit expression for the future trajectory and its interpretability. However, if the original model Eq. (\ref{general vehicle model}) embodies intrinsic nonlinearity and is subjected to a linearization process as specified by Eq. (\ref{eq:general linearization}), the precision is predominantly confined near the operating point at which linearization occurs. To preserve the fidelity of the vehicle's trajectory representation when confronting significant nonlinear dynamics, one may resort to numerical integration techniques such as the Euler method and the Runge-Kutta methods (\citep{burden2011numerical}). These methods involve discretizing the continuous model (breaking it into small steps) and iteratively finding approximate solutions at each step. Consequently, these methods generate a sequence of discrete trajectory points, which collectively approximate the vehicle's future states with, in general,  greater accuracy. As an illustrative case, given the model Eq. (\ref{general vehicle model}), the initial states $\chi(0)$, and the function $u(t)$, the explicit Runge-Kutta methods are formulated as follows:
\begin{equation}
    \chi_0=\chi(0) \label{RK1}
\end{equation}
\begin{equation}
    \chi_{l+1}=\chi_{l}+h\sum_j^vb_jk_j, \text{~~}l\in\{0,1,\cdots,N-1\}
\end{equation}
with

\begin{equation}
    k_{i,l}=hf(\chi_{l}+h\sum_{j=1}^{i-1}a_{i,j}k_{j,l},u(lh+c_ih)), \text{~~}i\in\{0,1,\cdots,v\} \label{RK6}
\end{equation}

%\begin{equation}
%    k_{1,l}=hf(\chi_{l},u(lh)) \label{RK2}
%\end{equation}
%\begin{equation}
%    k_{2,l}=hf(\chi_{l}+0.5k_{1,l},u(lh+0.5h)) \label{RK3}
%\end{equation}
%\begin{equation}
 %   k_{3,l}=hf(\chi_{l}+0.5k_{2,l},u(lh+0.5h)) \label{RK4}
%\end{equation}
%\begin{equation}
 %   k_{4,l}=hf(\chi_{l}+k_{3,l},u((l+1)h)) \label{RK5}
%\end{equation}

%\begin{equation}
%\chi_{l+1}=\chi_{l}+\frac{1}{6}(k_{1,l}+2k_{2,l}+2k_{3,l}+k_{4,l}), \text{~~}l\in\{0,1,\cdots,N-1\} \label{RK6}
%\end{equation}
\noindent where $\chi_l$ denotes the value of the state vector at discrete time step $l$, $v$ denotes the order of the specific Runge-Kutta method. $k_{i,l}$ are intermediately evaluated derivatives at step $l$. $h$ represents the temporal interval between successive discrete time steps, and $N$ denotes the total number of discrete steps required for the computation. $a_{i,j}, b_j, c_i$ are coefficients obtained by the Butcher tableau (\citep{butcher2007runge}).

As illustrated by Eqs. (\ref{RK1})-(\ref{RK6}), numerical integration methods construct a trajectory by iteratively calculating a series of discrete points that approximate the system's future states. This trajectory discretization enables the assessment of potential collisions without directly solving Eqs. (\ref{general vehicle collide}) and/or (\ref{general road collide}). Instead, the functions $g_v(\chi_{i,l}, \chi_{j,l})$ and $g_r(\chi_{i,l}, R_k)$ are evaluated at each discrete time step $l$. A change in the sign of $g_v(\cdot)$ and/or $g_r(\cdot)$ indicates an expected collision. Consequently, the earliest time step at which this sign change is observed provides an approximation for the remaining time before collision. 

However, the step-size $h$ is a critical factor in the detection of collision points. To mitigate the risk of overlooking potential collisions and to reduce the errors associated with discretization, it is imperative to choose a sufficiently small $h$. In order to calculate a future trajectory over extended time horizons, a substantial number of discrete steps need to be calculated by the numerical methods, imposing a higher computational demand. Moreover, when employing numerical methods that offer reduced truncation errors (e.g., the Runge-Kutta fourth-order method, the Runge-Kutta-Fehlberg method), additional function evaluations per step are required, leading to a further increment in computational intensity (\citep{burden2011numerical}).

\subsection{Tradeoffs between analytical and numerical solutions}
\label{sec4.4}
It should be noted that there exist some tradeoffs between the analytical solution presented in Section \ref{sec2} and the numerical solution discussed in Subsection \ref{sec4.3}. In this subsection, we provide some discussions on these tradeoffs for practitioners.

One inherent tradeoff centers on a key dichotomy: analytical methods provide deep insights at the cost of significant analytical effort, whereas numerical methods, although less demanding in analytical deductions, require extensive computation and offer limited insights. As shown by the major part of this paper, the analytical framework requires symbolic deductions of vehicle trajectories, these deductions can become intense when considering high-dimensional and high-fidelity models as presented in Subsections \ref{sec4.1}-\ref{sec4.2}. The advantage of this is that we obtain explicit vehicle trajectories and distance functions that are functions of time $t$, hence more insights can be provided by utilizing analytical analysis tools. Furthermore, the only real-time calculation required is to solve the collision equations (which can be solved either analytically or numerically, as discussed in previous subsections). On the other hand, numerical solutions barely require any analytical deduction. However, as mentioned in Subsection \ref{sec4.3}, numerical solutions generate discrete trajectory points in a point-by-point fashion in real-time, and all the trajectory points need to be traversed in real-time to check for potential collisions. Furthermore, to mitigate the risk of overlooking
potential collisions and to reduce the errors associated with discretization, it is preferred to choose a smaller time-step $h$. Therefore, with longer future duration considered (i.e., more trajectory points with a fixed time-step $h$) and smaller time-step used (i.e., more trajectory points in a fixed duration), the real-time computational burden become heavier, and can eventually cause difficulties on real-time implementations. Furthermore, numerical methods yield fewer insights since only a sequence of trajectory points is obtained.

Another consideration is the trade-off related to accuracy. It should be noted that this tradeoff does not always exist. For instance, for the conventional SSMs considered in this paper, the vehicle movement models are inherently linear and the representation of vehicle shape is relatively simple (i.e., vehicle length in one direction), and the analytical solutions of the proposed framework remain accurate as they are the exact solutions. The loss of accuracy of the analytical framework arises when the vehicle movement model is nonlinear, and linearization needs to be conducted for the purpose of an analytical solution, errors are introduced when ignoring higher-order terms as shown in Eq. (\ref{eq:general linearization}), however, the introduced error can be subtle if the vehicle movement model is not "strongly" nonlinear. On the other hand, with a sufficiently small time-step $h$, the numerical solutions remain accurate, while the disadvantages discussed in the previous paragraph hold.

\section{Simulation experiments \label{sec5}}
In this section, we analyze the accuracy of the proposed analytical framework when relying on linearized models (comparing to high-accuracy numerical solutions), compare the proposed extended 3D SSMs with conventional 1D SSMs, and provide a real-world application example to illustrate its practicality. 

As elucidated in Subsection \ref{sec4.3}, the act of linearizing a nonlinear model introduces errors. In Subsections \ref{sec5.1} through \ref{sec5.2}, our objective is to assess the accuracy of the analytical framework based on linearized vehicle movement models. Specifically, Subsection \ref{sec5.1} examines the SSM derived from the linearized 2-D kinematic model in the Cartesian coordinate system, as outlined in Subsection \ref{sec4.1}. Meanwhile, Subsection \ref{sec5.2} delves into the SSM obtained from the 2-D motion and 3-D dynamics model in the path coordinate system, as demonstrated in Subsection \ref{sec4.2}. Both of these subsections validate the analytical framework's results using the Runge-Kutta fourth-order method (\citep{burden2011numerical}). The Runge-Kutta fourth-order method possesses a truncation error of $O(h^4)$, and with a suitably small step size $h$, its outcomes can be considered as the ground truth. A comparison of a conventional 1D SSM and a proposed extended 3D SSM is shown in Subsection \ref{sec5.3} to demonstrate the necessity, accuracy, and reliability of higher-dimensional, higher-fidelity SSMs. Furthermore, in Subsection \ref{sec5.4}, we provide an illustration of an application that employs both the Cartesian coordinates model and the path coordinates model to comprehensively consider all potential collisions.

\subsection{2-D kinematic Cartesian coordinate model experiment \label{sec5.1}}
\begin{figure}[h]
    \centering
    \setlength{\abovecaptionskip}{0pt}
     {\includegraphics[width=0.4\textwidth]{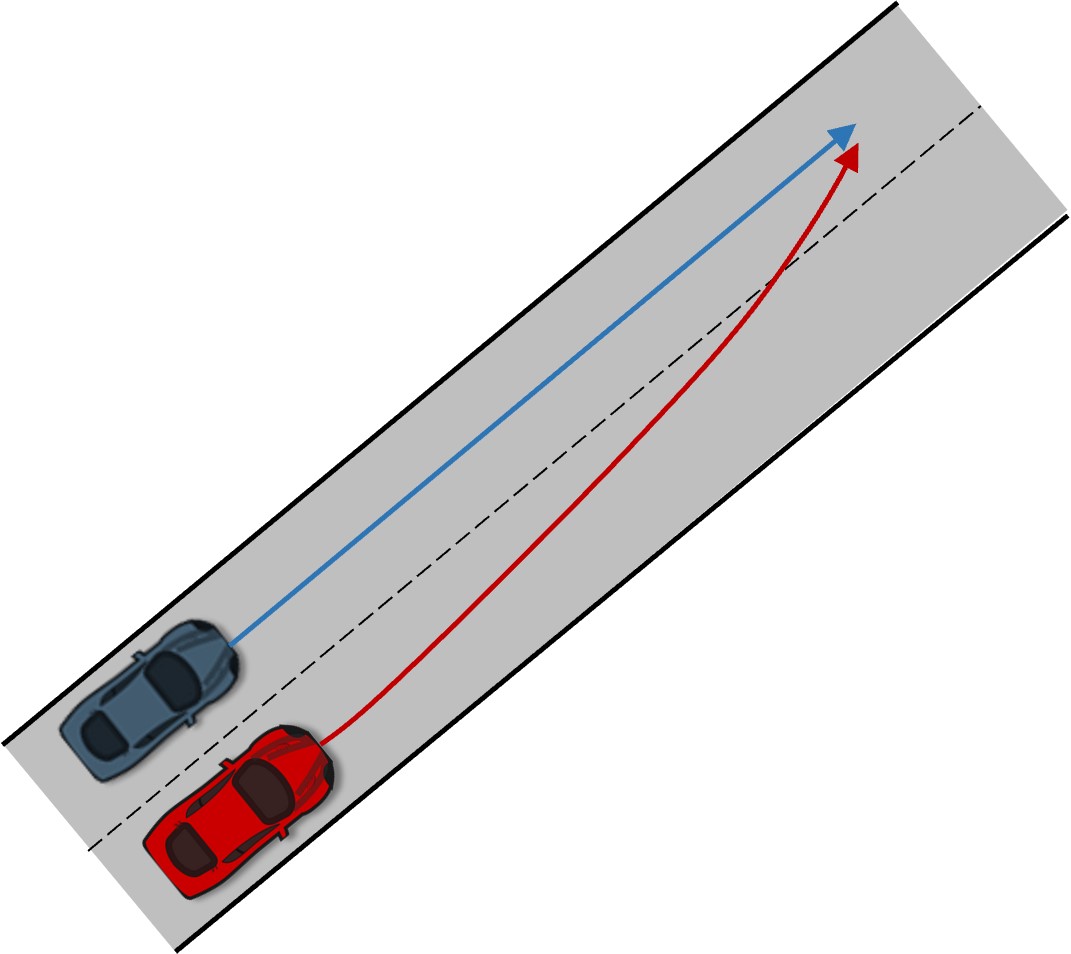}}
    \caption{Experiment 1 scenario}
    \label{experiment1}
\end{figure}

In this subsection, we conduct an evaluation of the accuracy of the SSM formulated in Subsection \ref{sec4.1}. For this purpose, we examine a specific driving scenario, as illustrated in Fig. \ref{experiment1}, wherein two vehicles initiate their trajectories in adjacent lanes. The first vehicle executes a lane-changing maneuver accompanied by a marginal reduction in speed, whereas the second vehicle maintains a consistent velocity within its original lane. An ensuing collision is precipitated by the convergence of the vehicles' trajectories, leading to an overlap in their spatial occupation.

We index the vehicle performing lane-changing as Vehicle 1, while the other vehicle is indexed as Vehicle 2. The parameters that we use are listed in Table \ref{experiment 1 parameters}. The superscript 0 indicates initial values at the simulation start time, which we define as when the simulation time, $T_r$, equals to zero. For example, $x_{1}^0$ is the initial $x$ coordinate of Vehicle 1 at $T_r=0$. 

\begin{figure}[h]
    \centering
    \setlength{\abovecaptionskip}{0pt}
    \subcaptionbox{}
    {\includegraphics[width=0.45\textwidth]{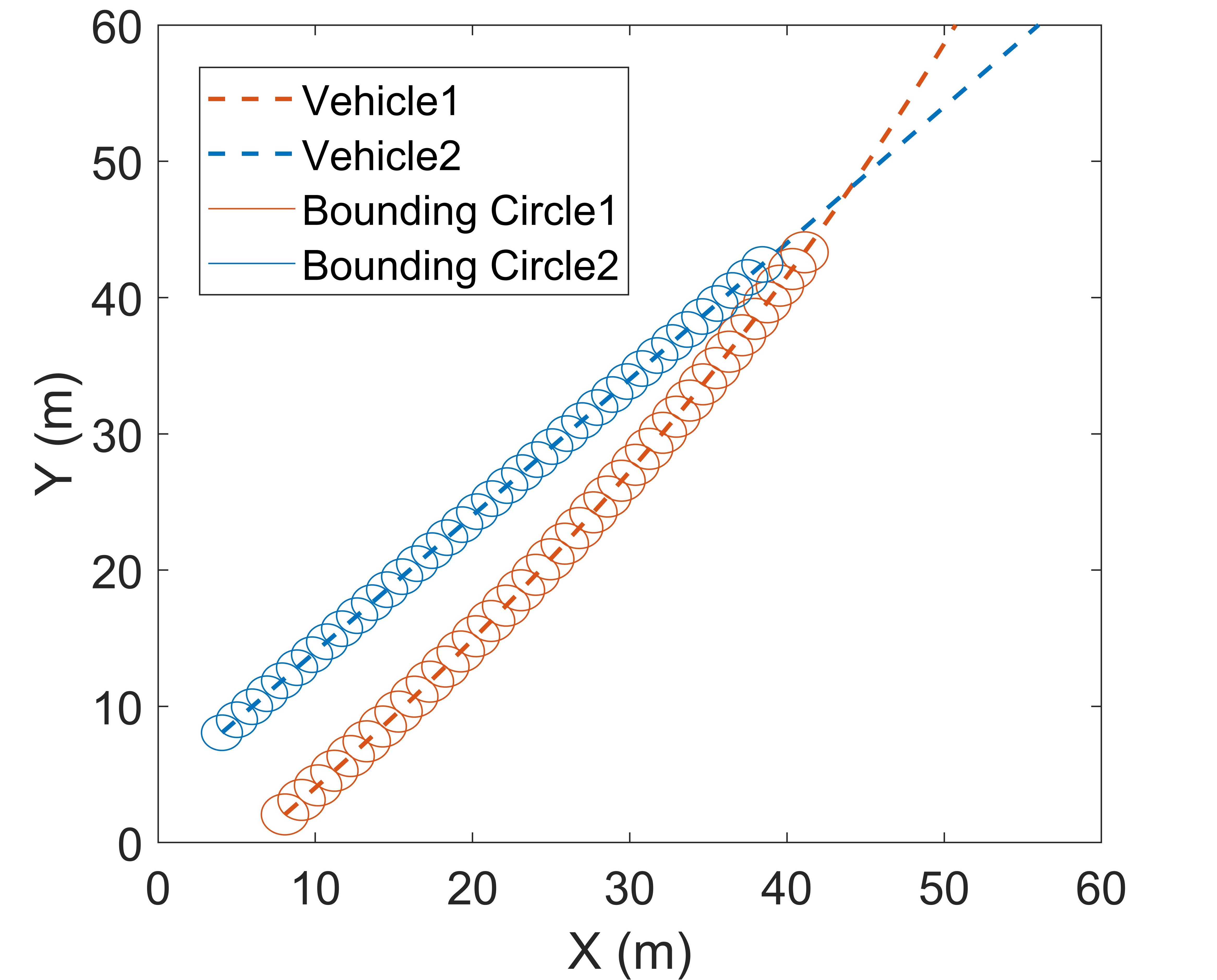}}
    \subcaptionbox{}{\includegraphics[width=0.45\textwidth]{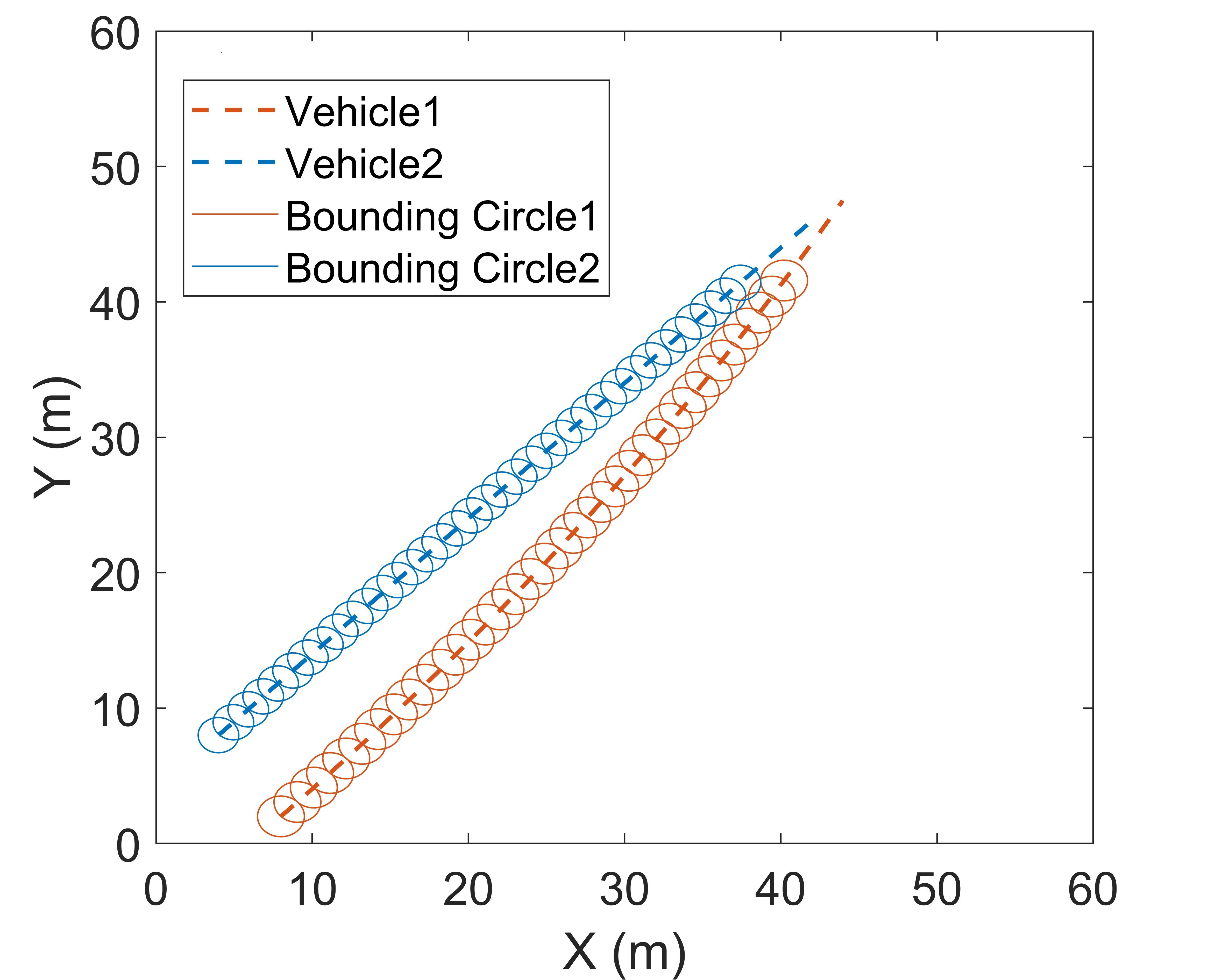}}
    \caption{Derived trajectories and bounding circles of experiment 1 at $T_r=0$: (a) analytical based on linearized model; (b) numerical based on Runge-Kutta}
    \label{experiment1 traj}
\end{figure}

\begin{table}[h]
\centering
\caption{Experiment 1 parameters}
\label{experiment 1 parameters}
\begin{tabular}{l l || l l}
\hline
parameter & value & parameter & value \\ \hline
$x^0_{1}$ & $8 (m)$ & $x^0_{2}$ & $4 (m)$\\ 
$y^0_{1}$ & $2 (m)$ & $y^0_{2}$ & $8 (m)$\\ 
$\theta^0_{1}$ & $\pi/4 (rad)$ & $\theta^0_{2}$ & $\pi/4 (rad)$\\
$v^0_{1}$ & $10 (m/s)$ & $v^0_{2}$ & $9 (m/s)$\\
$\delta^0_{1}$ & $0.01 (rad)$ & $\delta^0_{2}$ & $0 (rad)$\\
$a^0_{1}$ & $-0.1 (m/s^2)$ & $a^0_{2}$ & $0 (m/s^2)$\\
$L_1$ & $2.5 (m)$ & $L_2$ & $2 (m)$\\
$r_1$ & $1.5 (m)$ & $r_2$ & $1.3 (m)$\\
$h$ & $0.001 (s)$& $N$ & $6000$\\
\hline
\end{tabular}
\end{table}

To examine the variation in the SSM with decreasing relative distance between the two vehicles over simulaiton time $T_r$, we hold constant the accelerations and steering wheel angles for both vehicles. Consequently, $a^{T_r}_1=a^{0}_1$, $\delta^{T_r}_1=\delta^0_1$ for Vehicle 1, and similarly $a^{T_r}_2=a^{0}_2$, $\delta^{T_r}_2=\delta^0_2$ for Vehicle 2, apply for all $T_r>0$. To assess the inaccuracies resulting from the linearization of the model, we initially focus on the trajectory evolution estimated at $T_r=0$ as obtained by both the analytical and the numerical Runge-Kutta methods, shown in Fig. \ref{experiment1 traj}(a) and Fig. \ref{experiment1 traj}(b), respectively. Furthermore, bounding circles are drawn at intervals of 0.15 seconds along the trajectories to visualize the progression until the point where a collision is predicted by the corresponding method.

Fig. \ref{experiment1 traj} reveals that the estimations of trajectory evolution made by the analytical method, which relies on a linearized model, effectively grasp the directional trends of the vehicles' movements. Additionally, the analytical method provides a closed-form expression for the trajectories, enabling the calculation of the vehicles' trajectories at any future time $t>0$ by simply substituting the time value into the expression. In contrast, the numerical Runge-Kutta method is limited to estimating discrete trajectory points at the predefined $N$ time steps.

\begin{figure}[h!]
    \centering
    \setlength{\abovecaptionskip}{0pt}
    \subcaptionbox{}
    {\includegraphics[width=0.45\textwidth]{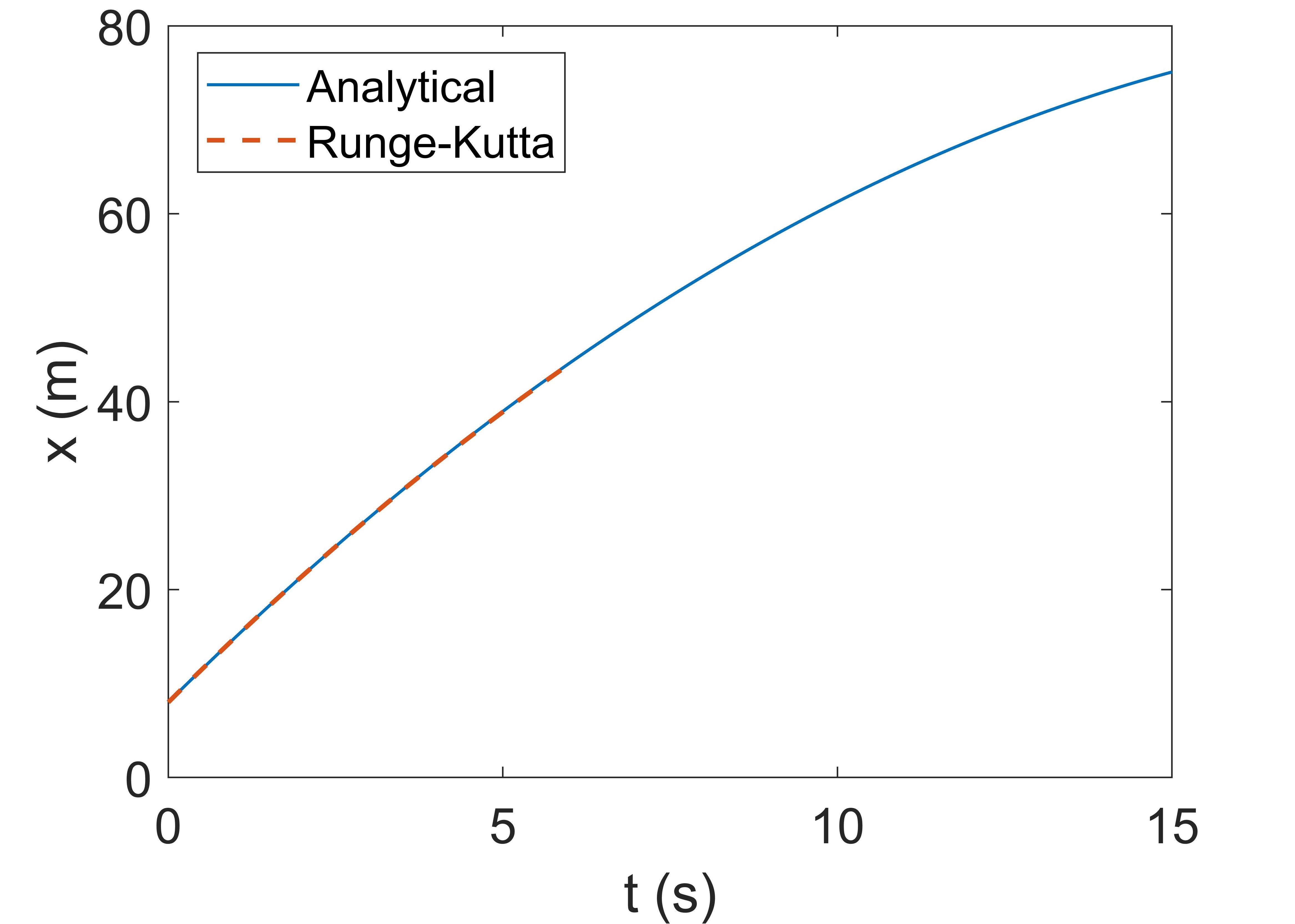}}
    \subcaptionbox{}{\includegraphics[width=0.45\textwidth]{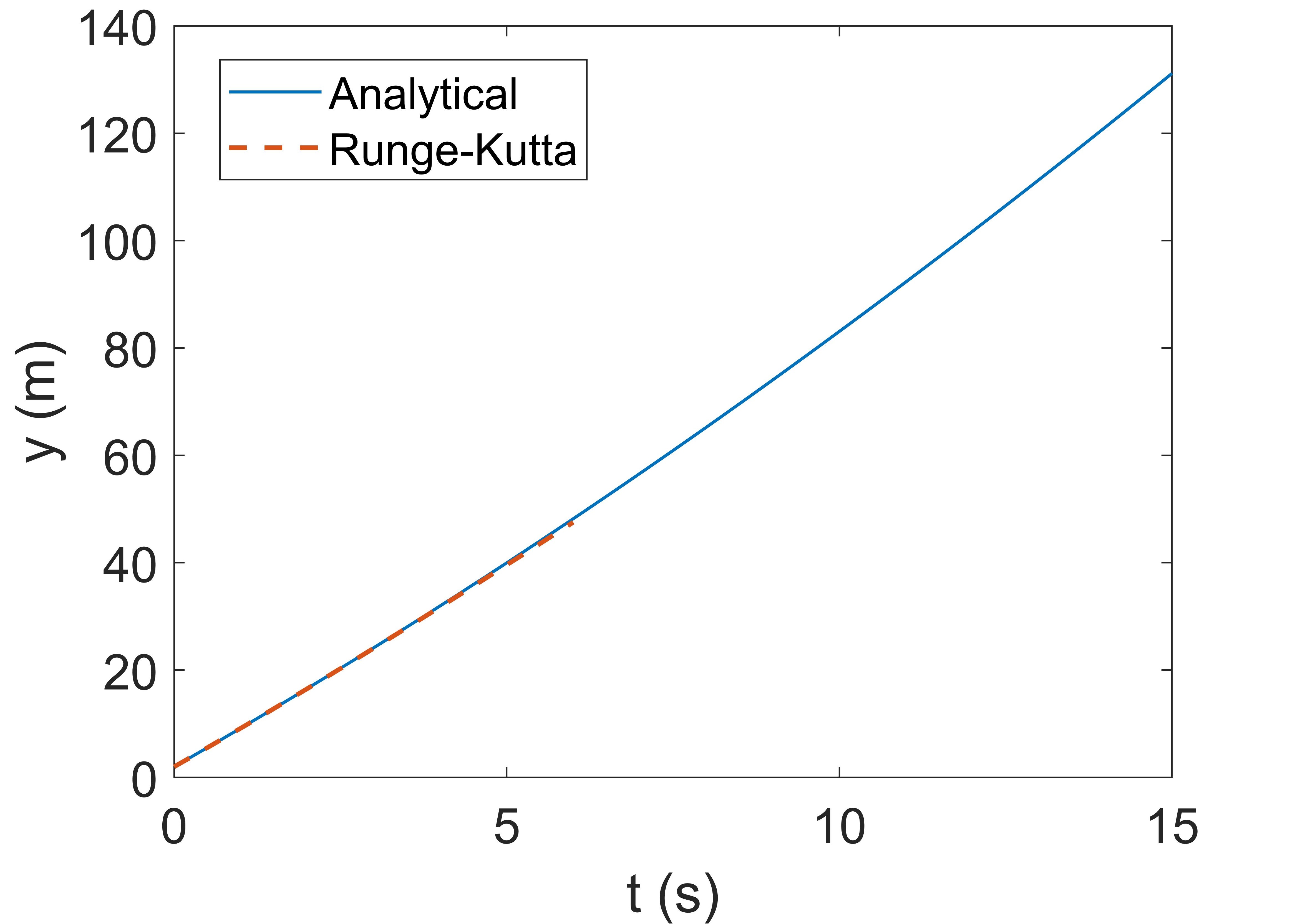}}
    \subcaptionbox{}{\includegraphics[width=0.45\textwidth]{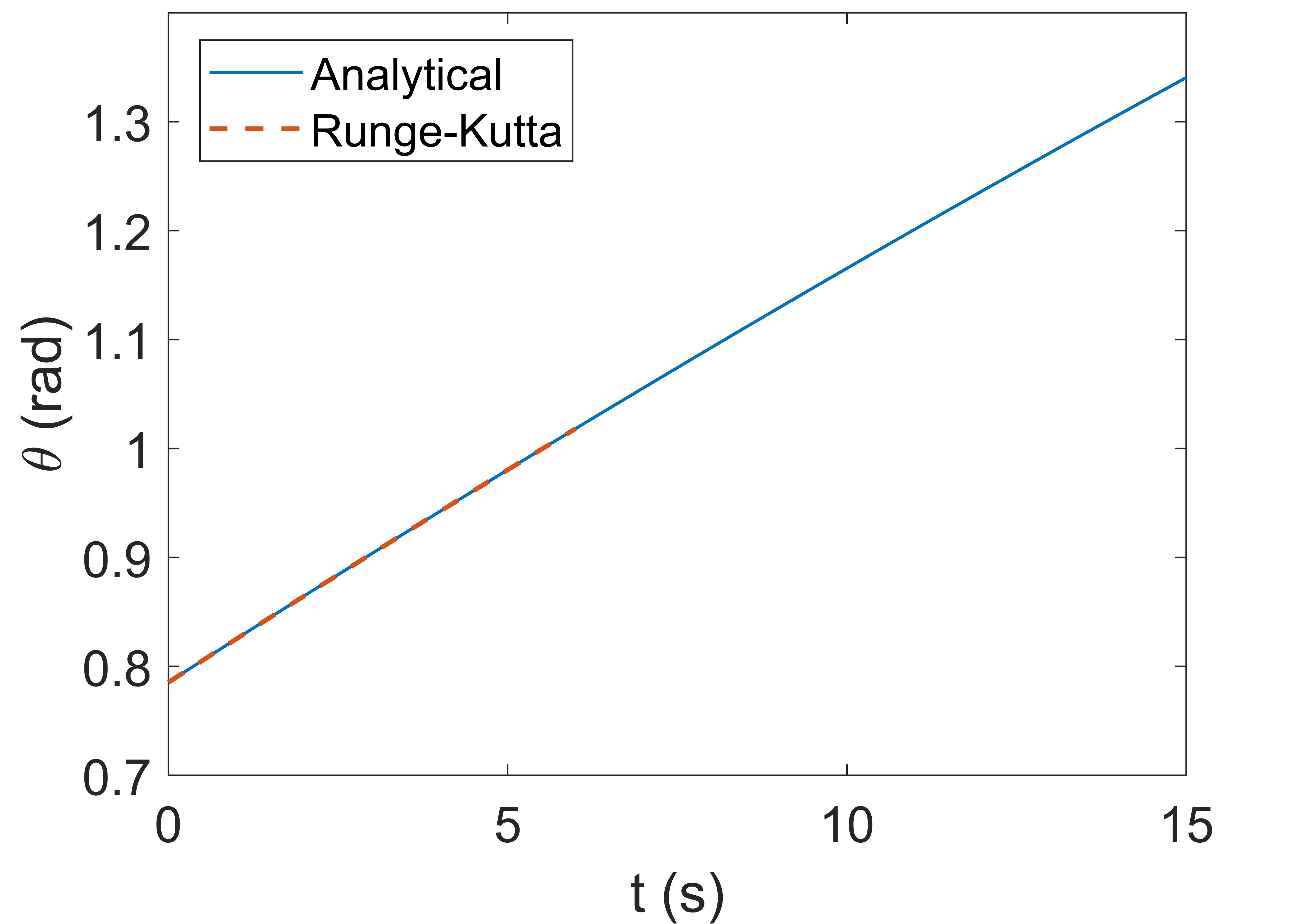}}
    \subcaptionbox{}{\includegraphics[width=0.45\textwidth]{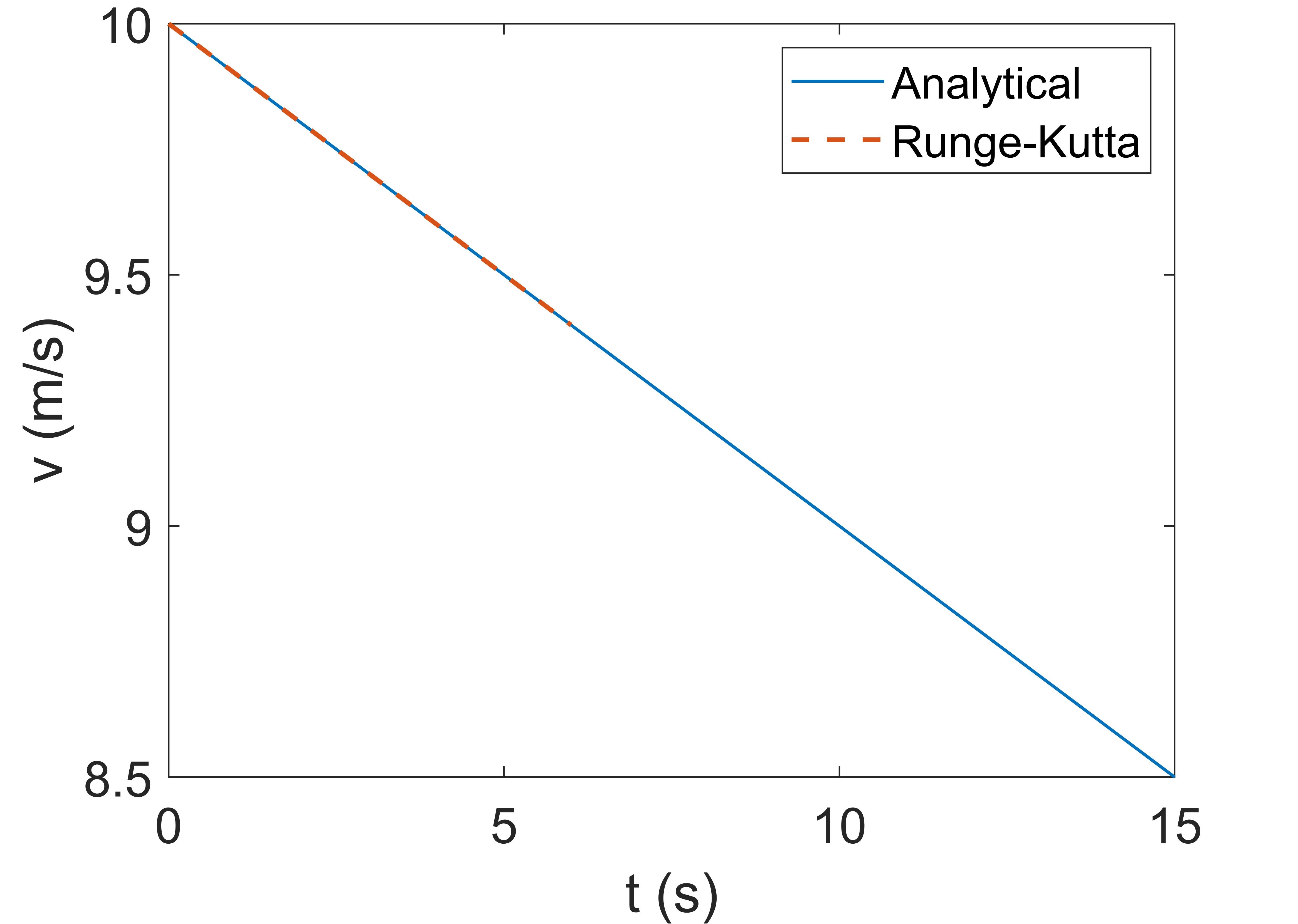}}
    \caption{Evolution of the estimated states of experiment 1 obtained by the analytical and Runge-Kutta method: (a) $x(t)$; (b) $y(t)$; (c) $\theta(t)$; (d) $v(t)$}
    \label{experiment1 state evolution}
\end{figure}

\begin{figure}[h!]
    \centering
    \setlength{\abovecaptionskip}{0pt}
    \subcaptionbox{}
    {\includegraphics[width=0.45\textwidth]{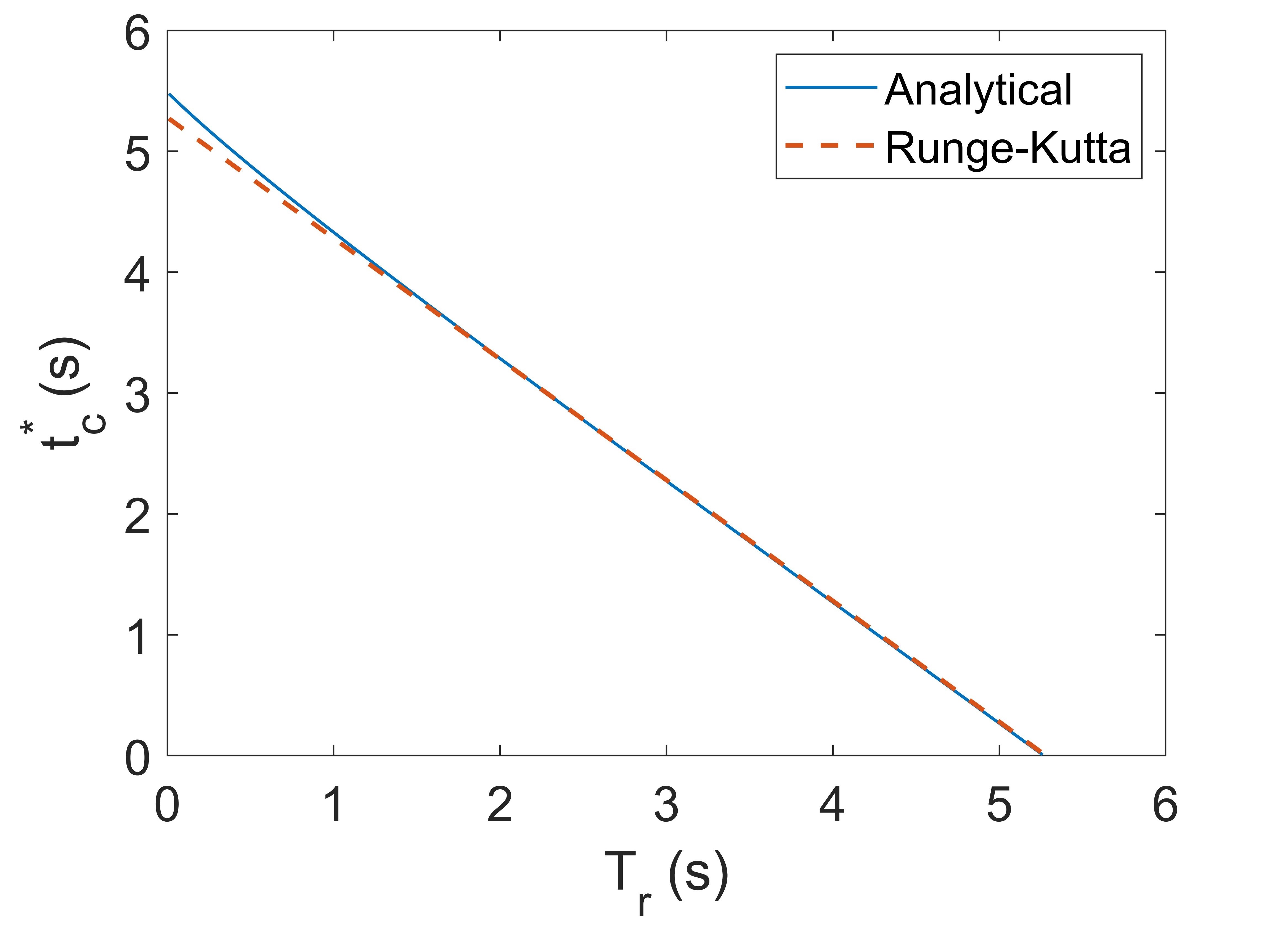}}
    \subcaptionbox{}{\includegraphics[width=0.45\textwidth]{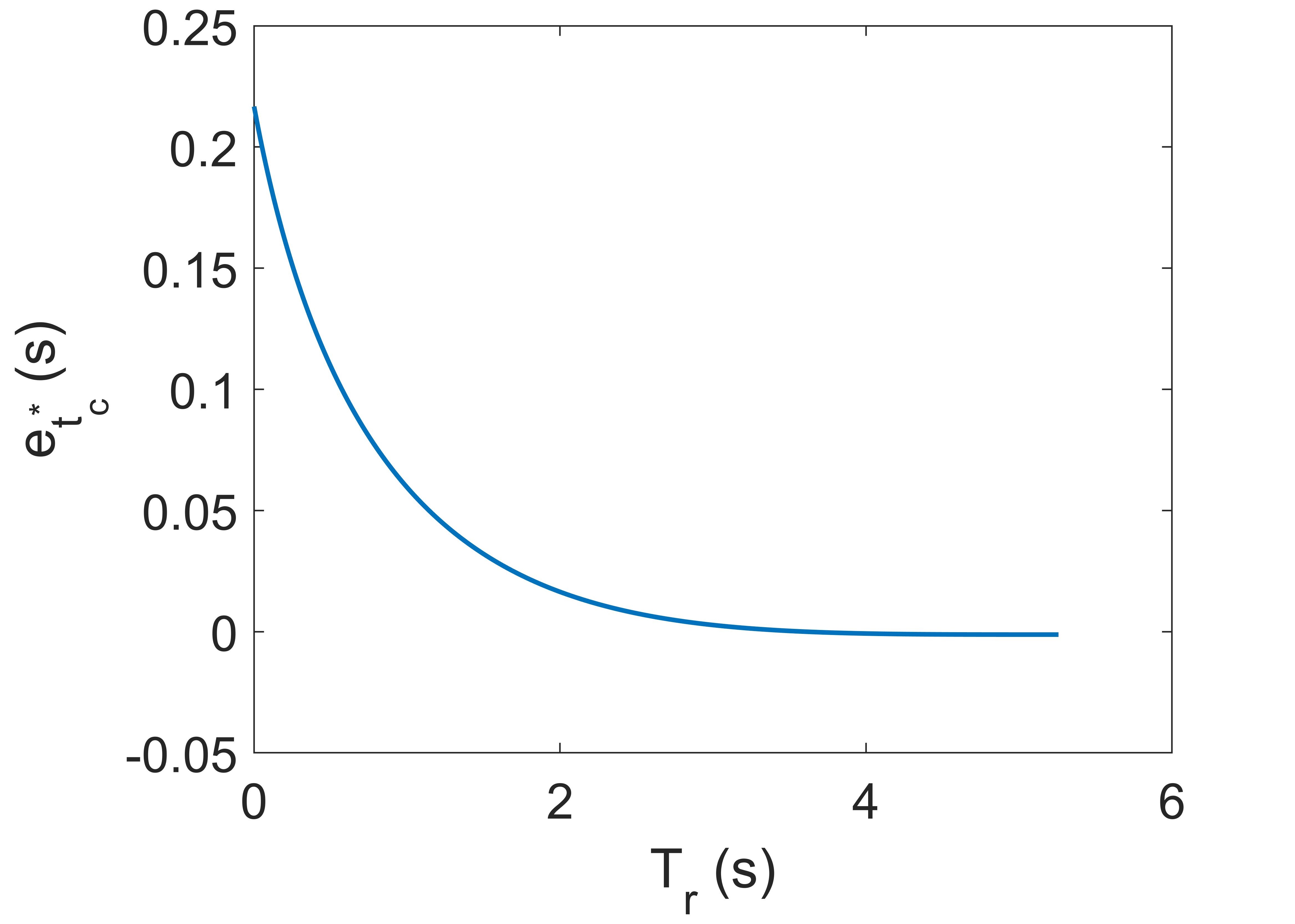}}
    \caption{Evolution of $t_c^*$ and $e_{t_c^*}$ of experiment 1: (a) $t_c^*$; (b) $e_{t_c^*}$}
    \label{evolution t_c}
\end{figure}

To better compare the estimated trajectory evolution from the analytical method with those from the numerical Runge-Kutta method, we examine the evolution of state variables $x(t)$, $y(t)$, $\theta(t)$, and $v(t)$ of Vehicle 1, estimated at $T_r=0$, as shown in Fig. \ref{experiment1 state evolution}(a)-(d). The Runge-Kutta method's outputs are plotted at the predefined $N$ steps, while the analytical method's results, which are available for all $t>0$, are illustrated for the first 15 seconds. This comprehensive representation is particularly noteworthy, as it reveals the analytical method's ability to accurately track the dynamic changes and nuances of all four state variables over time. This level of detail is essential for a thorough understanding of the vehicle's behavior under various driving conditions and scenarios.

To undertake a thorough quantitative analysis of the error generated by the linearization process and its impact on the vital estimation of collision time $t_c^*$ within the analytical framework, we direct attention to Fig. \ref{evolution t_c}(a). This figure illustrates the trajectory of  $t_c^*$ over the simulation time $T_r$. The illustration encompasses both the analytical framework and the numerical Runge-Kutta method as expounded in Subsection \ref{sec4.3}, tracking the evolution up to the point where a collision is simulated. Additionally, Fig. \ref{evolution t_c}(b) methodically details the discrepancy between the $t_c^*$ values derived from these two approaches, denoted as $e_{t_c^*}$. This measure, $e_{t_c^*}$, effectively quantifies the deviation in the collision time estimated by the analytical method, with the results from the Runge-Kutta method serving as a reference standard for precision.

In an analysis of Fig. \ref{evolution t_c}, it becomes evident that the largest discrepancy in $t_c^*$ is seen at the initial simulation step, $T_r=0$. This discrepancy is primarily attributed to the cumulative effect of linearization errors that arise during the evolution of the vehicles' state variables. Despite this, it is noteworthy that $e_{t_c^*}$ stays below 0.25 seconds, underscoring the relative accuracy of the estimations even at this early stage. As the simulation progresses and the distance between the two vehicles reduces, we observe a marked decrease in $e_{t_c^*}$, which diminishes steadily to a point where it can be considered negligible. This trend suggests that the linearization process, while introducing errors, becomes less influential as the vehicles approach one another, affirming the reliability of the analytical framework in dynamic, evolving traffic conditions.

\subsection{2-D motion and 3-D dynamics path coordinate model experiment \label{sec5.2}}
\begin{figure}[h]
    \centering
    \setlength{\abovecaptionskip}{0pt}
     \subcaptionbox{}{\includegraphics[width=0.4\textwidth]{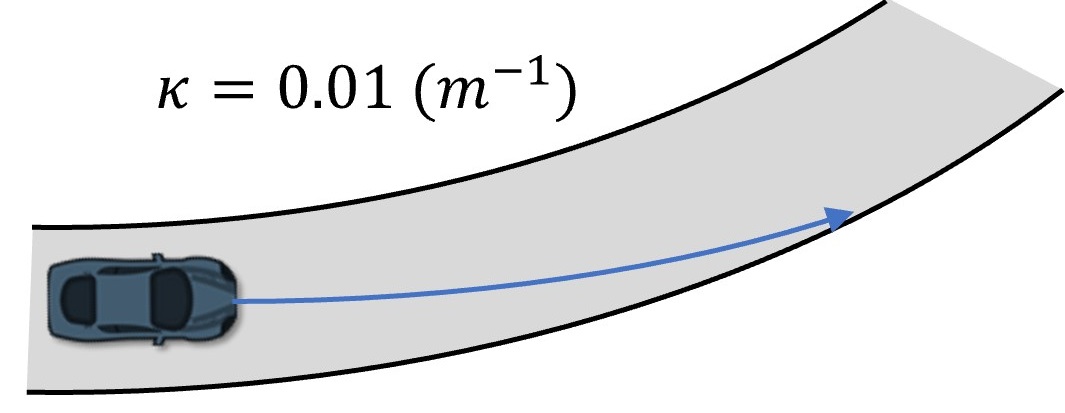}}
     \hspace{15mm}\subcaptionbox{}{\includegraphics[width=0.4\textwidth]{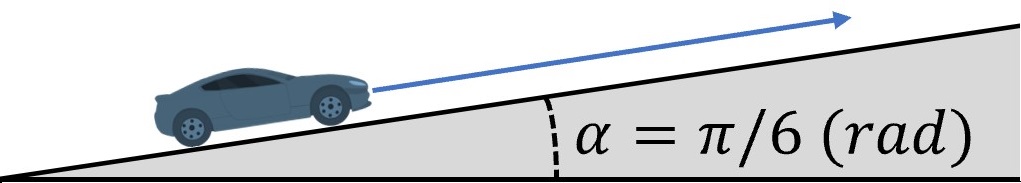}}
    \caption{Experiment 2 scenario: (a) top view; (b) side view}
    \label{experiment2}
\end{figure}

In this subsection, we undertake a detailed evaluation of the SSM presented in Subsection \ref{sec4.2}. Our assessment revolves around a specific driving scenario, as depicted in Fig. \ref{experiment2}. In this scenario, a vehicle travels on a horizontal curve with constant curvature and constant grade, the vehicle begins near the centerline of the lane. The scenario is characterized by a steering wheel angle that is not adequately large for the vehicle to successfully navigate through the curve. As the vehicle progresses along its trajectory, it becomes increasingly clear that the set steering angle is insufficient to maintain a safe path within the lane boundaries. This misalignment results in the vehicle's trajectory converging with the road boundary. The consequence of this convergence is an overlap between the vehicle's shape and the road edge, leading to a collision. The parameters that we use are listed in Table \ref{experiment 2 parameters}. The superscript 0 indicates initial values at $T_r=0$.

\begin{table}[b]
\centering
\caption{Experiment 2 parameters}
\label{experiment 2 parameters}
\begin{tabular}{l l || l l}
\hline
parameter & value& parameter & value\\ \hline
$s^0$ & $0 (m)$ & $C_d$ & $0.25$\\ 
$e_{cg}^0$ & $0.15 (m)$ & $S$ & $2 (m^2)$\\ 
$\theta^0_{e}$ & $-0.02 (rad)$ & $m$ & $1500 (kg)$\\
$v^0$ & $9 (m/s)$ & $r_{whl}$ & $0.25 (m)$\\
$\delta^0$ & $0.024 (rad)$ & $f_{roll}$ & $0.015$ \\
$T^0_{whl}$ & $2000 (Nm)$ & $w$ & $6 (m)$  \\
$L$ & $2.5 (m)$ & $h$ & $0.001 (s)$\\
$r$ & $1.5 (m)$ & $N$ & $6000$\\
$\rho$ & $1.2 (kg/m^3)$ \\

\hline
\end{tabular}
\end{table}

\begin{figure}[h]
    \centering
    \setlength{\abovecaptionskip}{0pt}
    \subcaptionbox{}
    {\includegraphics[width=0.49\textwidth]{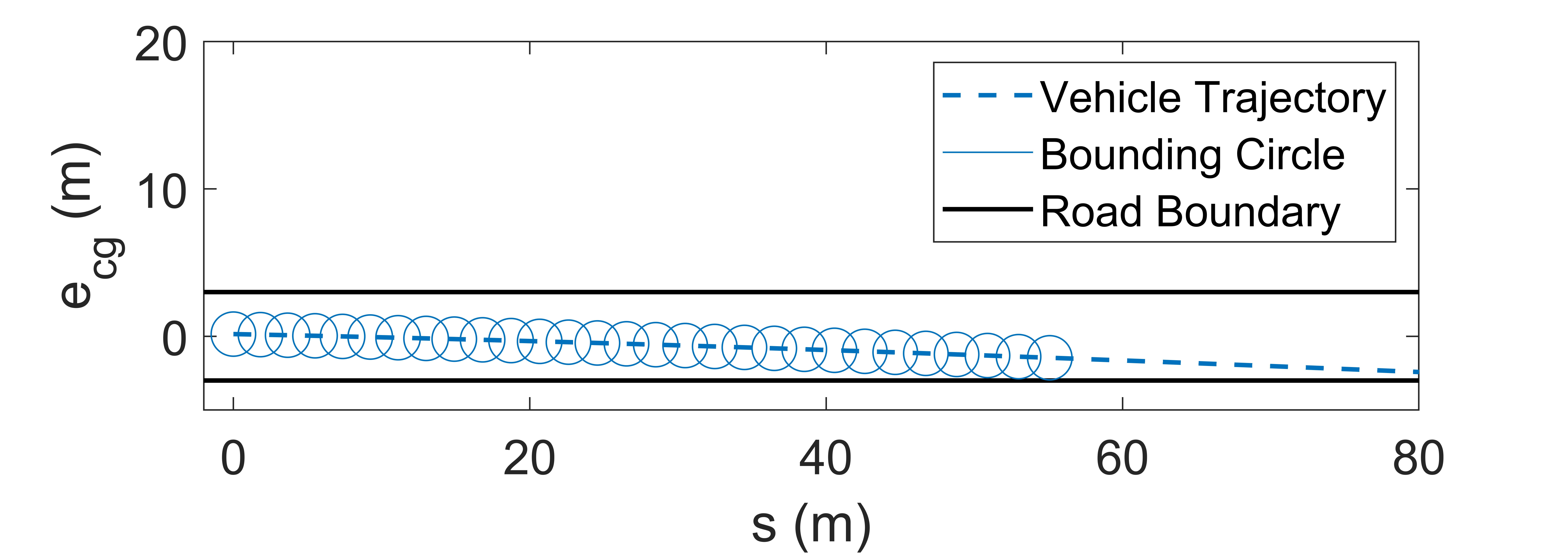}}
    \subcaptionbox{}{\includegraphics[width=0.49\textwidth]{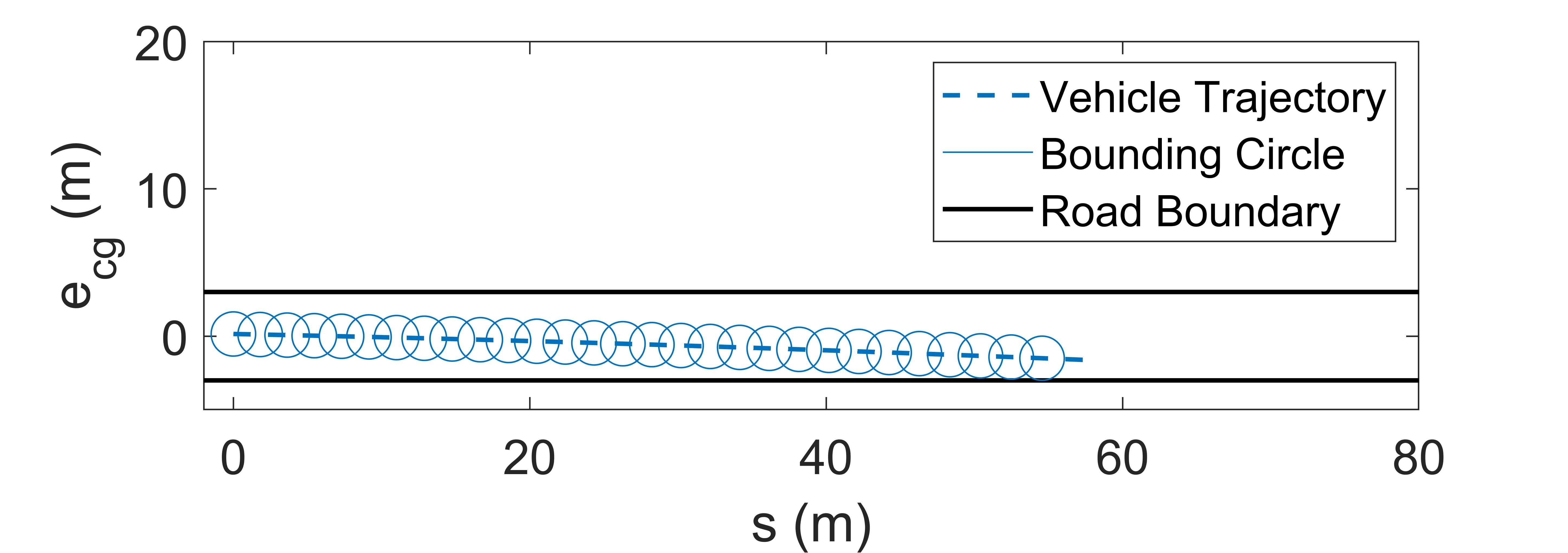}}
    \caption{Derived trajectories and bounding circles of experiment 2 at $T_r=0$: (a) analytical based on linearized model; (b) numerical based on Runge-Kutta}
    \label{experiment2 traj}
\end{figure}

To analyze how the SSM adapts with the decreasing relative distance between two vehicles over simulation time $T_r$, we maintain constant values for both the wheel torque and the steering wheel angle. This is expressed as $T_{whl}^{T_r}=T_{whl}^{0}$, $\delta^{T_r}=\delta^0$ for all $T_r>0$. To evaluate errors stemming from model linearization, we begin by illustrating the trajectories estimated at $T_r=0$ as obtained by both the analytical and the numerical Runge-Kutta methods in Fig. \ref{experiment2 traj}(a) and Fig. \ref{experiment2 traj}(b), respectively. Additionally, to aid in visualizing the trajectory development, bounding circles are plotted at 0.2-second intervals until a collision is anticipated by each method.

Fig. \ref{experiment2 traj} provides a clear indication that the estimation of trajectory evolution made using the analytical method, which is grounded in a linearized model, is successful in capturing the directional movements of the vehicle. This method not only traces the vehicle's trajectory with a high degree of accuracy but also offers a closed-form trajectory expression, it allows for the determination of the vehicle's position at any future moment $t>0$ by simply substituting the corresponding time value. This feature contrasts with the numerical Runge-Kutta method, which is inherently limited to calculating discrete trajectory points at predetermined $N$ time steps.

\begin{figure}[h]
    \centering
    \setlength{\abovecaptionskip}{0pt}
    \subcaptionbox{}
    {\includegraphics[width=0.45\textwidth]{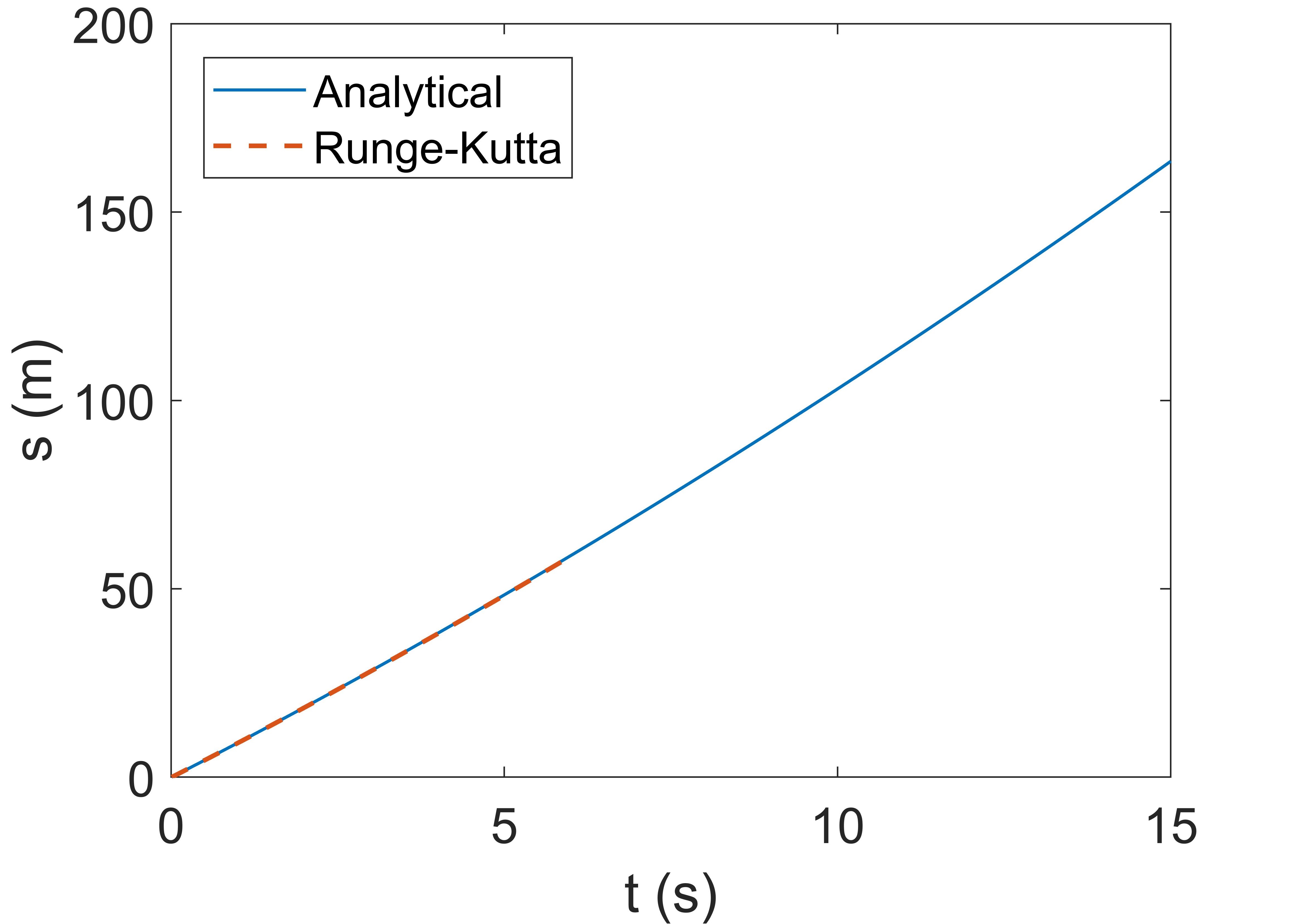}}
    \subcaptionbox{}{\includegraphics[width=0.45\textwidth]{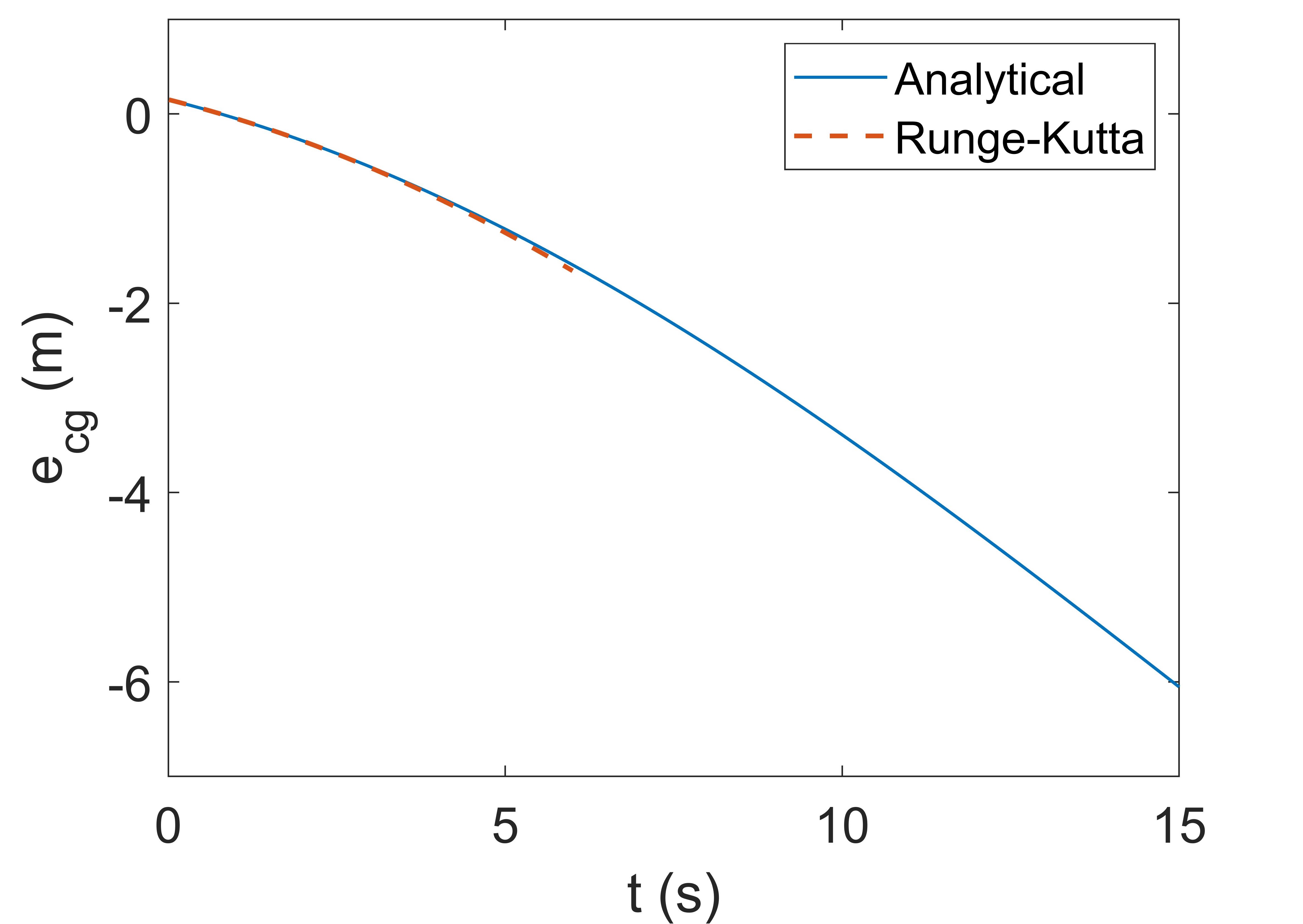}}
    \subcaptionbox{}{\includegraphics[width=0.45\textwidth]{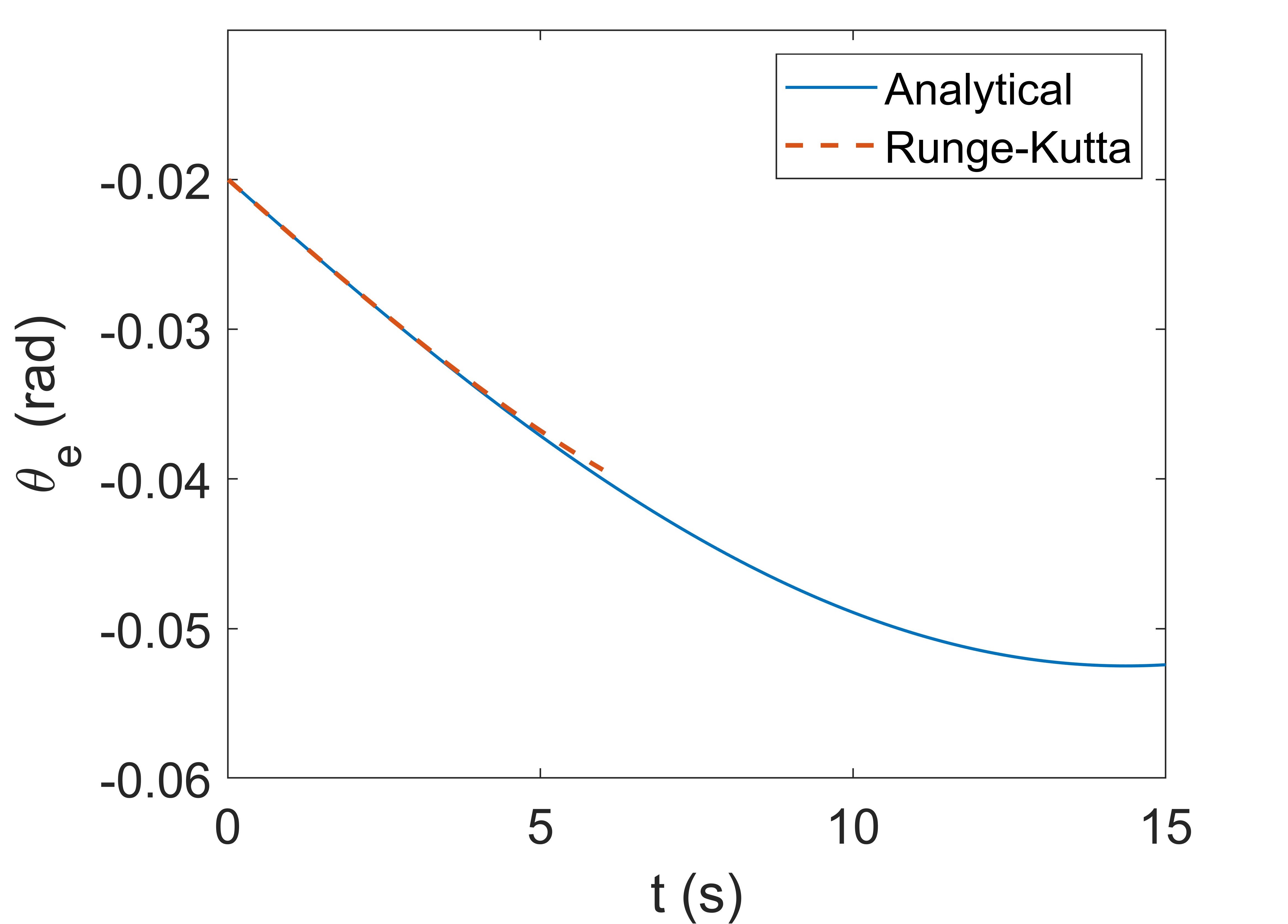}}
    \subcaptionbox{}{\includegraphics[width=0.45\textwidth]{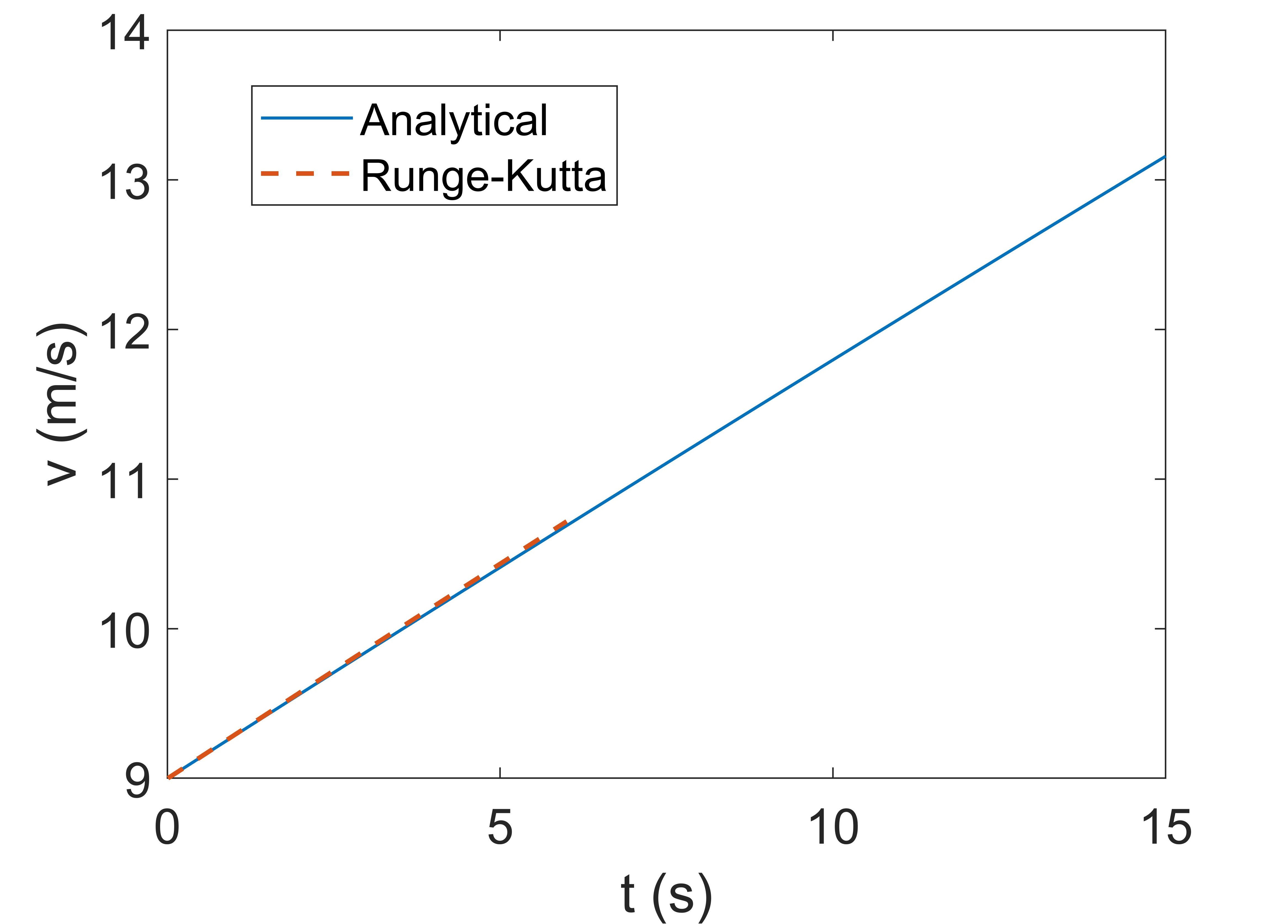}}
    \caption{Evolution of the estimated states of experiment 2 obtained by the analytical and Runge-Kutta method: (a) $s(t)$; (b) $e_{cg}(t)$; (c) $\theta_e(t)$; (d) $v(t)$}
    \label{experiment2 state evolution}
\end{figure}

To enable a detailed and direct comparison of the trajectory outputs from both the analytical method and the numerical Runge-Kutta approach, our analysis focuses on the progression of the vehicle's state variables $s(t)$, $e_{eg}(t)$, $\theta_e(t)$, and $v(t)$, estimated at $T_r=0$. These trajectories are displayed in Fig. \ref{experiment2 state evolution}(a)-(d). The Runge-Kutta method delineates the vehicle's state only at the predefined $N$ steps, whereas the analytical method's results, which are available for all $t>0$, are illustrated for the first 15 seconds. It is observed that the trajectories estimated by the analytical method accurately captures the evolution of all four states.

\begin{figure}[h!]
    \centering
    \setlength{\abovecaptionskip}{0pt}
    \subcaptionbox{}
    {\includegraphics[width=0.45\textwidth]{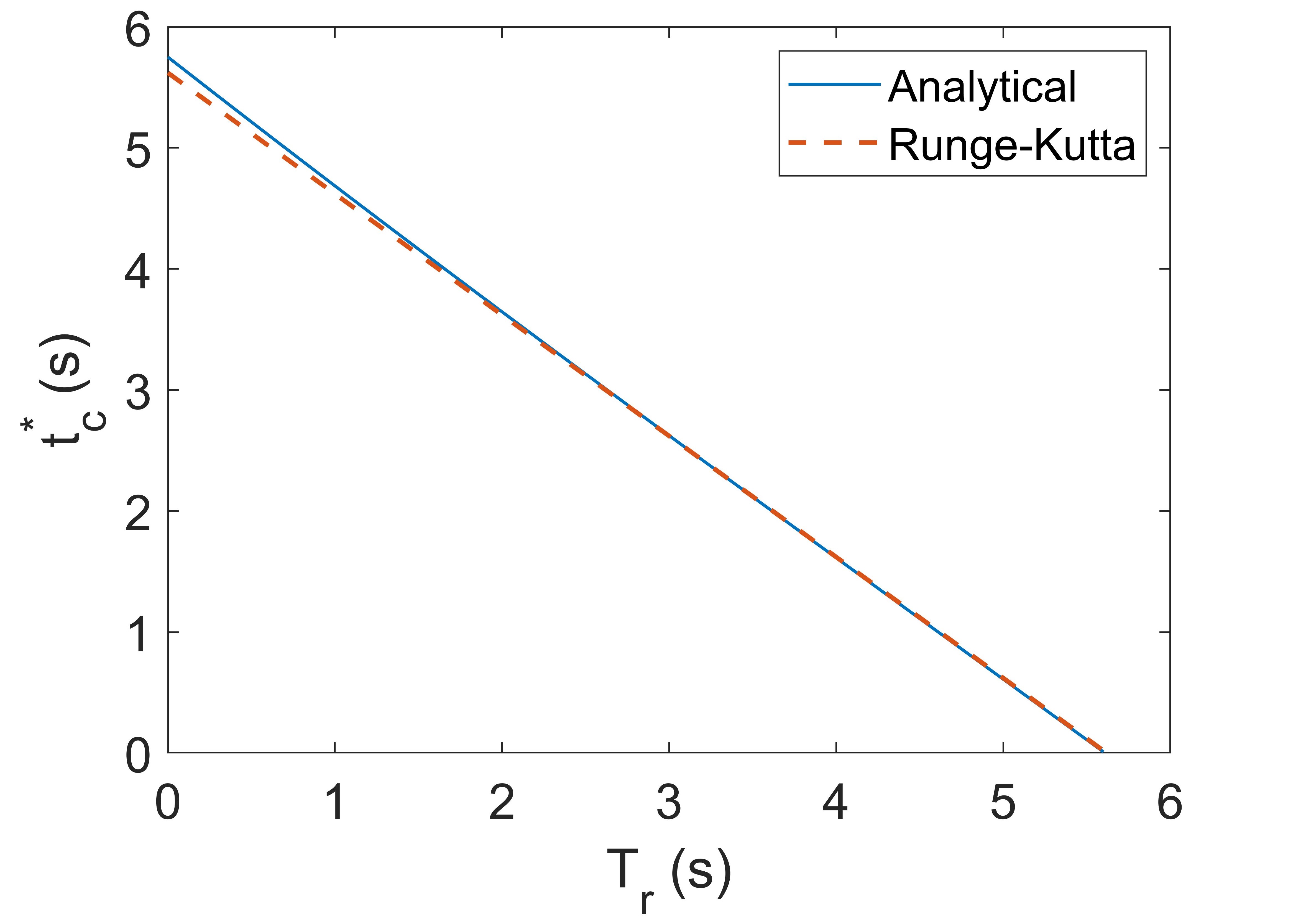}}
    \subcaptionbox{}{\includegraphics[width=0.45\textwidth]{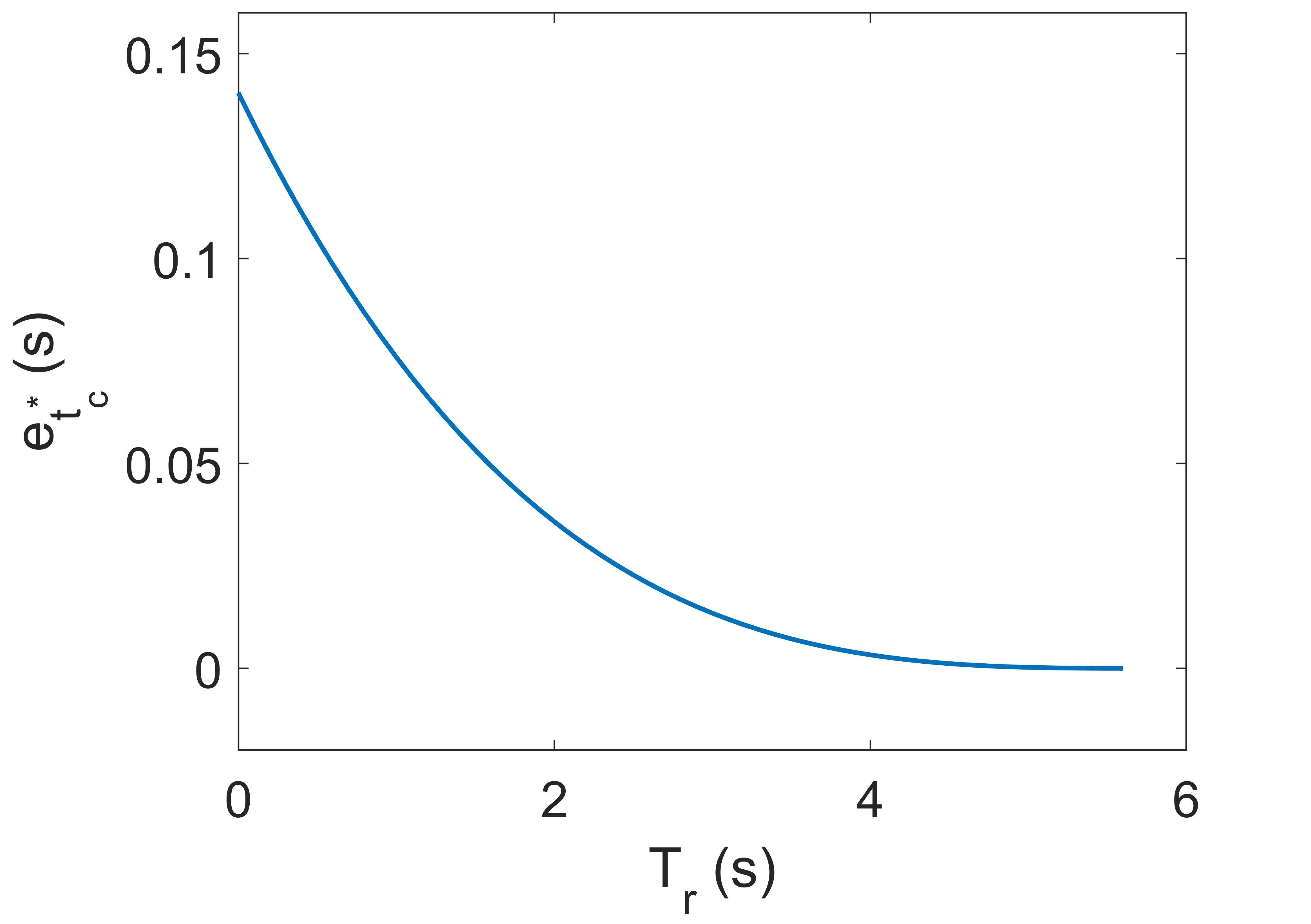}}
    \caption{Evolution of $t_c^*$ and $e_{t_c^*}$ of experiment 2: (a) $t_c^*$; (b) $e_{t_c^*}$}
    \label{evolution t_c 3D}
\end{figure}

To further conduct a quantitative analysis of the error resulting from the linearization process and its influence on the critical outcome $t_c^*$ of the analytical framework, Fig. \ref{evolution t_c 3D}(a) portrays the evolution of $t_c^*$ over the simulation time $T_r$. This comparison includes both the analytical framework and the numerical Runge-Kutta method detailed in Subsection \ref{sec4.3}, up to the point of collision in the simulation. Furthermore, Fig. \ref{evolution t_c 3D}(b) presents the discrepancy between the $t_c^*$ values derived from these two approaches, denoted as $e_{t_c^*}$. This measure, $e_{t_c^*}$, quantifies the error in $t_c^*$ as estimated by the analytical method, considering the Runge-Kutta method's outcomes as the benchmark for accuracy.

From Fig. \ref{evolution t_c 3D}, it is observed that the initial time step $T_r=0$ exhibits the largest error in $t_c^*$. This observation is attributed to the accumulation of linearization errors over the course of the vehicles' state evolution. Nonetheless, it is important to note that $e_{t_c^*}$ remains under 0.15 seconds. As the simulation progresses and the relative distance between the two vehicles decreases, $e_{t_c^*}$ gradually diminishes, eventually becoming negligible.

\subsection{3D and 1D SSMs comparison experiment \label{sec5.3}}

\begin{figure}[h]
    \centering
    \setlength{\abovecaptionskip}{0pt}
    \subcaptionbox{}
    {\includegraphics[width=0.45\textwidth]{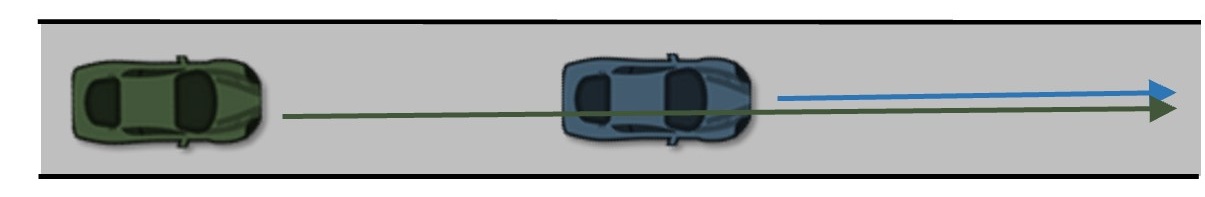}}
    \hspace{10mm}
    \subcaptionbox{}{\includegraphics[width=0.45\textwidth]{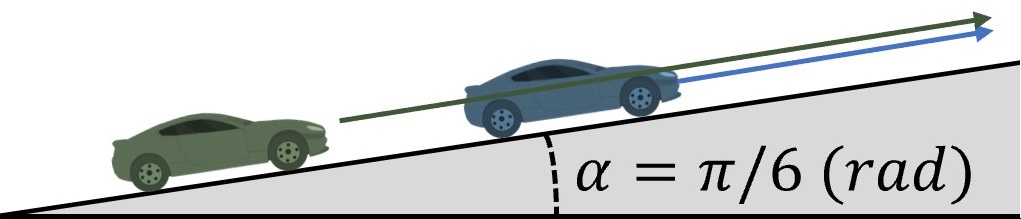}}
    \subcaptionbox{}{\includegraphics[width=0.45\textwidth]{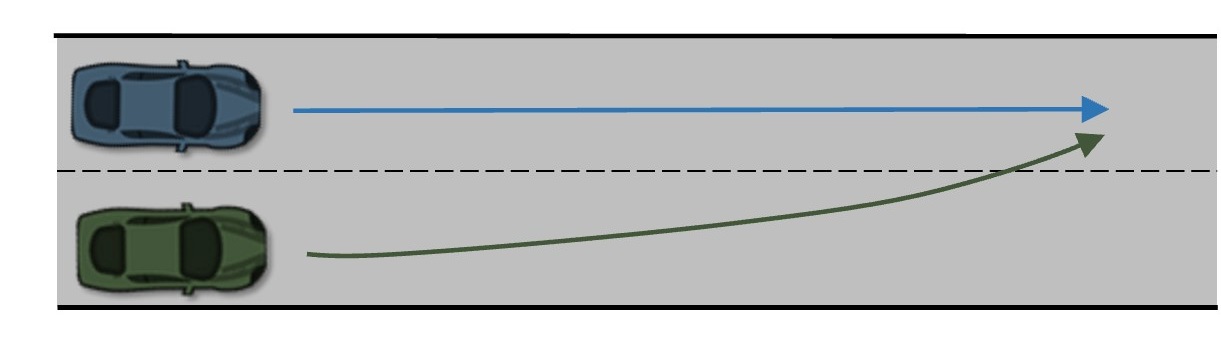}}
    \hspace{10mm}
    \subcaptionbox{}{\includegraphics[width=0.45\textwidth]{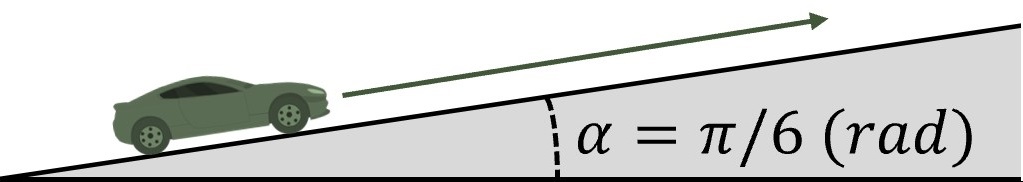}}
    \caption{Experiment 3 scenarios: (a) car-following top view; (b) car-following side view; (c) merging top view; (d) merging side view}
    \label{experiment3 scenario}
\end{figure}

In this subsection, we delve into the distinctions between conventional 1D SSMs and extended 3D SSMs obtained from the proposed framework. It is important to note that conventional 1D SSMs primarily address vehicle-to-vehicle conflicts and collisions resulting from movements in a single direction (\citep{wang2021review}). Their applicability is somewhat limited, particularly in scenarios involving vehicle-to-road conflicts, crossing conflicts, and curvy roads, where their suitability is not straightforward. To facilitate a comparative analysis between conventional 1D SSMs and our proposed 3D SSMs, we have intentionally devised two specific experimental scenarios. These scenarios serve as test cases, one centered around car-following conflicts and the other focused on merging conflicts, as depicted in Fig. \ref{experiment3 scenario}. In the car-following scenario Fig. \ref{experiment3 scenario}(a)-(b), the leading vehicle, Vehicle 2, maintains a constant speed, while the trailing vehicle, Vehicle 1, accelerates, eventually resulting in a rear-end collision. In the merging scenario Fig. \ref{experiment3 scenario}(c)-(d), Vehicle 2 maintains a constant speed within its initial lane, while the Vehicle 1 executes a lane-change at a constant speed, ultimately leading to a side collision. 

It is difficult to provide a reference value of risk due to the following reasons: (1) lack of data. To the best of the authors’ knowledge, no open-source dataset contains the information needed for the calculation of the exemplary high-dimensional and high-fidelity SSM (i.e., steering wheel angle and wheel torque); (2) even if such data were available, it is difficult to determine what the “true value of risk” should be. Therefore, the best we can do is to treat the numerical solution of the 3D SSM as the true value under specific assumption and condition. The assumption made in this subsection is that the vehicle movement model utilized by the 3D SSM is accurate enough to describe the dynamics of vehicles. This is reasonable because these models are extensively studied and well recognized in the automotive/mechanical engineering community, in which high accuracy is needed for vehicle models (\cite{rajamani2011vehicle}). Based on the aforementioned assumption and in the scenario where the steering wheel and wheel torque remains unchanged, the numerical 3D SSM solution can be treated as the true value for this specific case, allowing us to compare the 1D and 3D SSMs (at least in this specific scenario and obtain some indications).

To address these scenarios, we employ a 3D SSM that combines both lateral dynamics Eqs. (\ref{kinematic1})-(\ref{kinematic4}) and longitudinal dynamics Eq. (\ref{longitudinal dy}). In contrast, for the conventional 1D SSM, we utilize TTC, where the longitudinal position, speed, and length parameters are substituted with their lateral counterparts (position, speed, and length) for the merging scenario. The parameters used for both scenarios are listed in Table \ref{experiment 3 parameters}. At $T_r=0$, the initial states are assigned as $[x_1^0, y_1^0, \theta_1^0, v_1^0]^T=[0, 0, 0, 6]^T$, $[x_2^0, y_2^0, \theta_2^0, v_2^0]^T=[20, 0, 0, 5]^T$ for the car-following scenario, and $[x_1^0, y_1^0, \theta_1^0, v_1^0]^T=[0, 8, -0.05, 5]^T$, $[x_2^0, y_2^0, \theta_2^0, v_2^0]^T=[0, 0, 0, 5]^T$ for the merging scenario. Moreover, Vehicle 1 remains constant control inputs $\delta_1^{T_r}=0$, $T_{whl,1}^{T_r}=2100 Nm$ for the car-following scenario, and $\delta_1^{T_r}=-0.05 rad$, $T_{whl,1}^{T_r}=1890 Nm$ for the merging scenario.

\begin{table}[h]
\centering
\caption{Experiment 3 parameters}
\label{experiment 3 parameters}
\begin{tabular}{l l || l l}
\hline
parameter & value & parameter & value \\ \hline
$L_1$ & $2 (m)$ & $L_2$ & $2 (m)$\\
$r_1$ & $1.3 (m)$ & $r_2$ & $1.3 (m)$\\
 $C_d$ & $0.25$ & $S_1$ & $2 (m^2)$\\ 
$m_1$ & $1500 (kg)$ & $r_{whl,1}$ & $0.25 (m)$\\
$f_{roll,1}$ & $0.015$ &
$\rho$ & $1.2 (kg/m^3)$ \\
\hline
\end{tabular}
\end{table}

\begin{figure}[h]
    \centering
    \setlength{\abovecaptionskip}{0pt}
    \subcaptionbox{}
    {\includegraphics[width=0.45\textwidth]{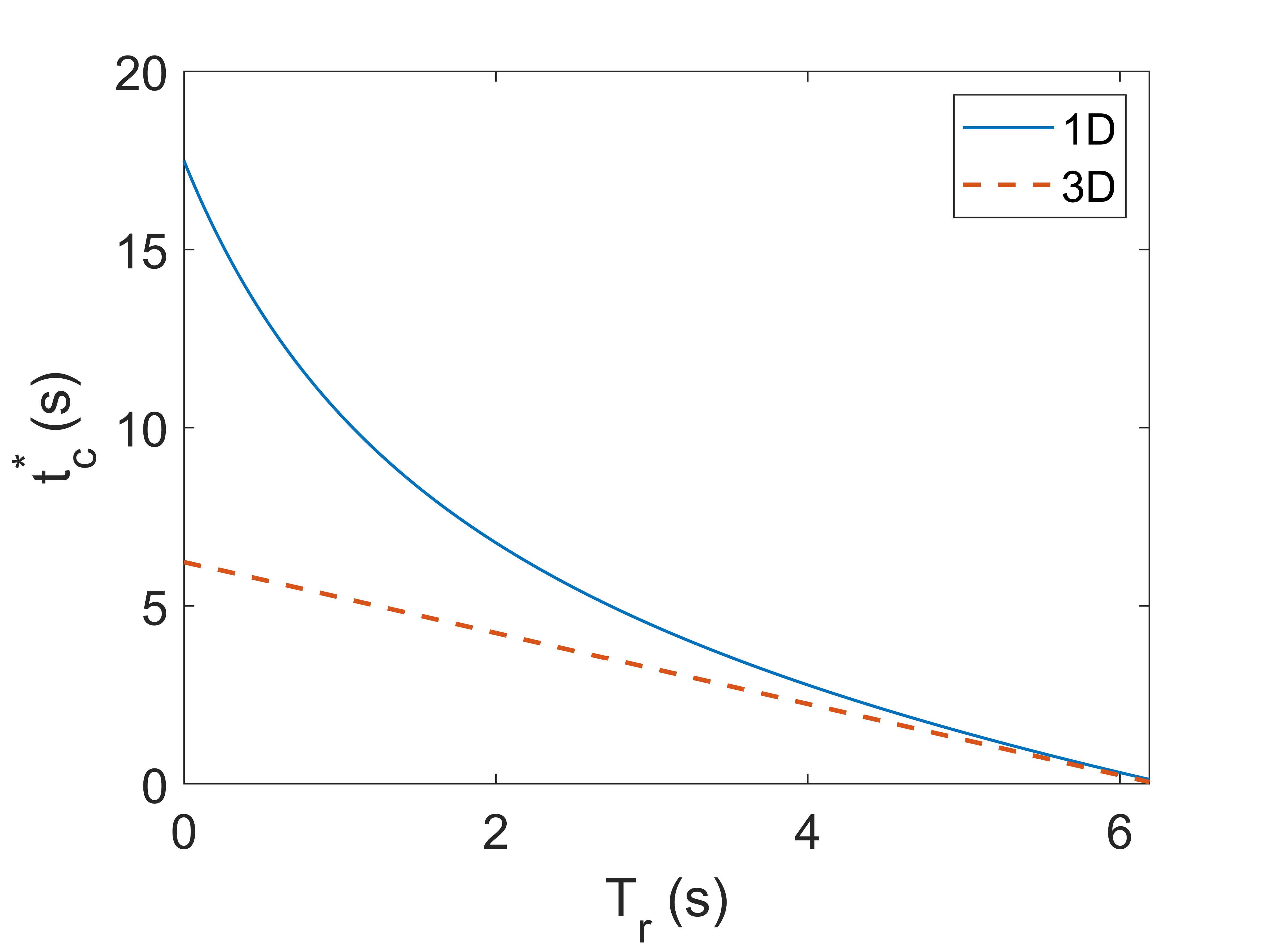}}
    \subcaptionbox{}{\includegraphics[width=0.45\textwidth]{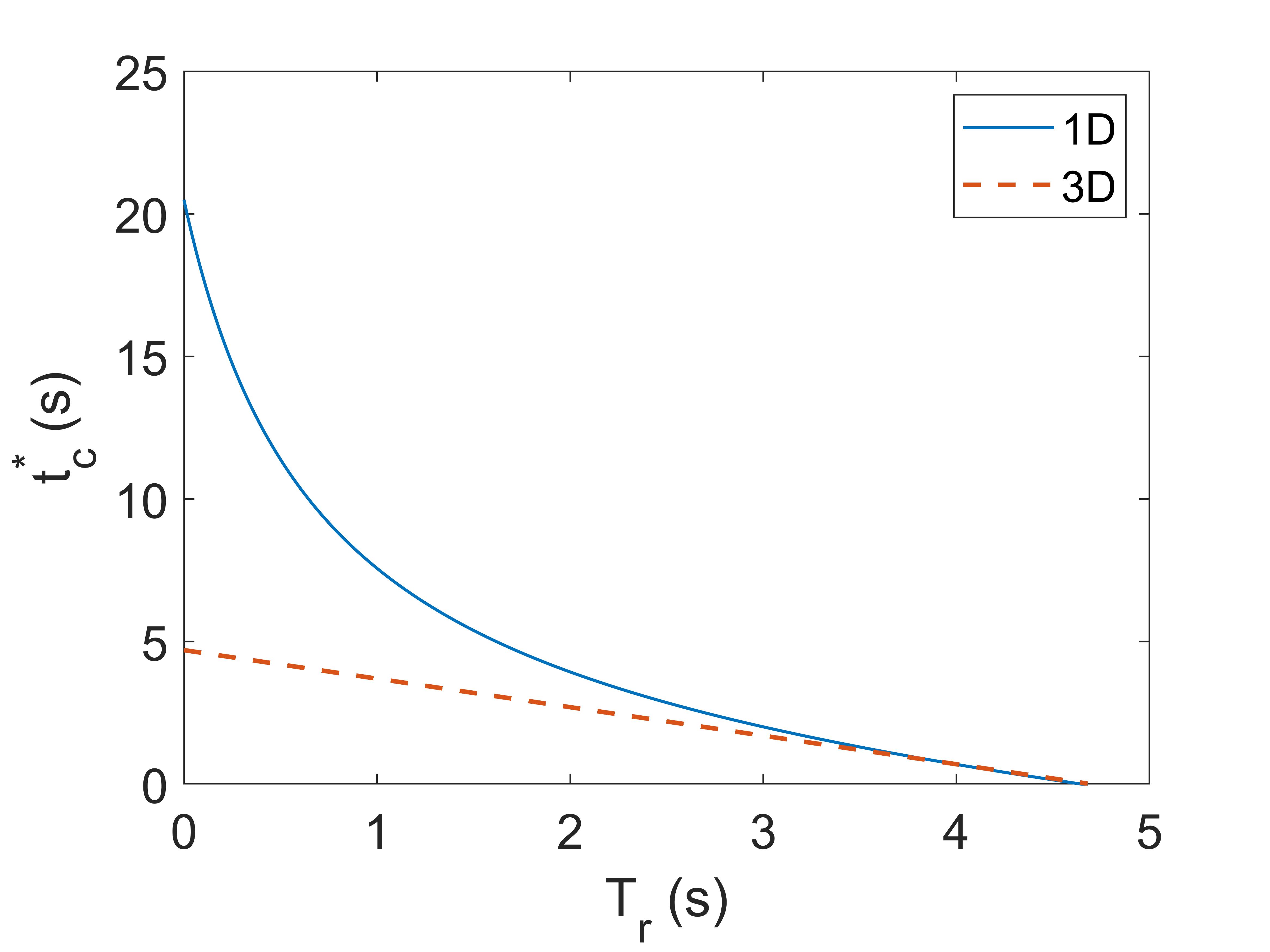}}
    \caption{Evolution of $t_c^*$ of experiment 3: (a) car-following scenario; (b) merging scenario}
    \label{evolution t_c 3D vs 1D}
\end{figure}

To quantitatively analyze the differences in the critical outcome $t_c^*$ between conventional 1D SSMs and extended 3D SSMs, we present the evolution of $t_c^*$ over the simulation time $T_r$ in Fig. \ref{evolution t_c 3D vs 1D}. This comparison encompasses both the 1D TTC and the extended 3D SSM aforementioned, considering the duration until the collision occurs during the simulation. In Fig. \ref{evolution t_c 3D vs 1D}(a) and Fig. \ref{evolution t_c 3D vs 1D}(b), we illustrate the evolution of $t_c^*$ for the car-following and merging scenarios, respectively.

Analysis of Fig. \ref{evolution t_c 3D vs 1D} reveals that the most significant discrepancy in the 1D TTC occurs at the initial time step, $T_r=0$. At this time instance, the 1D TTC calculates a $t_c^*=17.38s$, whereas the actual value is $6.22s$ in the car-following scenario, resulting in an error that is almost twice the true value. In the merging scenario, the 1D TTC estimates $t_c^*$ as $20.18s$ while the true value is $4.66s$, indicating an error that exceeds more than three times the actual value. This highlights that conventional 1D SSMs can yield highly inaccurate and unreliable results when assessing collision proximity. Furthermore, when a threshold of $t_c^*=1.5s$, which is sometimes considered as a critical threshold in the literature (\cite{deveaux2021extraction,van1993ti}), is reached, the error of TTC remains above $0.25s$ for the car-following scenario and exceeds $0.2s$ for the merging scenario. This suggests that deploying conventional 1D SSMs on real vehicles might lead to insufficient time for collision avoidance. Furthermore, when using 1D SSMs in simulations, they may substantially misjudge the count of conflicts with TTC less than a predefined threshold (e.g., 1.5s).

\subsection{Cartesian coordinate model and path coordinate model combination experiment \label{sec5.4}}

\begin{figure}[h]
    \centering
    \setlength{\abovecaptionskip}{0pt}
     \subcaptionbox{}{\includegraphics[width=0.49\textwidth]{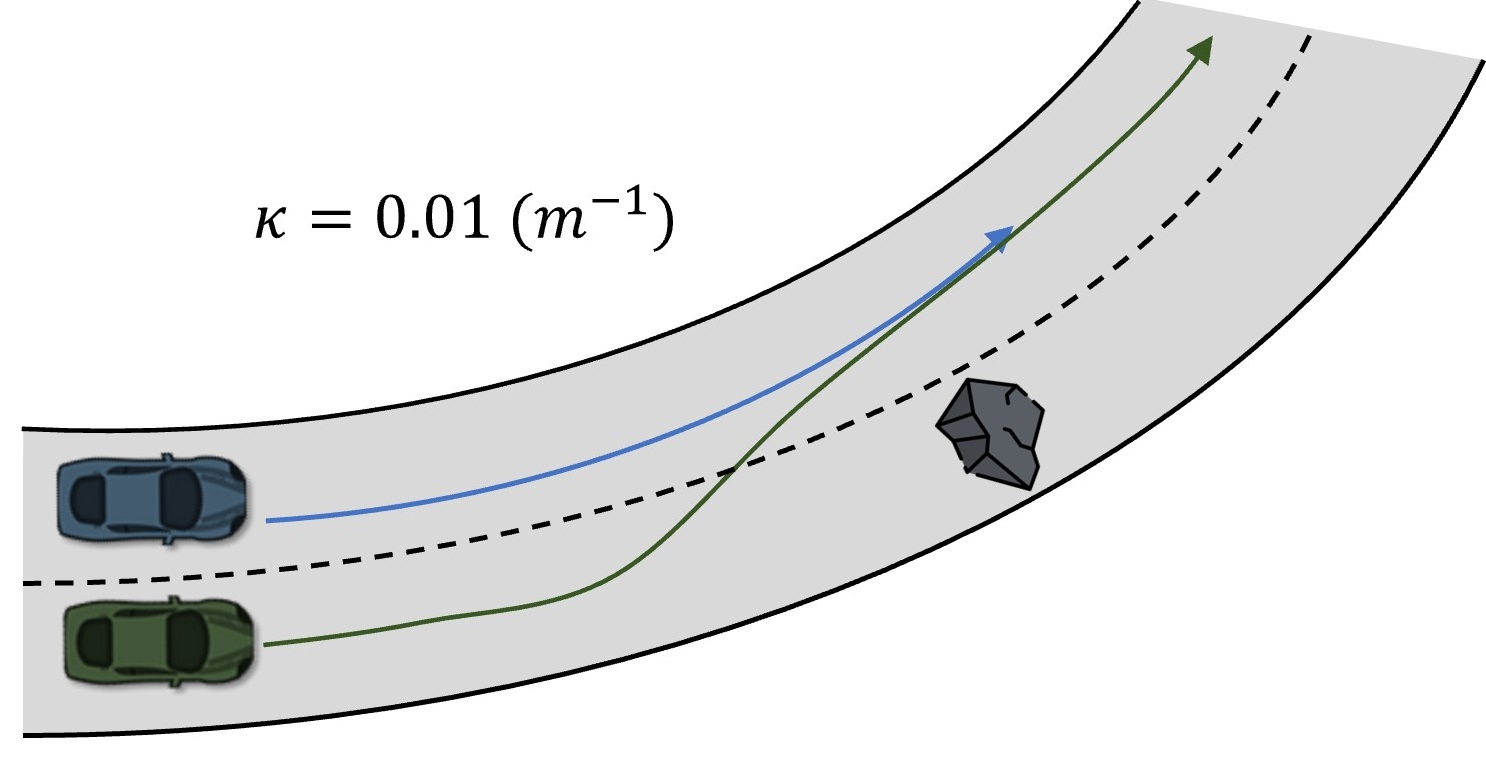}}
     %\hspace{15mm}
     \subcaptionbox{}{\includegraphics[width=0.49\textwidth]{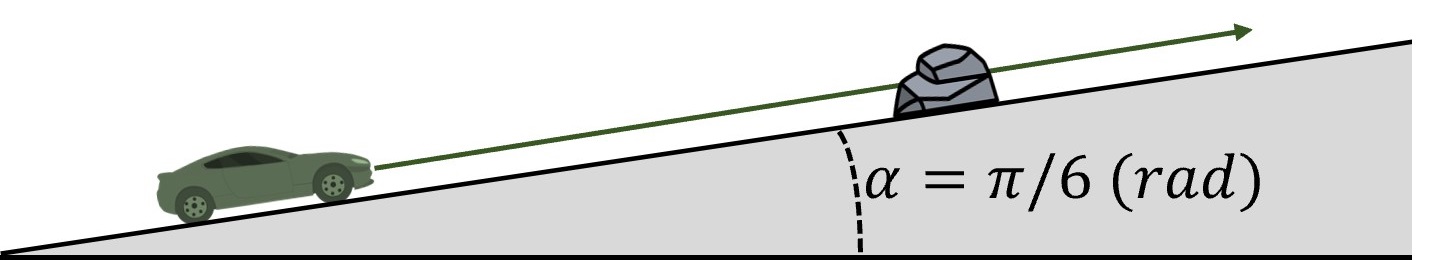}}
    \caption{Experiment 4 scenario: (a) top view; (b) side view}
    \label{experiment3}
\end{figure}

In this subsection, our objective is to showcase a real-world application that employs both the SSMs described in Subsections \ref{sec4.1} and \ref{sec4.2} to comprehensively assess all potential conflicts or ultimately collisions. Our demonstration centers around a specific driving scenario, as illustrated in Fig. \ref{experiment3}. In this scenario, two vehicles initiate their trajectories in adjacent lanes on a horizontal curve characterized by constant curvature and a uniform grade. The first vehicle executes a lane-changing maneuver while reducing speed to avoid an obstacle positioned on the road. Meanwhile, the second vehicle maintains a steady velocity within its original lane. Following the successful lane change, the first vehicle accelerates and safely continues its journey within the newly chosen lane.

We index the vehicle executing the lane change as $i=1$, while the other vehicle is indexed as $i=2$. The obstacle is bounded by a circumscribed circle with radius $r_{ob}$. The simulation environment is structured within the Cartesian coordinate system, and the parameters specific to this coordinate system are detailed in Table \ref{experiment 4 parameters}, where the superscript 0 signifies the initial values at $T_r=0$. The transformation from the lateral states of the Cartesian coordinates model, $[x(t),y(t),\theta(t)]^T$, to the lateral states of the path coordinates model, 
$[s(t),e_{cg}(t),\theta_e(t)]^T$, is elucidated as follows (\citep{reiter2023frenet}):

\begin{equation}
\begin{bmatrix}
{s}(t) \\
{e}_{\text{cg}}(t) \\
{\theta}_{e}(t)
\end{bmatrix}
=
\begin{bmatrix}
{s}(p_{\text{ref}}^*(t)) \\
(p_{\text{veh}}(t) - p_{\text{ref}}^*(t))^Te_{n}(t) \\
{\theta}_{ref}^*(t) - \theta(t)
\end{bmatrix}
\end{equation}
where at each time instant $t$, $p_{\text{ref}}^*(t) = [x_{\text{ref}}^*(t), y_{\text{ref}}^*(t)]^T$ represents the Cartesian coordinates of the point on the reference path (typically the centerline of a lane or road) that is nearest to the vehicle's C.G.. Here, $x_{\text{ref}}^*(t)$ and $y_{\text{ref}}^*(t)$ denote the respective Cartesian coordinates of this point. On the other hand, $p_{\text{veh}}(t) = [x(t), y(t)]^T$ denotes the Cartesian coordinates of the vehicle's C.G. at time $t$. The term $e_n(t)$ signifies the unit normal vector of the reference path at the location $p_{\text{ref}}^*(t)$. Lastly, $\theta_{\text{ref}}^*(t)$ represents the angle of the reference path's tangent line at $p_{\text{ref}}^*(t)$, relative to the $X$-axis of the Cartesian coordinate system.

\begin{table}[h]
\centering
\caption{Experiment 4 parameters}
\label{experiment 4 parameters}
\begin{tabular}{l l || l l}
\hline
parameter & value & parameter & value \\ \hline
$x^0_{1}$ & $0 (m)$ & $x^0_{2}$ & $0 (m)$\\ 
$y^0_{1}$ & $-2 (m)$ & $y^0_{2}$ & $2 (m)$\\ 
$\theta^0_{1}$ & $0 (rad)$ & $\theta^0_{2}$ & $0 (rad)$\\
$v^0_{1}$ & $8 (m/s)$ & $v^0_{2}$ & $5 (m/s)$\\
$\delta^0_{1}$ & $0.02 (rad)$ & $\delta^0_{2}$ & $0.02 (rad)$\\
$a^0_{1}$ & $0 (m/s^2)$ & $a^0_{2}$ & $0 (m/s^2)$\\
$L_1$ & $2 (m)$ & $L_2$ & $2 (m)$\\
$r_1$ & $1.3 (m)$ & $r_2$ & $1.3 (m)$\\
$r_{ob}$ & $1 (m)$ & $w$ & $8 (m)$\\
 $C_d$ & $0.25$ & $S_1$ & $2 (m^2)$\\ 
$m_1$ & $1500 (kg)$ & $r_{whl,1}$ & $0.25 (m)$\\
$f_{roll,1}$ & $0.015$ &
$T^0_{whl,1}$ & $1900 (Nm)$  \\
$\rho$ & $1.2 (kg/m^3)$ \\

\hline
\end{tabular}
\end{table}

A 20-second simulation has been executed for the experiment. Vehicle 2 maintains its lane position at a constant velocity, hence its movement is described by a constant steering angle, $\delta_2(T_r)=\delta_2^0$, and zero acceleration $a_2(T_r)=0$. On the other hand, the control inputs for Vehicle 1 are delineated as piece-wise constant functions as follows:

\begin{equation}
\delta_1(T_r) = 
\begin{cases} 
0.02 \text{ rad} & \text{if } 0 \leq T_r < 5 \\
0.029 \text{ rad} & \text{if } 5 \leq T_r < 10 \\
-0.004 \text{ rad} & \text{if } 10 \leq T_r < 15 \\
0.022 \text{ rad} & \text{otherwise}
\end{cases}
\end{equation}

\begin{equation}
T_{whl,1}(T_r) = 
\begin{cases} 
1900\ Nm & \text{if } 0 \leq T_r < 5 \\
1600\ Nm & \text{if } 5 \leq T_r < 10 \\
1850\ Nm & \text{if } 10 \leq T_r < 15 \\
1980\ Nm & \text{otherwise}
\end{cases}
\end{equation}

\begin{figure}[b!]
    \centering
    \setlength{\abovecaptionskip}{0pt}
    \subcaptionbox{}
    {\includegraphics[width=0.4\textwidth]{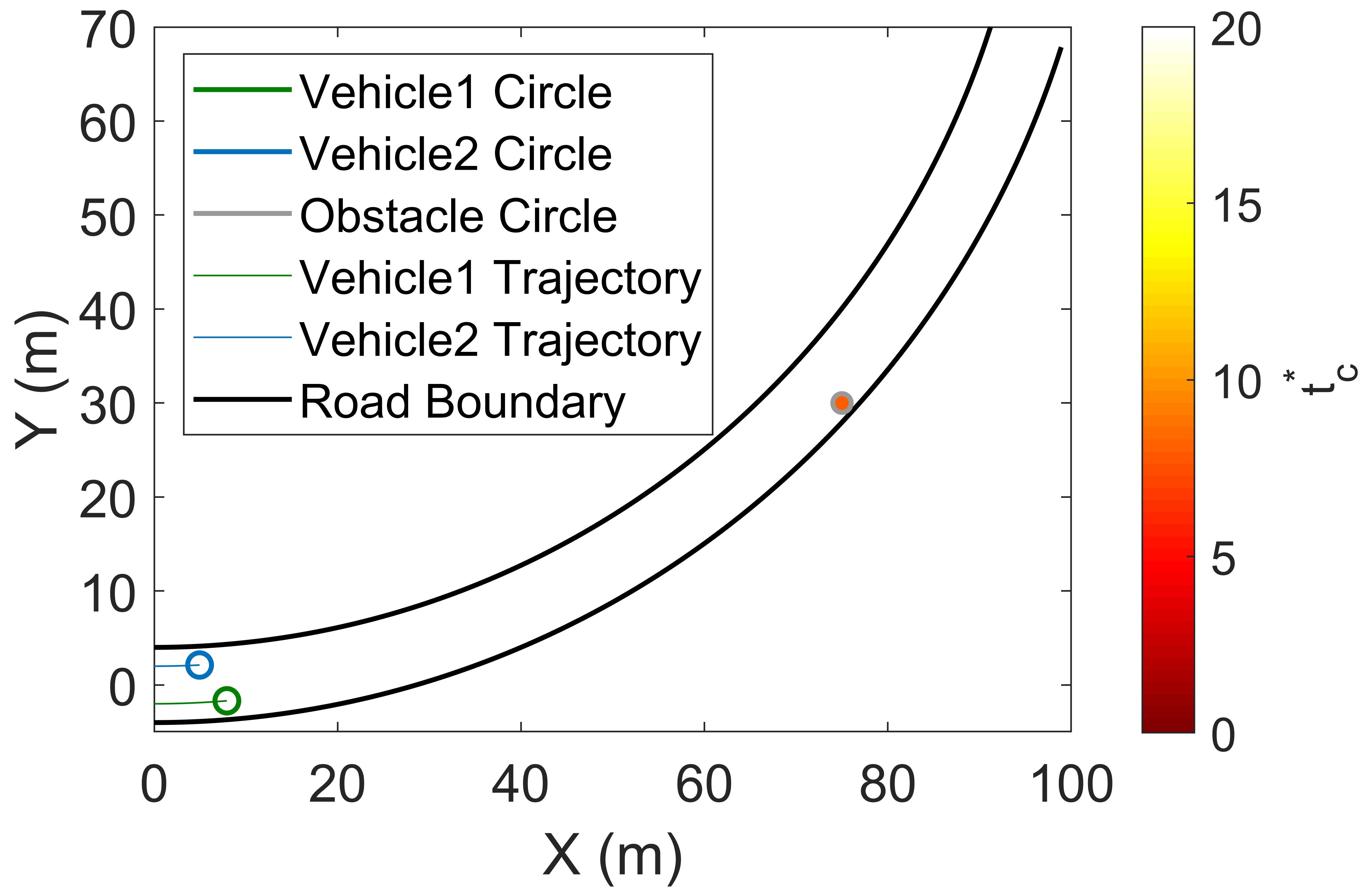}}
    \subcaptionbox{}
    {\includegraphics[width=0.4\textwidth]{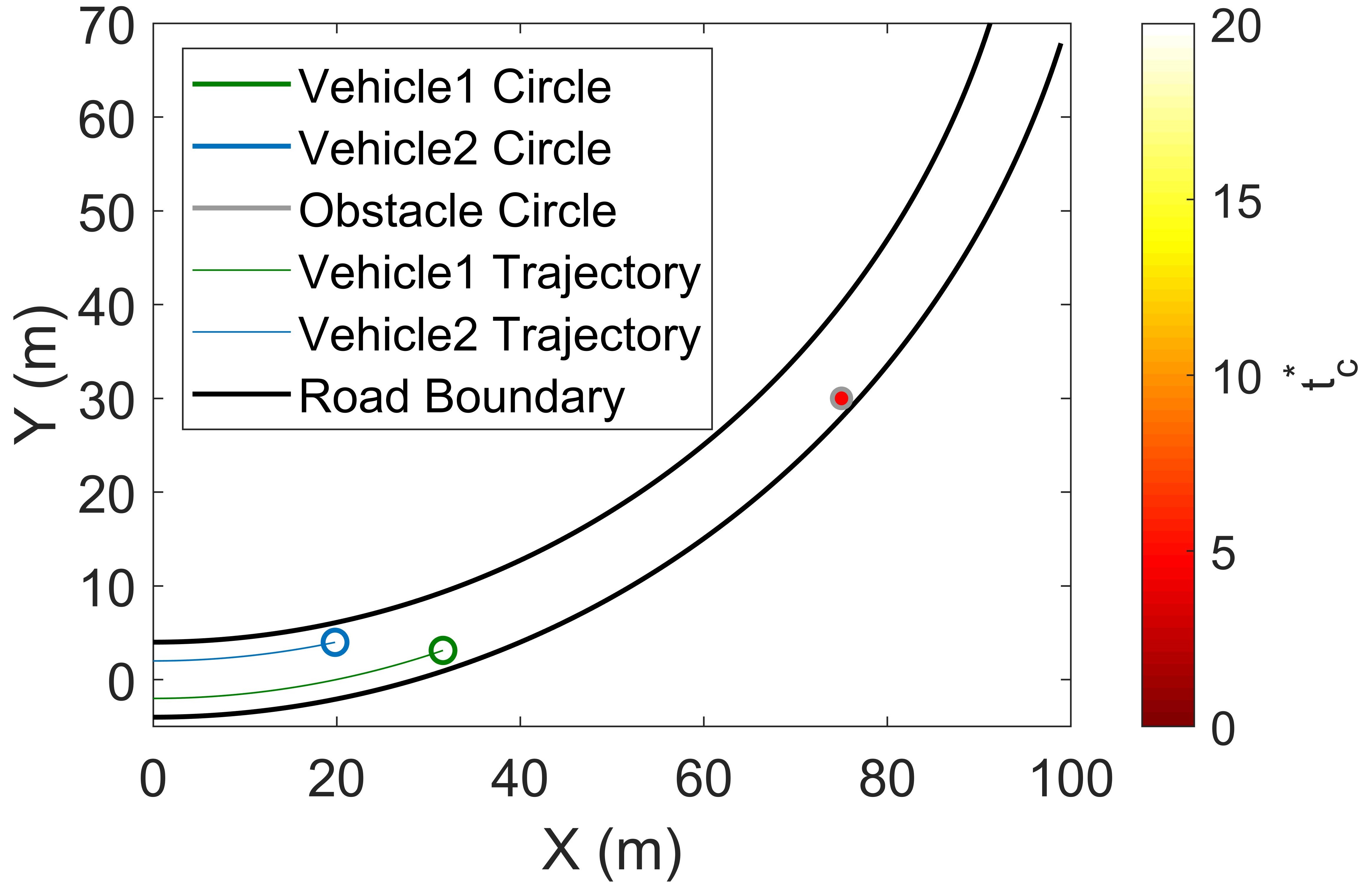}}
    \subcaptionbox{}{\includegraphics[width=0.4\textwidth]{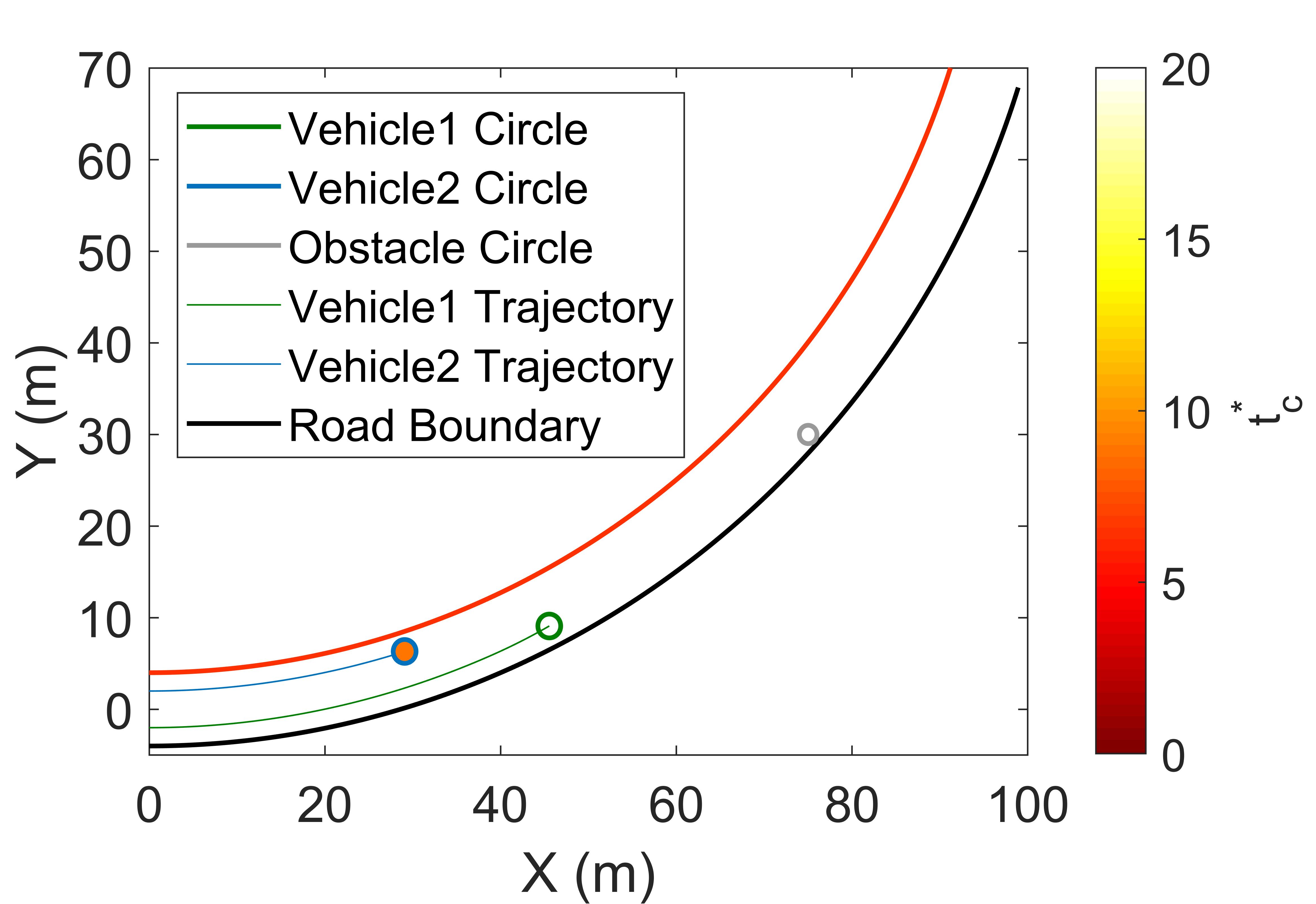}}
    \subcaptionbox{}
    {\includegraphics[width=0.4\textwidth]{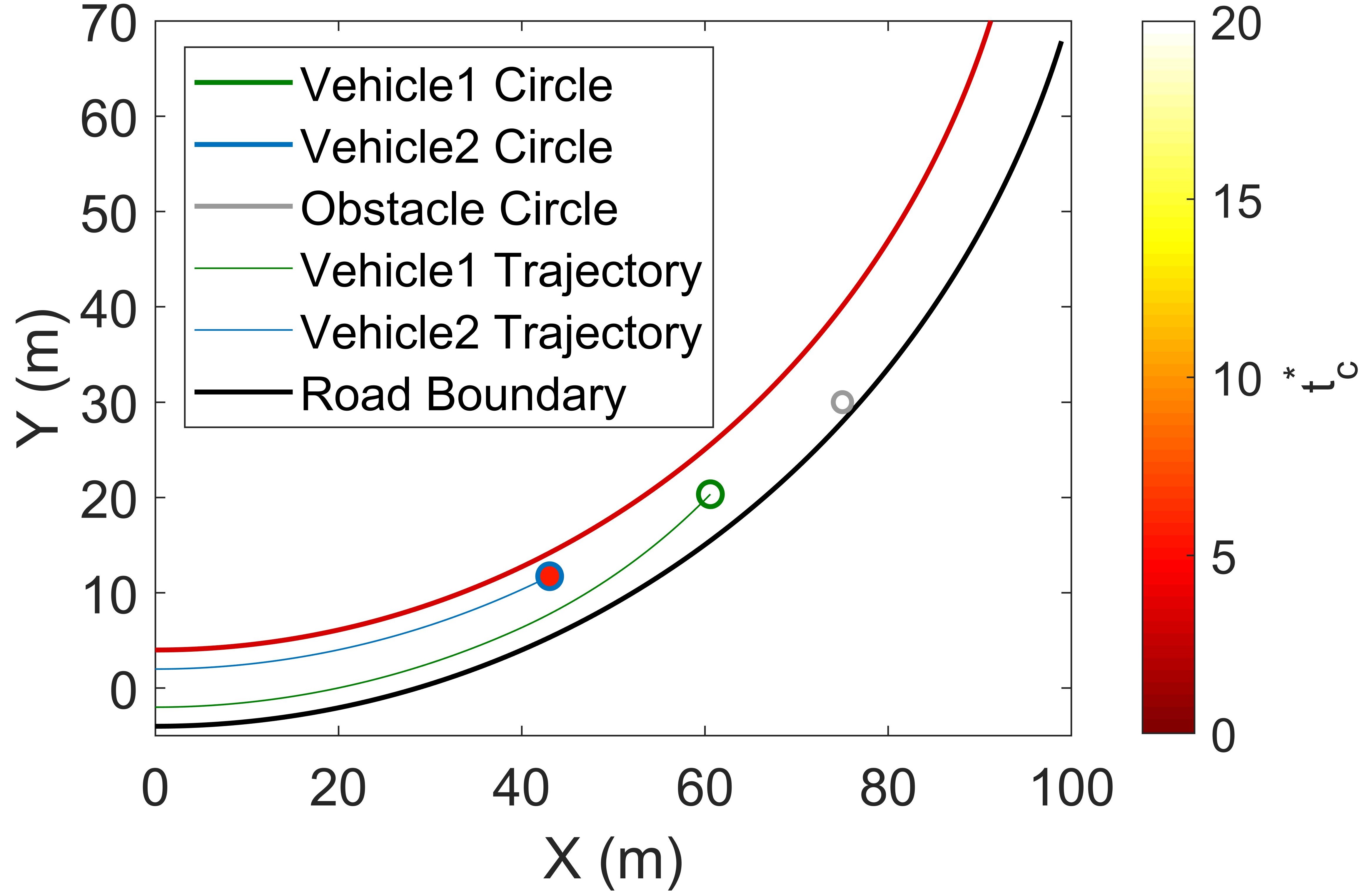}}
    \subcaptionbox{}{\includegraphics[width=0.4\textwidth]{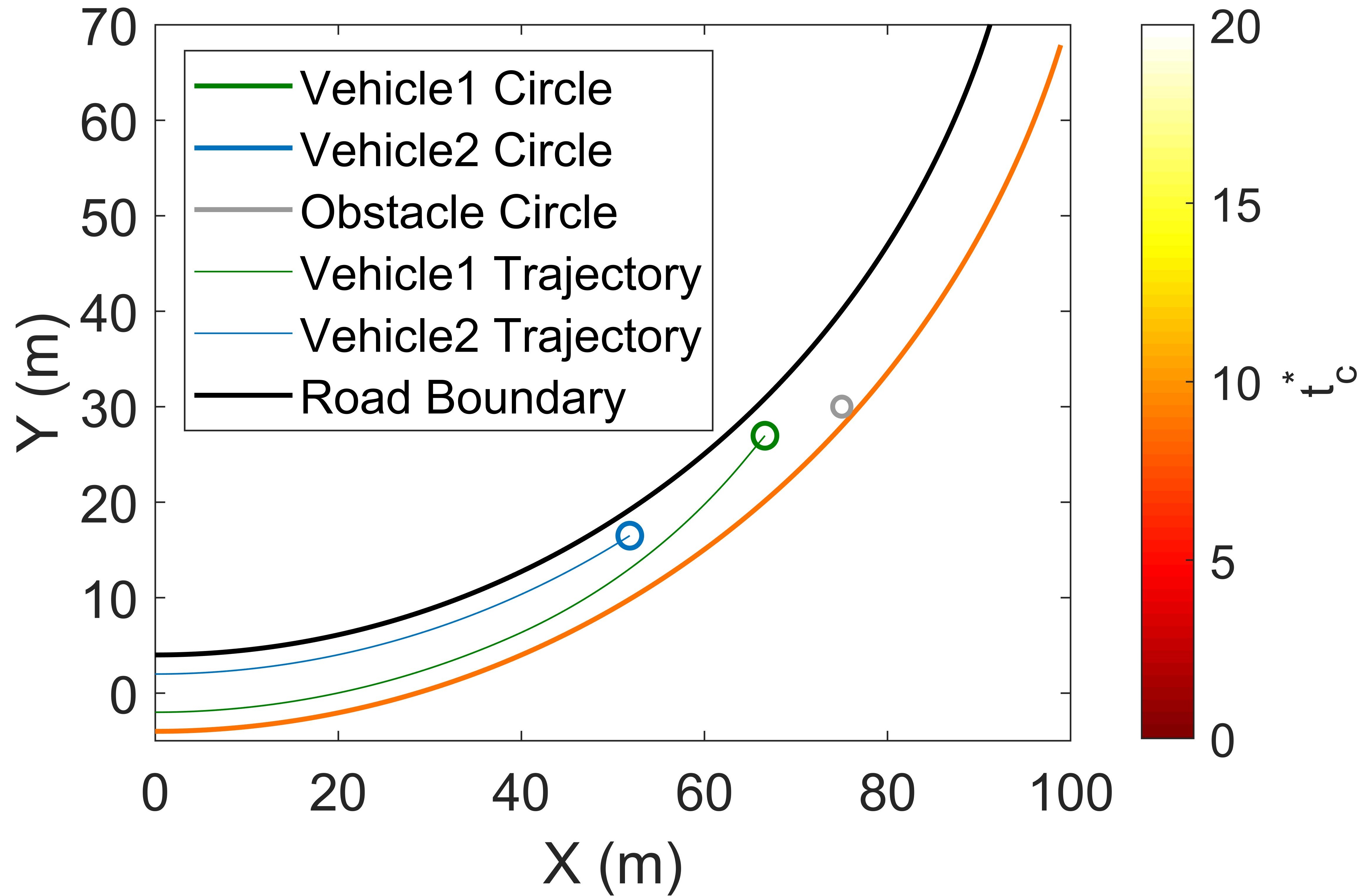}}
    \subcaptionbox{}
    {\includegraphics[width=0.4\textwidth]{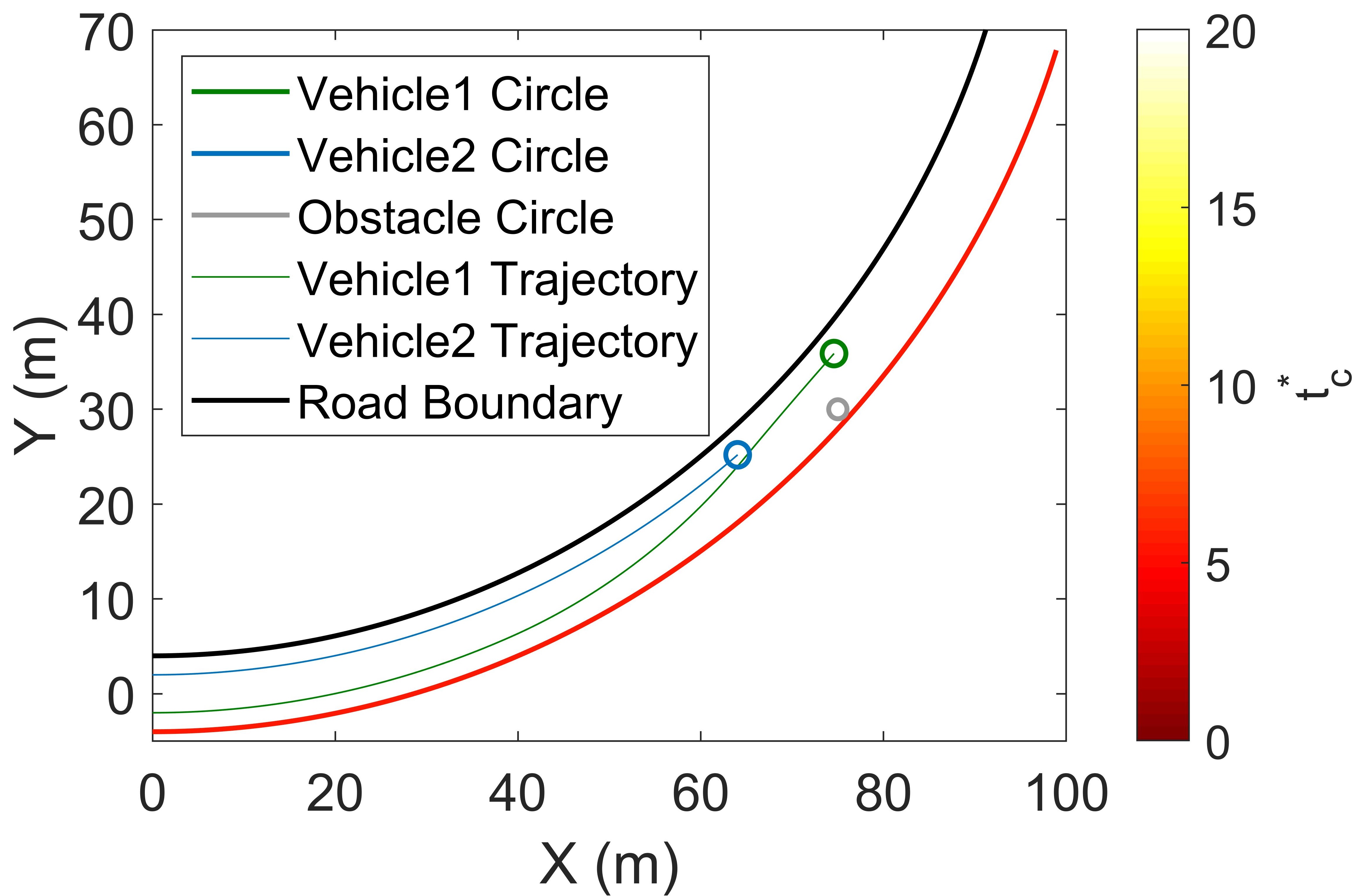}}
    \subcaptionbox{}{\includegraphics[width=0.4\textwidth]{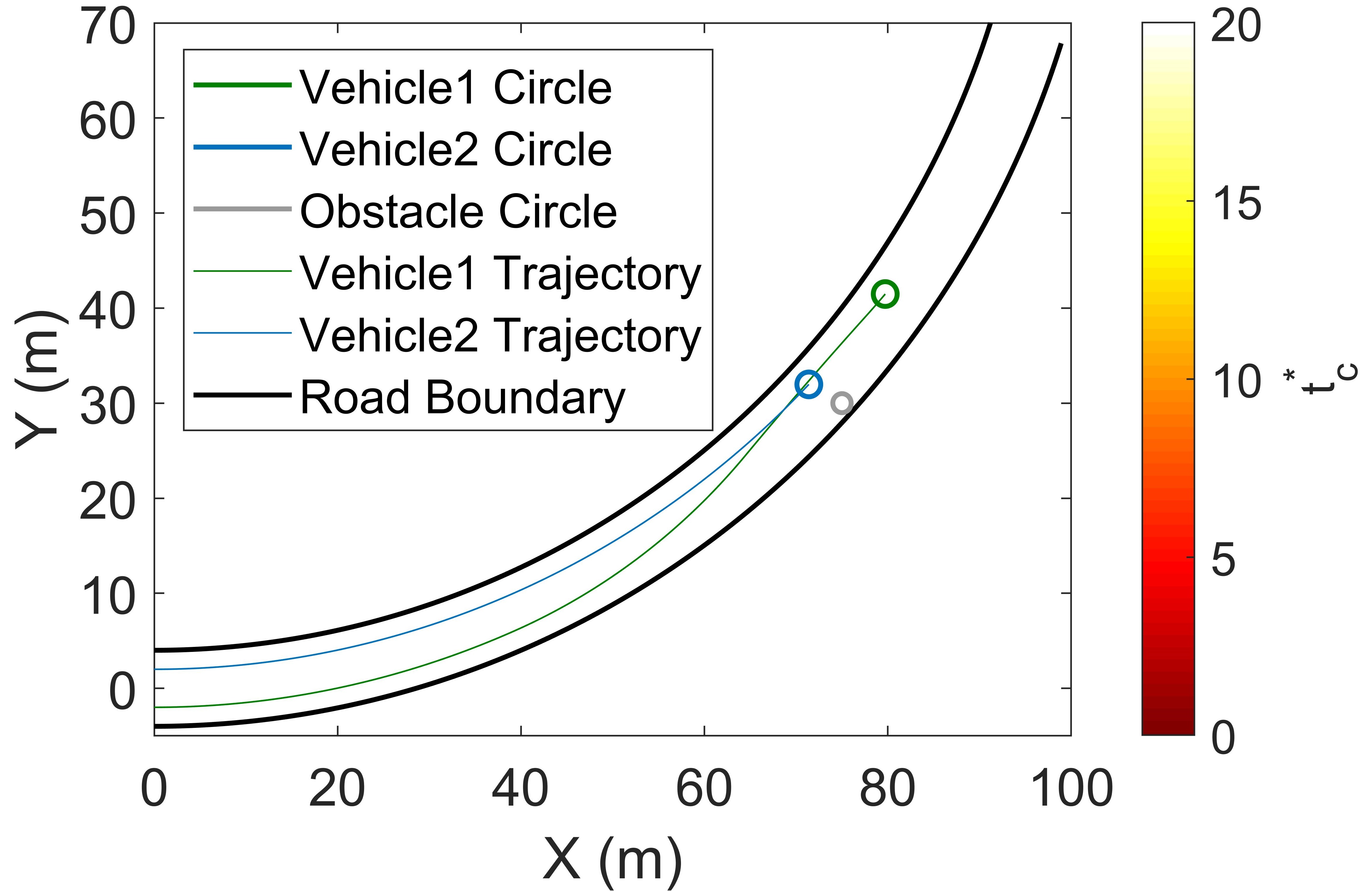}}
    \subcaptionbox{}{\includegraphics[width=0.4\textwidth]{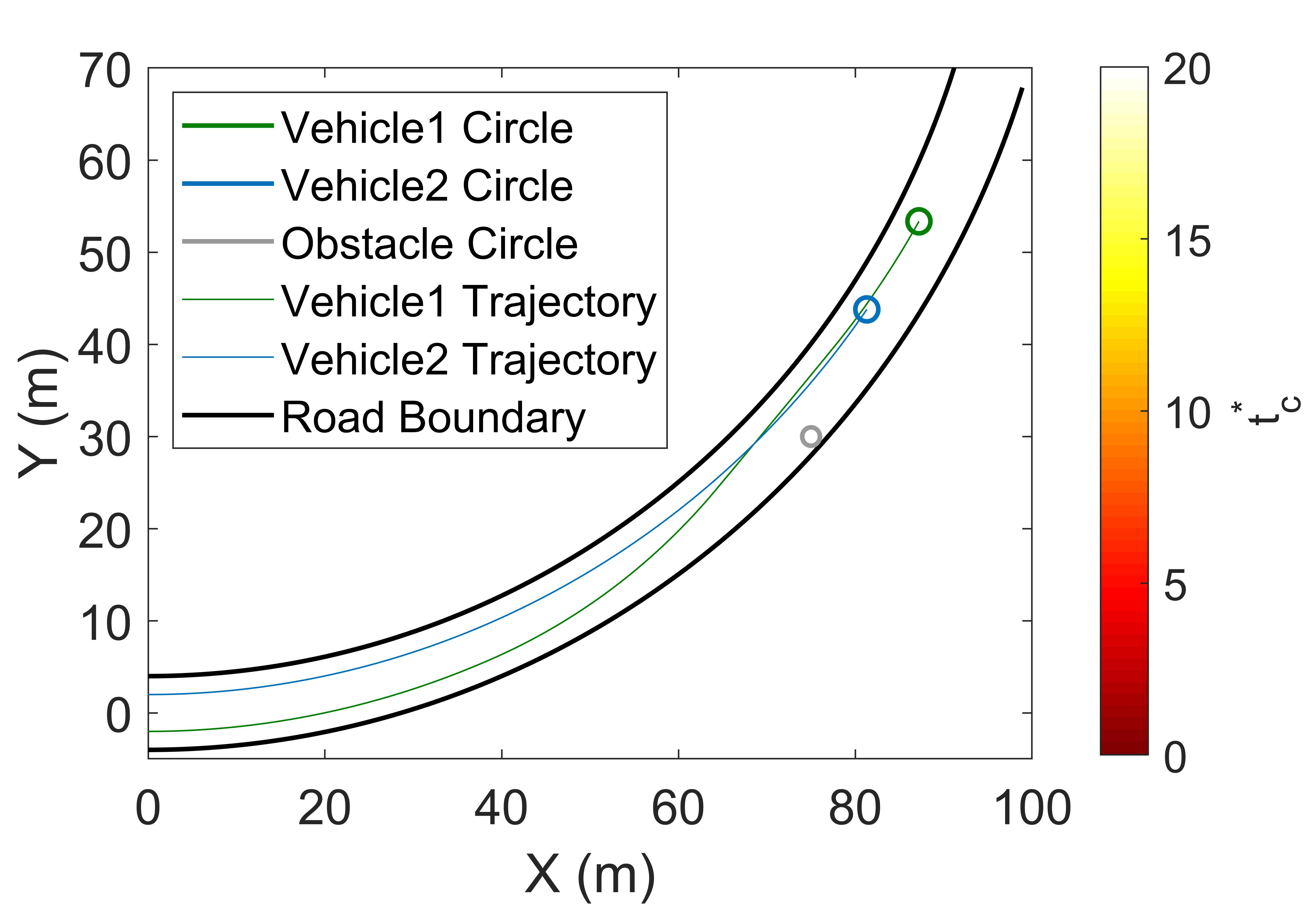}}
    \caption{Snapshots of experiment 4 with $t_c^*$ heat-map: (a) $T_r=1s$; (b) $T_r=4s$; (c)$T_r=6s$; (d) $T_r=9s$; (e) $T_r=11s$; (f) $T_r=14s$; (g) $T_r=16s$; (h) $T_r=19s$}
    \label{heat-map screenshots}
\end{figure}

In the initial 5 seconds, Vehicle 1 maintains its lane with a steering angle of $\delta_1(t)=0.02 rad$ and applies a wheel torque of $T_{whl,1}=1900 Nm$ , resulting in a gradual increase in speed. Upon detecting an obstacle at $T_r=5s$, Vehicle 1 sharply reduces the wheel torque to $T_{whl,1}=1600 Nm$ to decelerate, while manipulating the steering angle to $\delta_1=0.029 rad$ for a lane-change. At $T_r=10s$, Vehicle 1 adjusts its steering angle to $\delta_1=-0.004 rad$ to converge into the new lane, while the wheel torque is increased to $1850 Nm$ which leads to a marginal deceleration. Finally, at $T_r=15s$, having successfully navigated past the obstacle, Vehicle 1 adopts a steering angle of $\delta_1=0.022 rad$ to stabilize within the new lane and increases wheel torque to $T_{whl,1}=1980 Nm$ to accelerate.

Throughout the simulation, snapshots are captured to vividly demonstrate the implementation of the proposed framework, as illustrated in Fig. \ref{heat-map screenshots}. The subfigures, labeled Fig. \ref{heat-map screenshots} (a)-(h), present these snapshots at the designated time  instances $T_r=1s$, $T_r=4s$, $T_r=6s$, $T_r=9s$, $T_r=11s$, $T_r=14s$, $T_r=16s$, and $T_r=19s$, respectively. Each subfigure outlines the bounding circles of Vehicle 1, Vehicle 2, and the obstacle, alongside the delineation of the road's edges. Vehicle 1 is regarded as the ego vehicle, and its potential collision risks are carefully assessed. At each time step, $t_c^*$ is evaluated for the collisions involving Vehicle 1: with Vehicle 2, with the obstacle, and with the road boundaries. Heat-maps are then generated within each subfigure to represent the values of $t_c^*$. If a collision risk is present between Vehicle 1 and Vehicle 2, Vehicle 2's bounding circle is shaded according to the corresponding $t_c^*$, similarly, collision risks between Vehicle 1 and the obstacle are indicated by filling the obstacle's bounding circle with the relevant $t_c^*$ color. Road boundaries at risk of collision with Vehicle 1 are likewise highlighted with the color indicative of their respective $t_c^*$.

Fig. \ref{heat-map screenshots}(a)-(b) captures Vehicle 1 at $T_r=1s$ and $T_r=4s$, proceeding along its original lane, with a collision risk pertaining only to the obstacle. From $T_r=1s$ to $T_r=4s$, the vehicle1-to-obstacle collision $t_c^*$ value reduces from $9.05s$ to $6.05s$, resulting in a darkening of the filling color of the obstacle. As depicted in Fig. \ref{heat-map screenshots}(c), at $T_r=6s$, Vehicle 1 has been engaged in the lane-change for one second. This shift replaces the previous collision risk with a new one involving the upper road boundary due to lateral movement, and another involving Vehicle 2 as a result of deceleration, resulting in $t_c^*$ values of $7.41s$ for the road boundary and $9.68s$ for Vehicle 2. In Fig. \ref{heat-map screenshots}(d), at $T_r=9s$, the lane-change is underway for four seconds. The $t_c^*$ values for Vehicle 1's potential collisions with the road and Vehicle 2 have decreased to $4.40s$ and $6.72s$, respectively, which is reflected by the intensified coloration of the upper road boundary and Vehicle 2 compared to Fig. \ref{heat-map screenshots}(c). Fig. \ref{heat-map screenshots}(e), at $T_r=11s$, illustrates Vehicle 1 adjusting its steering to merge into the new lane, making the lower road boundary the sole collision risk with $t_c^*=9.19s$. As shown in Fig. \ref{heat-map screenshots}(f), at $T_r=14s$, after steering in the opposite direction for four seconds, the $t_c^*$ for a collision with the road is reduced to $6.19s$, darkening the color of the lower road boundary, compared to Fig. \ref{heat-map screenshots}(f). An observation is that the lower boundary in Fig. \ref{heat-map screenshots}(e)-(f) has relatively lighter color comparing to the upper boundary in Fig. \ref{heat-map screenshots}(c)-(d), this is because when Vehicle 1 merges into the left lane, a larger steering wheel angle is used for changing lanes, if the same steering wheel angle remains, Vehicle 1 will collide with the upper boundary within a smaller time duration (hence smaller $t_c^*$). On the other hand, when trying to stabilize into the left lane, a much smaller steering wheel angle to the right is needed, hence result in larger $t_c^*$ for the lower boundary. Finally, Figure \ref{heat-map screenshots}(g)-(h) at $T_r=16s$ and $T_r=19s$, show that Vehicle 1 have navigated past the obstacle, hence the collision risk with the obstacle no longer exists. Furthermore, Vehicle 1 accelerates by a large wheel torque, with the same torque remained, Vehicle 1 will have a larger speed than Vehicle 2 before Vehicle 2 catches up with it, hence there’s no risk of colliding with Vehicle 2. Moreover, Vehicle 1 stabilizes in the new lane and travels along the lane, hence no risk of colliding with the road boundaries exists.

\section{Conclusions \label{sec6}}

This paper has presented a comprehensive framework for SSMs, applicable across a diverse range of highway geometries. The framework adeptly captures vehicle dynamics with varying levels of dimension and fidelity, proving highly adaptable to dynamic and real-world traffic scenarios. By employing a generic vehicle movement model, the framework can estimate future vehicle trajectory evolution under different complexities and dimensions.
The framework introduces a generic mathematical condition to represent potential collisions, based on the spatial overlap of a vehicle with other objects. This enables the identification of collision risks at non-negative time instances, with the smallest time instance indicating the imminent time to a collision. The versatility of the framework is showcased by its capacity to incorporate measures equivalent to traditional 1-D time-based SSMs and deceleration-based SSMs. Additionally, it demonstrates the extraction of traditional energy-based SSMs. Furthermore, the framework introduces more advanced, higher-dimensional, and higher-fidelity SSMs.

The accuracy of the framework  in scenarios involving linearization of the vehicle movement model, comparing to high-accuracy numerical solutions, has been validated through simulation experiments. These experiments have shown high precision in estimating future vehicle trajectory evolution and the remaining time before collisions, affirming the framework's reliability and applicability. The necessity and reliability of higher-dimensional, higher-fidelity SSMs are further highlighted through a comparison of conventional 1D SSMs and extended 3D SSMs, it is also observed that, under the assumption that the vehicle model is accurate, and for the scenario that the wheel torque and steering wheel angle remains invariant for a small duration, conventional 1D SSMs can yield highly
inaccurate and unreliable results when assessing collision proximity and substantially misjudge the count of conflicts with a TTC less than a predefined threshold; the count of TTCs below 1.5s could be off by as much
as 20\% (in the scenarios analyzed in this paper). Moreover, the framework's practical implementation is exemplified through a real-world scenario application, highlighting its capability in actively analyzing potential collisions in dynamic traffic conditions. This demonstration of the framework's effectiveness in real-world scenarios underscores its potential as a valuable tool for traffic safety analysis and management.

If the vehicle model embodies intrinsic nonlinearity and the linearization brings in too much error, and if practitioners would like a highly accurate solution to the nonlinear vehicle model, they could resolve this by utilizing the numerical methods, as shown by Subsection 4.3. As discussed in Subsection \ref{sec4.4}, there are tradeoffs between analytical solutions and numerical solutions, which practitioners should pay attention to and choose the suitable one.

In summary, the proposed framework represents a significant advancement in the field of active safety analysis. Its flexibility, accuracy, and practical applicability make it a promising approach for future research and application in the domain of traffic safety and management. One particularly intriguing application lies in integrating the proposed framework into digital twin environments. By importing data from datasets such as NGSIM, Waymo, and Wejo into digital twins, our framework can facilitate the analysis of safety aspects concerning human-driven vehicles, autonomous vehicles, and connected vehicles. Future work involves incorporating understandings of specified vehicle/driver behaviors into the proposed framework for behavior embedded safety analysis.\\

\subparagraph{\textbf{Acknowledgement:}}
This research is funded by Federal Highway Administration Exploratory Advanced Research 693JJ323C000010.
\bigskip
\subparagraph{\textbf{Disclaimer:}}
The results presented in this document do not necessarily reflect those from the Federal Highway Administration.

\printcredits

\bibliographystyle{cas-model2-names}

\bibliography{main}


\end{document}